\documentclass[12pt,aps,superscriptaddress,noshowpacs,nofootinbib]{revtex4}
\usepackage{epsfig}
\usepackage{graphicx}
\usepackage{amssymb,amsmath}

\begin{document}

\title{\bf 
Measurement of the $e^+e^-\to\pi^+\pi^-\pi^0$ cross section in the
energy region from 0.56 to 1.1 GeV with the SND detector}

\author{M.~N.~Achasov} 
\affiliation{Budker Institute of Nuclear Physics, SB RAS, Novosibirsk, 630090, Russia} 
\affiliation{Novosibirsk State University, Novosibirsk, 630090, Russia} 
\author{A.~E.~Alizzi}
\affiliation{Budker Institute of Nuclear Physics, SB RAS, Novosibirsk, 630090, Russia}
\affiliation{Novosibirsk State University, Novosibirsk, 630090, Russia}
\author{A.~Yu.~Barnyakov} 
\affiliation{Budker Institute of Nuclear Physics, SB RAS, Novosibirsk, 630090, Russia}
\author{K.~I.~Beloborodov}
\affiliation{Budker Institute of Nuclear Physics, SB RAS, Novosibirsk, 630090, Russia} 
\affiliation{Novosibirsk State University, Novosibirsk, 630090, Russia} 
\author{A.~V.~Berdyugin} 
\affiliation{Budker Institute of Nuclear Physics, SB RAS, Novosibirsk, 630090, Russia} 
\affiliation{Novosibirsk State University, Novosibirsk, 630090, Russia} 
\author{D.~E.~Berkaev} 
\affiliation{Budker Institute of Nuclear Physics, SB RAS, Novosibirsk, 630090, Russia} 
\affiliation{Novosibirsk State University, Novosibirsk, 630090, Russia} 
\author{A.~G.~Bogdanchikov} 
\affiliation{Budker Institute of Nuclear Physics, SB RAS, Novosibirsk, 630090, Russia} 
\author{A.~A.~Botov} 
\affiliation{Budker Institute of Nuclear Physics, SB RAS, Novosibirsk, 630090, Russia} 
\author{V.~S.~Denisov} 
\affiliation{Budker Institute of Nuclear Physics, SB RAS, Novosibirsk, 630090, Russia} 
\author{T.~V.~Dimova} 
\affiliation{Budker Institute of Nuclear Physics, SB RAS, Novosibirsk, 630090, Russia} 
\affiliation{Novosibirsk State University, Novosibirsk, 630090, Russia} 
\author{V.~P.~Druzhinin} 
\email{druzhinin@inp.nsk.su}
\affiliation{Budker Institute of Nuclear Physics, SB RAS, Novosibirsk, 630090, Russia} 
\affiliation{Novosibirsk State University, Novosibirsk, 630090, Russia} 
\author{R.~A.~Efremov}
\affiliation{Budker Institute of Nuclear Physics, SB RAS, Novosibirsk, 630090, Russia} 
\affiliation{Novosibirsk State Technical University, Novosibirsk,
630073, Russia}
\author{E.~A.~Eminov}
\affiliation{Budker Institute of Nuclear Physics, SB RAS, Novosibirsk, 630090, Russia} 
\author{L.~B.~Fomin}
\affiliation{Budker Institute of Nuclear Physics, SB RAS, Novosibirsk, 630090, Russia} 
\author{L.~V.~Kardapoltsev}
\affiliation{Budker Institute of Nuclear Physics, SB RAS, Novosibirsk, 630090, Russia} 
\affiliation{Novosibirsk State University, Novosibirsk, 630090, Russia} 
\author{A.~A.~Kattsin}
\affiliation{Budker Institute of Nuclear Physics, SB RAS, Novosibirsk, 630090, Russia} 
\author{A.~G.~Kharlamov} 
\affiliation{Budker Institute of Nuclear Physics, SB RAS, Novosibirsk, 630090, Russia} 
\affiliation{Novosibirsk State University, Novosibirsk, 630090, Russia} 
\author{I.~A.~Koop}
\affiliation{Budker Institute of Nuclear Physics, SB RAS, Novosibirsk, 630090, Russia} 
\affiliation{Novosibirsk State University, Novosibirsk, 630090, Russia} 
\author{A.~A.~Korol} 
\affiliation{Budker Institute of Nuclear Physics, SB RAS, Novosibirsk, 630090, Russia} 
\affiliation{Novosibirsk State University, Novosibirsk, 630090, Russia} 
\author{D.~P.~Kovrizhin} 
\affiliation{Budker Institute of Nuclear Physics, SB RAS, Novosibirsk, 630090, Russia} 
\author{A.~S.~Kupich} 
\affiliation{Budker Institute of Nuclear Physics, SB RAS, Novosibirsk, 630090, Russia} 
\affiliation{Novosibirsk State University, Novosibirsk, 630090, Russia} 
\author{A.~P.~Kryukov} 
\affiliation{Budker Institute of Nuclear Physics, SB RAS, Novosibirsk, 630090, Russia} 
\author{N.~A.~Melnikova} 
\affiliation{Budker Institute of Nuclear Physics, SB RAS, Novosibirsk, 630090, Russia} 
\author{N.~Yu.~Muchnoi} 
\affiliation{Budker Institute of Nuclear Physics, SB RAS, Novosibirsk, 630090, Russia} 
\affiliation{Novosibirsk State University, Novosibirsk, 630090, Russia} 
\author{A.~E.~Obrazovsky} 
\affiliation{Budker Institute of Nuclear Physics, SB RAS, Novosibirsk, 630090, Russia} 
\author{A.~A.~Oorzhak}
\affiliation{Budker Institute of Nuclear Physics, SB RAS, Novosibirsk, 630090, Russia}
\affiliation{Novosibirsk State University, Novosibirsk, 630090, Russia}
\author{I.~V.~Ovtin}
\affiliation{Budker Institute of Nuclear Physics, SB RAS, Novosibirsk, 630090, Russia}
\affiliation{Novosibirsk State University, Novosibirsk, 630090, Russia}
\author{E.~V.~Pakhtusova} 
\affiliation{Budker Institute of Nuclear Physics, SB RAS, Novosibirsk, 630090, Russia} 
\author{E.~A.~Perevedentsev}
\affiliation{Budker Institute of Nuclear Physics, SB RAS, Novosibirsk, 630090, Russia}
\affiliation{Novosibirsk State University, Novosibirsk, 630090, Russia}
\author{I.~A.~Polomoshnov}
\affiliation{Budker Institute of Nuclear Physics, SB RAS, Novosibirsk, 630090, Russia}
\affiliation{Novosibirsk State University, Novosibirsk, 630090, Russia}
\author{K.~V.~Pugachev} 
\affiliation{Budker Institute of Nuclear Physics, SB RAS, Novosibirsk, 630090, Russia} 
\affiliation{Novosibirsk State University, Novosibirsk, 630090, Russia} 
\author{Yu.~A.~Rogovsky} 
\affiliation{Budker Institute of Nuclear Physics, SB RAS, Novosibirsk, 630090, Russia} 
\affiliation{Novosibirsk State University, Novosibirsk, 630090, Russia} 
\author{V.~A.~Romanov} 
\affiliation{Budker Institute of Nuclear Physics, SB RAS, Novosibirsk, 630090, Russia} 
\affiliation{Novosibirsk State University, Novosibirsk, 630090, Russia} 
\author{S.~I.~Serednyakov} 
\affiliation{Budker Institute of Nuclear Physics, SB RAS, Novosibirsk, 630090, Russia} 
\affiliation{Novosibirsk State University, Novosibirsk, 630090, Russia} 
\author{Yu.~M.~Shatunov} 
\affiliation{Budker Institute of Nuclear Physics, SB RAS, Novosibirsk, 630090, Russia} 
\author{D.~A.~Shtol} 
\affiliation{Budker Institute of Nuclear Physics, SB RAS, Novosibirsk, 630090, Russia} 
\author{Z.~K.~Silagadze} 
\affiliation{Budker Institute of Nuclear Physics, SB RAS, Novosibirsk, 630090, Russia} 
\affiliation{Novosibirsk State University, Novosibirsk, 630090, Russia} 
\author{K.~D.~Sungurov}
\affiliation{Budker Institute of Nuclear Physics, SB RAS, Novosibirsk, 630090, Russia} 
\affiliation{Novosibirsk State University, Novosibirsk, 630090, Russia} 
\author{M.~V.~Timoshenko}
\affiliation{Budker Institute of Nuclear Physics, SB RAS, Novosibirsk, 630090, Russia} 
\author{I.~K.~Surin} 
\affiliation{Budker Institute of Nuclear Physics, SB RAS, Novosibirsk, 630090, Russia} 
\author{Yu.~V.~Usov} 
\affiliation{Budker Institute of Nuclear Physics, SB RAS, Novosibirsk, 630090, Russia} 
\author{I.~M.~Zemlyansky} 
\affiliation{Budker Institute of Nuclear Physics, SB RAS, Novosibirsk, 630090, Russia} 
\author{V.~N.~Zhabin} 
\affiliation{Budker Institute of Nuclear Physics, SB RAS, Novosibirsk, 630090, Russia} 
\affiliation{Novosibirsk State University, Novosibirsk, 630090, Russia} 
\author{Yu.~M.~Zharinov} 
\affiliation{Budker Institute of Nuclear Physics, SB RAS, Novosibirsk, 630090, Russia}
\author{V.~V.~Zhulanov}
\affiliation{Budker Institute of Nuclear Physics, SB RAS, Novosibirsk, 630090, Russia}
\affiliation{Novosibirsk State University, Novosibirsk, 630090, Russia}
\author{P.~V.~Zhulanova}
\affiliation{Budker Institute of Nuclear Physics, SB RAS, Novosibirsk, 630090, Russia}
\affiliation{Novosibirsk State University, Novosibirsk, 630090, Russia}

\begin{abstract} 
The precise measurement of the $e^+e^-\to\pi^+\pi^-\pi^0$ cross section
is performed in the center-of-mass energy range $E = 560$--1100 MeV
using a data sample of 66 pb$^{-1}$ collected in the experiment with
the SND detector at the VEPP-2000 $e^+e^-$ collider. The systematic
uncertainty of the cross section measurement is 0.9\% at the maximum of the
$\omega$ resonance and 1.2\% at the maximum of the $\phi$ resonance.
The leading-order hadronic contribution to the muon magnetic anomaly calculated
using the $e^+e^-\to\pi^+\pi^-\pi^0$ cross section measured by SND
from 0.62 to 1.975 GeV is $(45.95\pm0.06\pm 0.47)\times 10^{-10}$.
From the fit to the cross section data with the vector meson dominance 
model, the parameters of the $\omega$, $\rho$, and $\phi$ resonances
are obtained. The obtained values of ${\cal B}(\omega\to e^+e^-){\cal
B}(\omega\to3\pi)$, mass and width of the $\omega$ meson, 
and ${\cal B}(\rho\to 3\pi)$
have accuracy better than the current world average values.
\end{abstract}

\maketitle
\setcounter{footnote}{0}

\section{Introduction}
This article is devoted to the measurement of the $e^+e^- \to
\pi^+\pi^-\pi^0$ cross section in the center-of-mass energy range of
$0.56<E<1.1$~GeV in the experiment with the SND detector at the
VEPP-2000 collider~\cite{vepp2000}. In this energy range, the cross section is
dominated by the contributions of the $\omega$ and $\phi$ 
resonances.\footnote{Throughout this paper, the notation $\rho$, $\omega$, and
$\phi$ is used for $\rho(770)$, $\omega(782)$, and $\phi(1020)$.}
From the approximation of the cross section in the vector meson
dominance  (VMD) model, the parameters of these resonances, as well as
the branching fraction of the isospin-violating decay
$\rho\to\pi^+\pi^-\pi^0$, can be extracted. Below 1 GeV, the $e^+e^-
\to \pi^+\pi^-\pi^0$ cross section is the second largest cross section
after $e^+e^- \to \pi^+\pi^-$. Therefore, its measurement is important
for the Standard Model calculation of the muon anomalous magnetic
moment (see, for example, Ref.~\cite{DHMZ}).

The most accurate measurements of the cross section for the $e^+e^-
\to \pi^+\pi^-\pi^0$ process in the energy range under study were
performed in the CMD-2~\cite{cmd-3pi-1,cmd-3pi-2} and
SND~\cite{snd-3pi-1,snd-3pi-2} experiments at the VEPP-2M $e^+e^-$
collider, as well as by the radiative return method in the
BABAR~\cite{BABAR-3pi} and Belle~II~\cite{belle-3pi} experiments. There
are significant discrepancies between the data from these experiments.
For example, in the region near the maximum of the $\omega$ resonance
the cross section measured by Belle~II is approximately 8\% larger
than the BABAR cross section, with declared systematic uncertainties
of 2.3\% and 1.3\%, respectively. Obviously, a new measurement of the
$e^+e^- \to \pi^+\pi^-\pi^0$ cross section with an accuracy of no
worse than 1.5\% is required to clarify the experimental situation.

\section{SND detector}
\begin{figure}
\centering
\includegraphics [width = 0.7\textwidth]{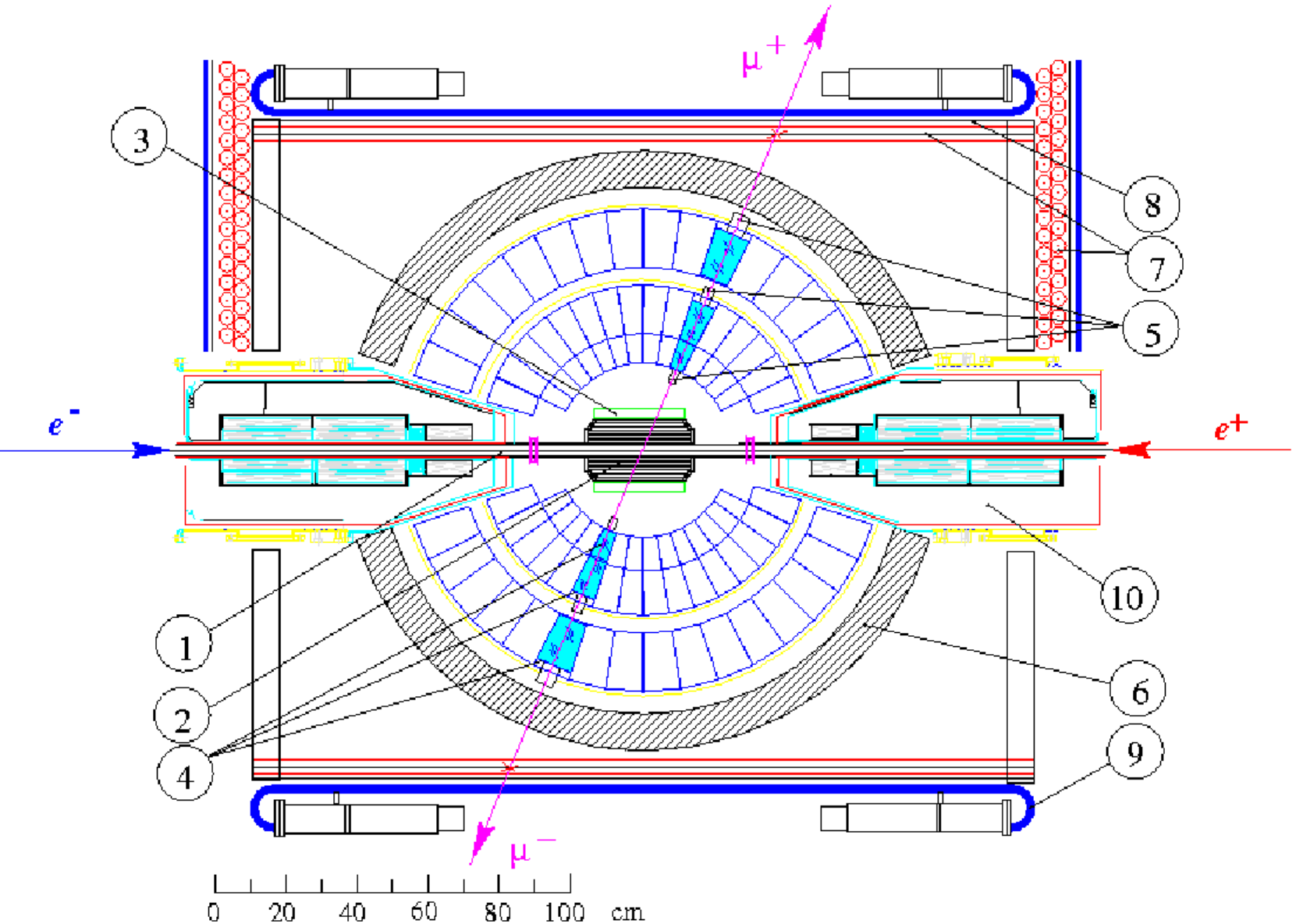}
\caption{SND detector, section along the beams: (1) beam pipe,
(2) tracking system, (3) aerogel Cherenkov counters, (4) NaI (Tl)
crystals, (5) vacuum phototriodes, (6) iron absorber, (7) proportional
tubes,
(8) iron absorber, (9) scintillation counters, (10) VEPP-2000 focusing
solenoids.}
\label{fig1}
\end{figure}

The Spherical Neutral Detector (SND)~\cite{SNDdet1,SNDdet2,SNDdet3,SNDdet4} 
(Fig.~\ref{fig1}) is a general-purpose non-magnetic detector operating at the 
VEPP-2000 $e^+e^-$ collider~\cite{vepp2000} with the c.m. energy $E = 0.32 -
2$ GeV. It consist of a tracking system, a system of aerogel threshold 
Cherenkov counters, an electromagnetic calorimeter, and a muon system.
The tracking system is a nine-layer drift chamber and a single-layer 
proportional chamber with cathode strip readout located in a common gas volume.
Its angular resolution is $0.45^\circ$ for azimuthal angle and 
$0.8^\circ$ for polar angle. The reconstructed charged track must contains at 
least 4 hits in the drift chamber. The tracking system solid angle
for charged particles originating from the collider interaction region is 94\%
of $4\pi$.  The Cherenkov counters are located around the tracking system.
They use aerogel with the refractive index of 1.05 allowing for the pion
identification at momenta below 436 MeV/$c$. This system covers 60\% of the
total solid angle. The three-layer spherical electromagnetic
calorimeter, based on  NaI (Tl) crystals, covers 95\% of the solid angle.
It is used, in particular, to identify photons, which are defined as
clusters of adjacent crystals in the calorimeter with energy depositions
exceeding 20 MeV and no associated charged tracks.
The calorimeter energy resolution for 
photons is $4.2\%/\sqrt[4]{E(\mbox{GeV})}$,
and the angular resolution is about 1.5$^{\circ}$. The muon system 
is located outside the calorimeter and consists of proportional tubes and 
scintillation counters. 

The Monte-Carlo (MC) event generator for the signal process $e^+e^- \to
\pi^+\pi^-\pi^0$ uses a model with an intermediate $\rho\pi$ state,
which works well at the $\omega$ and $\phi$ 
resonances~\cite{BESIII-om,KLOE-phi}. Generators for
signal and background hadronic ($e^+e^- \to \pi^+\pi^-\pi^0$, $K^+K^-$, 
$K_SK_L$, $\omega\pi^0$, $\eta\gamma$, $\pi^0\gamma$) events takes into 
account initial-state radiative corrections~\cite{radcor}.
In the generators, we assume that the energy carried by photons
emitted from the initial state, generated according to
Ref.~\cite{radcor}, belongs to a single photon. The angular distribution 
of this initial state radiation (ISR) photon is generated according to 
Ref.~\cite{BM}. The ﬁnal-state radiation (FSR) for the
signal process is simulated using the PHOTOS package~\cite{photos}.
The background $e^+e^- \to \pi^+\pi^-(\gamma)$ events are simulated with the
MCGPJ event generator~\cite{MCGPJ}. To simulate backgrounds from the QED 
processes $e^+e^-\to e^+e^-\gamma(\gamma)$ and $e^+e^-\to
\mu^+\mu^-\gamma(\gamma)$, the event generator
BabaYaga@NLO~\cite{BabaYaga} is used. Interactions of the particles 
produced in $e^{+}e^{-}$ annihilation with the detector materials are 
simulated using the GEANT4 software~\cite{geant}. The simulation takes into 
account variation of experimental conditions during data taking, in particular 
dead detector channels and beam-induced background. 
To take into account the effect of superimposing the beam background on
the $e^+e^-$ annihilation events, the simulation uses special background 
events recorded during data taking with a random trigger. These events are 
superimposed on the simulated events, leading to the appearance of additional
tracks and photons in them.

In this article, a data sample, collected  with the SND detector at the 
VEPP-2000 $e^+e^-$ collider in 2018 in the energy range 
$E = 560\mbox{--}1100$ MeV, is analyzed. The integrated luminosity
accumulated at 102 energy points is about 66 pb$^{-1}$.

During data taking, the average beam energy and the energy spread were 
measured by a dedicated system using the Compton back-scattering of laser 
photons on the electron beam~\cite{compton}.
The systematic uncertainty of the energy measurement by this method
was estimated in Ref.~\cite{compton} by comparison with the measurement
by the resonance depolarization method at $E_{\rm b} = 510$
and 460 MeV, where $E_{\rm b}$ is the beam energy. It is found to be
$\Delta E_{\rm b}/E_{\rm b} = 6\times 10^{-5}$. 
Energy measurements performed at a given energy point are averaged with 
weights proportional to the integrated luminosity. The obtained
values for average energy, energy spread $\sigma_{E}$ and their errors
are listed in Table~\ref{tab1}. The error of the measured energy
($\delta{E}$) includes the statistical error and the uncertainty due to
the beam energy drift during data taking. The systematic error of
$\sigma_{E}$ does not exceed 5\%.

The integrated luminosity is measured using $e^+e^-\to e^+e^-$ events.
The procedure for selecting $e^+e^-$ events, subtracting
$\pi^+\pi^-$, $\mu^+\mu^-$ backgrounds, and analyzing various
sources of systematic uncertainties in determining the number
of $e^+e^-$ events is described in detail in Ref.~\cite{snd2pi}.
The distribution of the integrated luminosity over
energy points is given in Table~\ref{tab1}.
Only the statistical error is quoted. 
The systematic uncertainty does not exceed 0.7\%. 

It should be noted that the analysis presented in this article was not
blinded. It was carried out in the sequence described below in 
Sec.~\ref{evsel}--\ref{xsfit}.
That is, all corrections to the detection efficiency of the
process under study, as well as the final cross-section result,
were obtained before comparison with previous measurements and were not
changed after the comparison.
\begin{table}
\caption{\footnotesize The center-of-mass energy ($E$), its
spread ($\sigma_E$), and integrated luminosity ($IL$).
\label{tab1}}
\begin{ruledtabular}
\fontsize{9pt}{9pt}\selectfont
\begin{tabular}{cccccc}
$E$, MeV   & $\sigma_E$, keV & $IL$, ${\rm pb}^{-1}$ & $E$, MeV   &$\sigma_E$, keV & $IL$, ${\rm pb}^{-1}$ \\
\hline
$ 560.237\pm 0.007$ & $152\pm  8$ & $  52.7\pm 0.2$ & $ 787.811\pm 0.007$ & $282\pm  6$ & $1204.9\pm 2.4$\\
$ 570.215\pm 0.005$ & $134\pm  6$ & $  76.1\pm 0.3$ & $ 788.023\pm 0.011$ & $267\pm 10$ & $ 213.9\pm 0.7$\\
$ 580.145\pm 0.005$ & $136\pm  6$ & $  87.2\pm 0.3$ & $ 790.133\pm 0.016$ & $245\pm 15$ & $ 220.4\pm 0.7$\\
$ 590.206\pm 0.009$ & $129\pm 14$ & $  61.6\pm 0.3$ & $ 791.856\pm 0.009$ & $315\pm  6$ & $ 655.3\pm 1.5$\\
$ 600.151\pm 0.005$ & $152\pm  6$ & $  91.6\pm 0.3$ & $ 792.108\pm 0.014$ & $251\pm 13$ & $ 194.9\pm 0.7$\\
$ 610.068\pm 0.006$ & $139\pm  7$ & $  64.6\pm 0.3$ & $ 794.047\pm 0.019$ & $289\pm 18$ & $ 221.2\pm 0.7$\\
$ 620.167\pm 0.008$ & $175\pm  7$ & $  82.4\pm 0.3$ & $ 796.067\pm 0.019$ & $266\pm 17$ & $ 237.0\pm 0.7$\\
$ 630.085\pm 0.010$ & $160\pm 13$ & $  82.6\pm 0.3$ & $ 797.955\pm 0.016$ & $225\pm 17$ & $ 163.8\pm 0.6$\\
$ 640.019\pm 0.007$ & $170\pm  5$ & $  83.1\pm 0.3$ & $ 800.668\pm 0.011$ & $231\pm 11$ & $ 226.3\pm 0.7$\\
$ 650.612\pm 0.011$ & $207\pm 13$ & $  82.6\pm 0.3$ & $ 810.583\pm 0.008$ & $222\pm  8$ & $ 338.3\pm 0.9$\\
$ 660.328\pm 0.022$ & $188\pm 21$ & $  83.0\pm 0.3$ & $ 820.284\pm 0.009$ & $238\pm  7$ & $ 353.1\pm 1.0$\\
$ 670.201\pm 0.015$ & $203\pm 15$ & $  83.7\pm 0.3$ & $ 829.859\pm 0.015$ & $268\pm 13$ & $ 226.2\pm 0.7$\\
$ 679.781\pm 0.270$ & $237\pm 30$ & $2942.8\pm 4.8$ & $ 840.025\pm 0.013$ & $257\pm 10$ & $ 196.1\pm 0.7$\\
$ 689.880\pm 0.260$ & $240\pm 20$ & $ 286.6\pm 0.7$ & $ 850.553\pm 0.013$ & $239\pm 13$ & $ 268.3\pm 0.8$\\
$ 699.860\pm 0.170$ & $250\pm 30$ & $ 242.3\pm 0.6$ & $ 860.252\pm 0.012$ & $238\pm 10$ & $ 174.4\pm 0.7$\\
$ 709.969\pm 0.008$ & $238\pm  8$ & $ 190.8\pm 0.6$ & $ 870.291\pm 0.010$ & $272\pm  9$ & $ 267.9\pm 0.9$\\
$ 719.930\pm 0.050$ & $274\pm 10$ & $2363.6\pm 3.6$ & $ 880.194\pm 0.015$ & $273\pm 14$ & $ 217.2\pm 0.8$\\
$ 723.798\pm 0.023$ & $307\pm 19$ & $ 144.8\pm 0.5$ & $ 889.788\pm 0.022$ & $255\pm 20$ & $ 239.4\pm 0.8$\\
$ 727.733\pm 0.020$ & $286\pm 14$ & $ 124.1\pm 0.5$ & $ 899.816\pm 0.034$ & $271\pm 12$ & $ 298.0\pm 0.9$\\
$ 731.994\pm 0.017$ & $310\pm 13$ & $ 116.9\pm 0.5$ & $ 909.399\pm 0.018$ & $383\pm 13$ & $1376.8\pm 2.8$\\
$ 736.202\pm 0.031$ & $310\pm  9$ & $ 167.9\pm 0.6$ & $ 910.112\pm 0.011$ & $260\pm 10$ & $ 173.3\pm 0.7$\\
$ 739.948\pm 0.013$ & $225\pm 13$ & $ 139.0\pm 0.5$ & $ 920.282\pm 0.012$ & $362\pm  7$ & $ 916.9\pm 2.1$\\
$ 744.089\pm 0.012$ & $202\pm 13$ & $ 129.1\pm 0.5$ & $ 920.685\pm 0.039$ & $260\pm 11$ & $ 223.9\pm 0.8$\\
$ 748.048\pm 0.012$ & $218\pm 11$ & $ 148.5\pm 0.5$ & $ 921.726\pm 0.019$ & $326\pm 15$ & $ 282.9\pm 0.9$\\
$ 750.019\pm 0.050$ & $295\pm 10$ & $5484.9\pm 6.2$ & $ 929.637\pm 0.012$ & $356\pm  8$ & $1461.5\pm 3.0$\\
$ 751.989\pm 0.014$ & $251\pm 10$ & $ 118.9\pm 0.5$ & $ 930.427\pm 0.020$ & $295\pm 17$ & $ 271.1\pm 0.9$\\
$ 756.093\pm 0.014$ & $219\pm 12$ & $ 126.7\pm 0.5$ & $ 940.342\pm 0.035$ & $356\pm 57$ & $ 194.5\pm 0.7$\\
$ 760.024\pm 0.010$ & $246\pm 10$ & $ 184.2\pm 0.6$ & $ 950.000\pm 0.200$ & $356\pm200$ & $ 286.0\pm 0.9$\\
$ 760.188\pm 0.008$ & $278\pm  7$ & $ 711.0\pm 1.5$ & $ 960.424\pm 0.026$ & $285\pm 48$ & $ 219.7\pm 0.8$\\
$ 761.852\pm 0.012$ & $275\pm  8$ & $ 166.1\pm 0.6$ & $ 970.651\pm 0.013$ & $282\pm 25$ & $ 324.5\pm 1.0$\\
$ 763.938\pm 0.011$ & $211\pm 10$ & $ 173.9\pm 0.6$ & $ 980.618\pm 0.019$ & $278\pm 37$ & $ 145.2\pm 0.7$\\
$ 767.763\pm 0.011$ & $254\pm 11$ & $ 259.3\pm 0.8$ & $ 984.201\pm 0.040$ & $377\pm 30$ & $ 482.9\pm 1.3$\\
$ 771.846\pm 0.010$ & $235\pm 10$ & $ 224.1\pm 0.7$ & $ 990.308\pm 0.033$ & $274\pm 40$ & $ 185.8\pm 0.8$\\
$ 773.788\pm 0.012$ & $240\pm 11$ & $ 154.5\pm 0.6$ & $1000.350\pm 0.220$ & $274\pm 58$ & $ 594.0\pm 1.5$\\
$ 775.057\pm 0.013$ & $244\pm 12$ & $ 142.2\pm 0.5$ & $1001.900\pm 0.040$ & $347\pm  9$ & $ 631.6\pm 1.6$\\
$ 775.059\pm 0.011$ & $305\pm  9$ & $1063.6\pm 2.1$ & $1005.998\pm 0.030$ & $361\pm 10$ & $1664.4\pm 3.2$\\
$ 776.019\pm 0.010$ & $229\pm  9$ & $ 167.1\pm 0.6$ & $1009.610\pm 0.030$ & $362\pm 10$ & $ 722.1\pm 1.8$\\
$ 776.944\pm 0.013$ & $260\pm 13$ & $ 143.8\pm 0.5$ & $1015.742\pm 0.030$ & $391\pm 11$ & $ 627.3\pm 1.7$\\
$ 777.925\pm 0.009$ & $316\pm  7$ & $1270.2\pm 2.5$ & $1016.843\pm 0.021$ & $351\pm 16$ & $1651.8\pm 3.4$\\
$ 778.017\pm 0.011$ & $243\pm 11$ & $ 210.1\pm 0.7$ & $1017.959\pm 0.020$ & $374\pm 18$ & $1253.2\pm 2.7$\\
$ 779.023\pm 0.009$ & $243\pm  8$ & $ 198.8\pm 0.7$ & $1019.101\pm 0.011$ & $380\pm  8$ & $2441.6\pm 4.7$\\
$ 779.994\pm 0.010$ & $233\pm 10$ & $ 183.7\pm 0.6$ & $1019.955\pm 0.021$ & $409\pm  7$ & $2617.6\pm 5.0$\\
$ 780.960\pm 0.006$ & $207\pm  6$ & $ 171.5\pm 0.6$ & $1020.932\pm 0.040$ & $406\pm  9$ & $1396.3\pm 3.0$\\
$ 781.023\pm 0.004$ & $311\pm  2$ & $3706.4\pm 6.4$ & $1022.114\pm 0.030$ & $372\pm  8$ & $1223.3\pm 2.7$\\
$ 781.898\pm 0.008$ & $203\pm  8$ & $ 155.5\pm 0.6$ & $1022.930\pm 0.040$ & $382\pm 13$ & $ 815.3\pm 2.0$\\
$ 782.760\pm 0.003$ & $323\pm  2$ & $5042.7\pm 8.6$ & $1027.752\pm 0.030$ & $388\pm 13$ & $ 658.6\pm 1.7$\\
$ 782.807\pm 0.009$ & $251\pm  8$ & $ 234.9\pm 0.7$ & $1033.822\pm 0.030$ & $373\pm 15$ & $ 533.7\pm 1.5$\\
$ 784.019\pm 0.030$ & $239\pm 21$ & $ 198.7\pm 0.7$ & $1039.826\pm 0.040$ & $431\pm 19$ & $ 585.6\pm 1.6$\\
$ 784.997\pm 0.014$ & $260\pm 14$ & $ 224.3\pm 0.7$ & $1049.812\pm 0.070$ & $448\pm 23$ & $ 627.6\pm 1.7$\\
$ 785.023\pm 0.005$ & $327\pm  4$ & $4690.8\pm 8.0$ & $1060.030\pm 0.050$ & $415\pm 24$ & $ 601.5\pm 1.7$\\
$ 786.090\pm 0.012$ & $245\pm 11$ & $ 239.2\pm 0.7$ & $1099.980\pm 0.040$ & $472\pm 15$ & $1407.2\pm 3.1$\\
\end{tabular}
\end{ruledtabular}
\end{table}

\section{Selection of $e^+e^-\to\pi^+\pi^-\pi^0$ events \label {evsel}}
\begin{figure}
\centering
\includegraphics[width=0.70\linewidth]{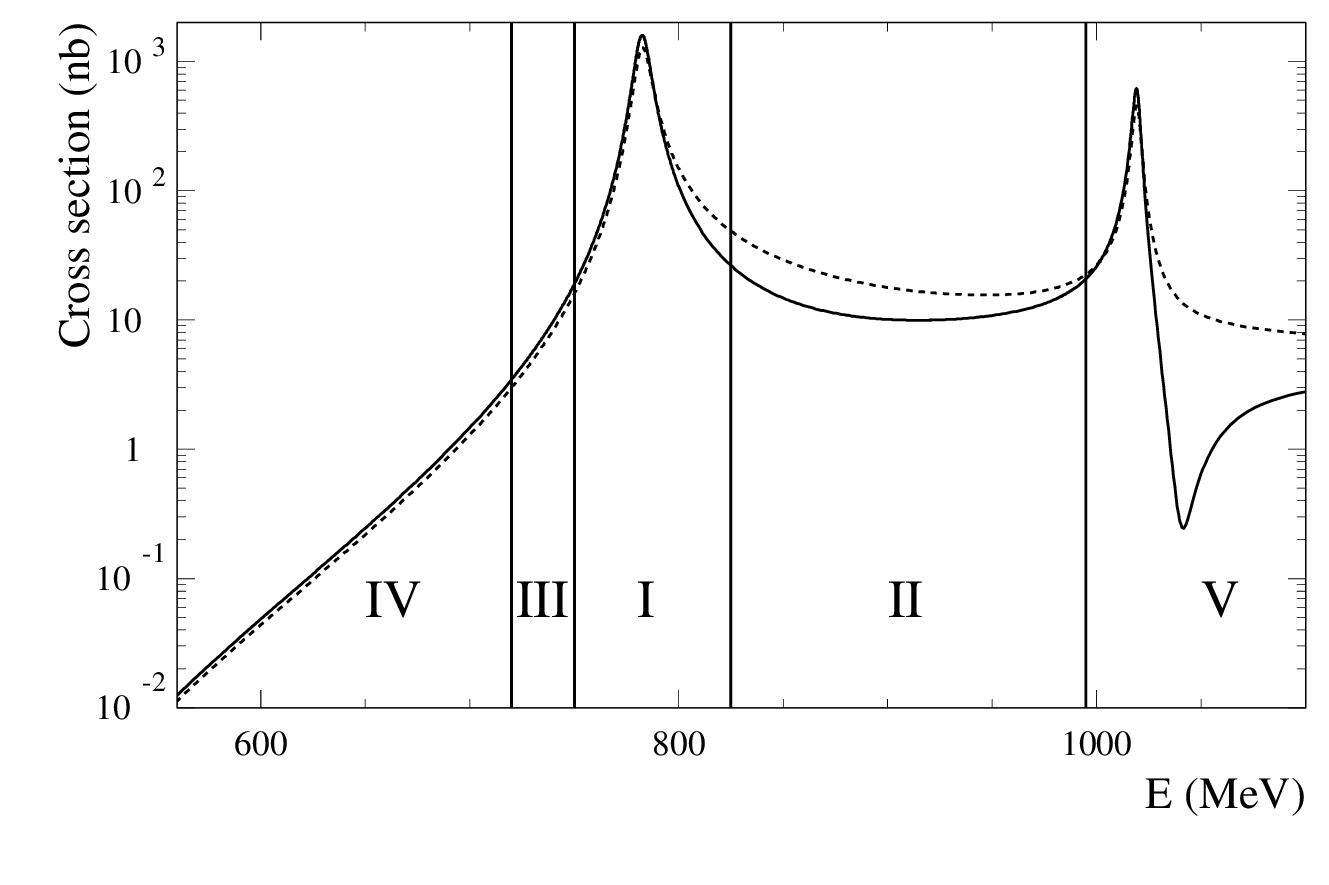}
\caption{The Born (solid curve) and visible (dashed curve) cross sections for
the process $e^+e^-\to\pi^+\pi^-\pi^0$ in the energy region under study.
The vertical lines separate five intervals in which
different event selection conditions are used.
\label{fig2}}
\end{figure}
Events of the process under study contain two charged pions and two
photons from $\pi^0$ decay. About 28\% of signal events contain spurious
photons arising mainly from pion nuclear interaction in the
calorimeter and beam background. In about 10\% of events, the beam
background and delta electrons lead to the appearance of additional
charged tracks. Therefore, for cross section measurement
we select events with two or more detected charged tracks ($n_{\rm
ch}>1$) and two or more detected photons ($n_\gamma>1$). At least two charged
particles must originate from the collider interaction region, i.e. satisfy
the conditions: $d_i < 1$ cm,
$|z_i| < 15$ cm, $i = 1, 2$, and $|z_1-z_2| < 5$ cm, where $d_i$ is
the distance between the track and the beams axis, and the $z_i$ is the 
$z$-coordinate of the track point closest to the beam axis. The ﬁt to a common
vertex is performed using the parameters of the two charged tracks.
If there are more than two charged tracks in an event satisfying the above criteria, 
two of them with the best $\chi^2$ of the vertex ﬁt are selected.
The found vertex is used to reﬁne the measured angles of charged particles and
photons.

For selected events, a kinematic ﬁt to the $\pi^+\pi^-\gamma\gamma$ hypothesis
is performed with the four constraints of energy and momentum balance. The 
quality of the ﬁt is characterized by the parameter $\chi^2_{3\pi}$.
The photon parameters after the kinematic ﬁt are used to
calculate their invariant mass ($M_{\gamma\gamma}$). It is required
to be in the range 70--200 MeV. If there are more than two photons in an event,
all two-photon combinations are tested and one with $70<M_{\gamma\gamma}<200$
MeV and the smallest $\chi^2_{3\pi}$ is selected.

Figure~\ref{fig2} show the Born and visible cross sections for
the process $e^+e^-\to\pi^+\pi^-\pi^0$ obtained in Sec.~\ref{xsfit}.
The visible cross section is the cross section actually observed in
the experiment. It takes into account radiative corrections and
is related to the Born cross section $\sigma(E)$ as follows:
\begin{equation}
\label{viscrs}
\sigma_{vis}(E) = \int\limits_{0}^{x_{max}} F(x,E) \sigma(E\sqrt{1-x})dx~,
\end{equation}
where $F(x,E)$ is a so-called radiator function~\cite{radcor}
describing the probability to emit extra photons with the total energy 
$E_{\rm ISR}=xE/2$ from the initial state, $x_{max} = 1 - (2m_{\pi^+}+m_{\pi^0})^2/E^2$. 

It is seen that in the energy region under study the 
$e^+e^-\to\pi^+\pi^-\pi^0$ cross section changes by five orders of magnitude.
We divide the energy region into five intervals, indicated in Fig.~\ref{fig2}
by vertical lines, with very different background conditions, and use 
different selection criteria in them. The selection criteria are
listed in Table~\ref{tab2}. 
\begin{table}
\caption{The selection criteria and efficiency corrections applied in energy
intervals I--V.
\label{tab2}}
\begin{ruledtabular}
\begin{tabular}{cccc}
    & Interval, MeV & Selection conditions & Efficiency correction\\
\hline
I   & 749 -- 825 & $\chi^2_{3\pi}<100$ & $1.047\pm 0.005$\\   
II  & 825 -- 995 & $\chi^2_{3\pi}<30$  & $1.056\pm 0.013$\\   
III & 724 -- 749 & (II) and $(|\Delta\varphi| >2.3^\circ)$
and $(\psi<140^\circ)$ & $1.055\pm 0.010$ \\ 
IV  & 560 -- 722 & (III) and $(E_{\rm EMC}/E < 0.75)$ 
and $(\mu{\rm veto}=0)$ and $(n_{\rm ch}=2)$ & $1.020\pm 0.010$\\
V   & 995 -- 1045  & (III) and $(\chi^2_{4\pi}>500)$ &  $1.054\pm 0.008$\\
V   & 1045 -- 1100 & (III) and $(\chi^2_{4\pi}>500)$ &  $1.054\pm 0.013$\\
\end{tabular}
\end{ruledtabular}
\end{table}

\begin{figure}
\centering
\includegraphics[width=0.40\linewidth]{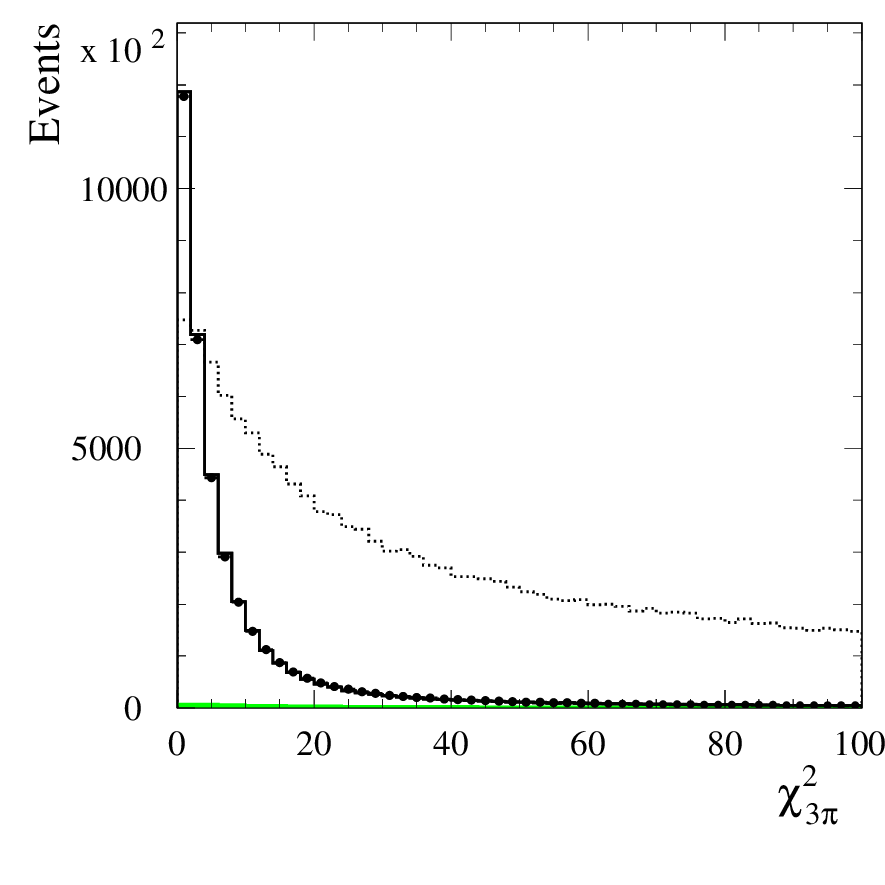}
\includegraphics[width=0.40\linewidth]{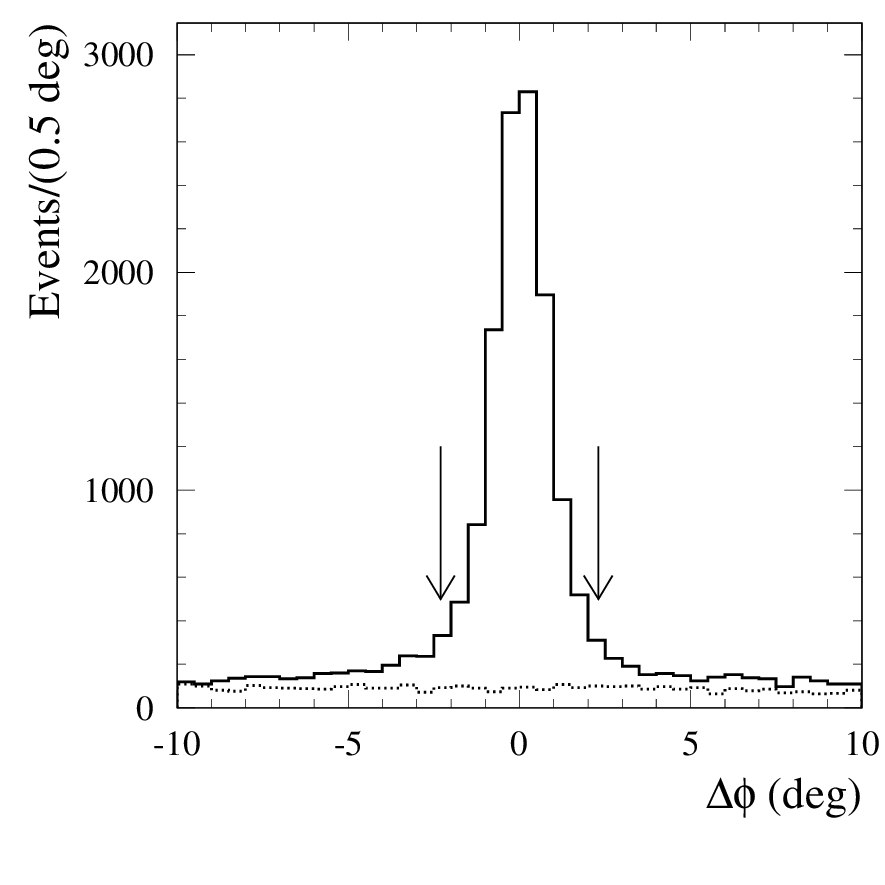}
\caption{Left panel: The $\chi^2_{3\pi}$ distributing for data (point
with error bars) and simulated events at $E=782.76$ MeV. The solid
histogram is a sum of simulated signal and background distributions.
The filled histogram represents background contribution. The dashed histogram
is the background distribution multiplied by a factor of 25.
Right panel: The $\Delta\varphi$ distribution for data events at
$E=719.93$ MeV (solid histogram). The dotted histogram represents the
signal simulation. The arrows indicate the selection condition.
\label{fig3}}
\end{figure}

The loosest condition $\chi^2_{3\pi}<100$ is applied in the $\omega$-resonance
energy interval (I). The $\chi^2_{3\pi}$ distribution at the $\omega$ maximum
($E=782.76$ MeV) is shown in Fig.~\ref{fig3} (left). The filled histogram
represents background contribution, which is about 4\%. The dashed
histogram is the background distribution multiplied by a factor of 25.
The number of signal and background events are determined from the fit to
the $M_{\gamma\gamma}$ spectrum, in which signal events peak near
$\pi^0$ mass (see Sec.~\ref{mggfit}).
About 95\% of background events come from the processes 
$e^+e^-\to e^+e^-(\gamma\gamma)$, $\mu^+\mu^-(\gamma\gamma)$, 
and $\pi^+\pi^-(\gamma\gamma)$, about 2\% arise from $e^+e^-\to \eta\gamma$
and $\pi^0\gamma$, and about 3\% are due to cosmic-rays.

In the energy interval between $\omega$ and $\phi$ resonances (II), the
tighter condition $\chi^2_{3\pi}<30$ suppresses background and also
improves the $\pi^0$ line-shape in the $M_{\gamma\gamma}$ spectrum, which
is distorted by the radiative return to the $\omega$-resonance.

To enhance the signal above the background, additional conditions are
required in intervals III and IV. Figure~\ref{fig3} (right) shows the
$\Delta\varphi=|\varphi_1-\varphi_2|-180^\circ$ distribution for
data events and simulated $e^+e^-\to\pi^+\pi^-\pi^0$ events. The peak
near $\Delta\varphi=0$ is mainly due to $e^+e^-\to \pi^+\pi^-(\gamma)$
background events. The dotted histogram represents the signal simulation.
The requirement $\Delta\varphi>2.3^\circ$ suppresses the $e^+e^-\to
\pi^+\pi^-(\gamma)$ background by a factor of about 3, while the signal loss is
3\%. 

\begin{figure}
\centering
\includegraphics[width=0.40\linewidth]{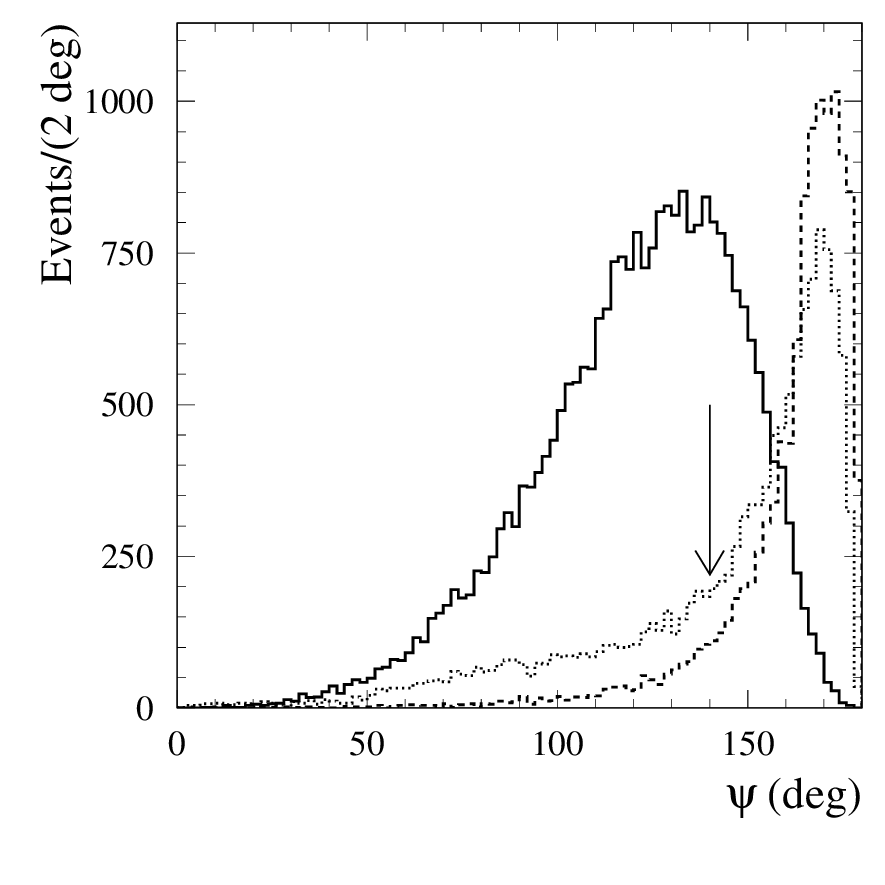}
\includegraphics[width=0.40\linewidth]{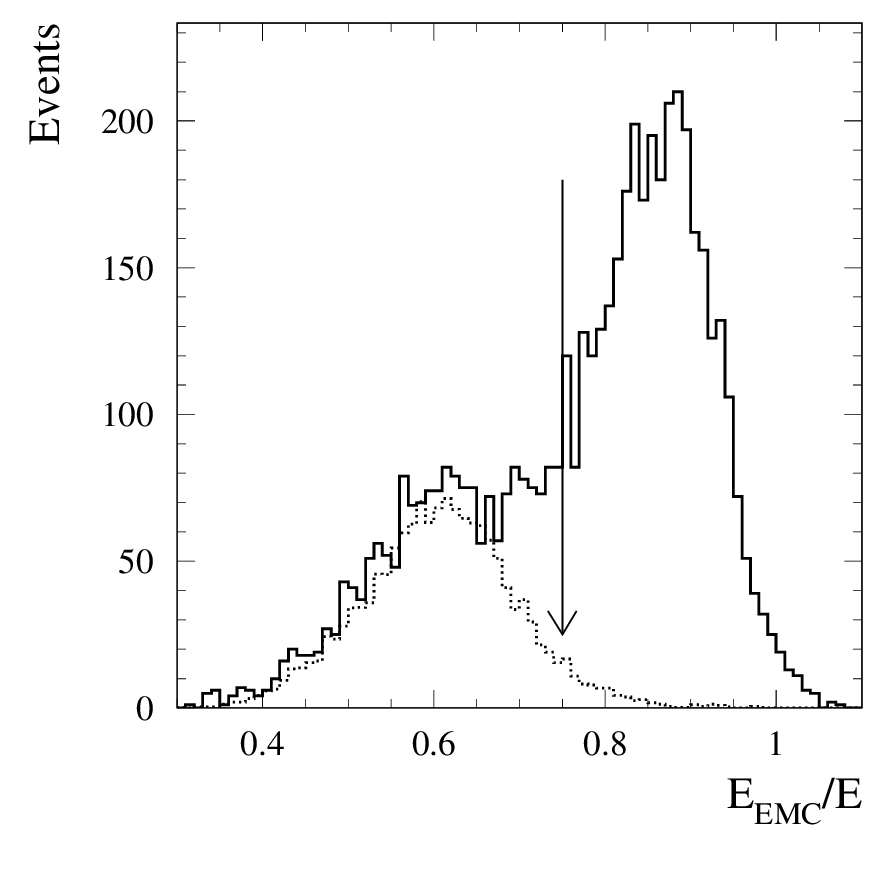}
\caption{Left panel: The distribution of the open angle
between charged pions for simulated signal (solid histogram),
$\pi^+\pi^-(\gamma)$ (dashed histogram), and $e^+e^-(\gamma\gamma)$
(dotted histogram) events at $E=719.93$ MeV. The arrow indicates the
condition $\psi<140^\circ$.
Right panel: The $E_{\rm EMC}/E$ distribution for data events at
$E=679.78$ MeV (solid histogram). The dotted histogram represents the
signal simulation. The arrow indicates the selection condition.
\label{fig4}}
\end{figure}

Figure~\ref{fig4} (left) shows the distribution of the open angle
between charged pions ($\psi$) for simulated signal and background
$e^+e^-(\gamma\gamma)$ and $\pi^+\pi^-(\gamma)$ events at $E=719.93$ MeV
after applying the cut on $\Delta\varphi$. The condition
$\psi<140^\circ$ suppresses the $\pi^+\pi^-(\gamma)$ and $e^+e^-(\gamma
\gamma)$ backgrounds by approximately 10 and 3 times, respectively,
while the signal loss is about 30\%.

\begin{figure}
\centering
\includegraphics[width=0.40\linewidth]{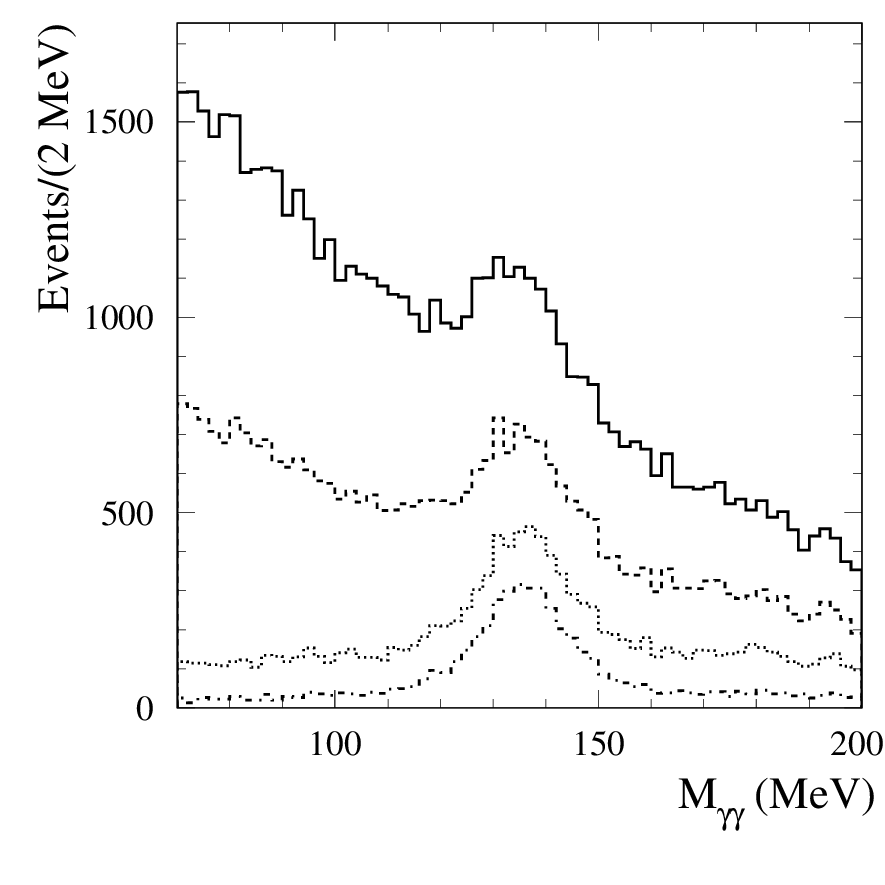}
\includegraphics[width=0.40\linewidth]{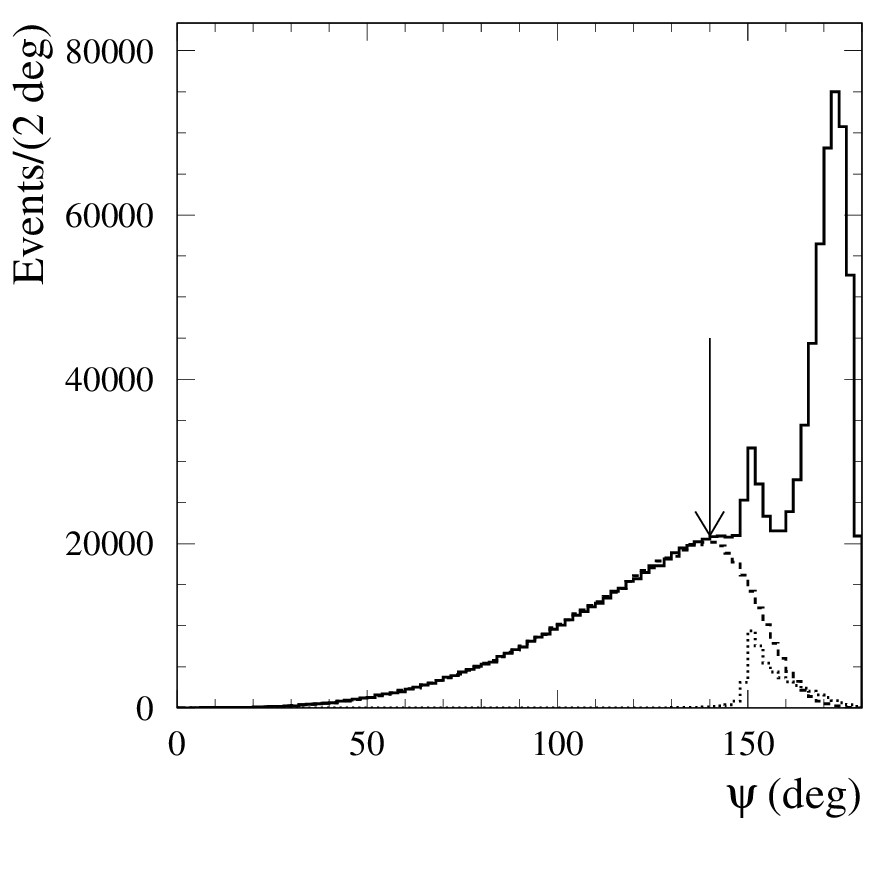}
\caption{Left panel: The $M_{\gamma\gamma}$ spectrum for data events
at $E=719.93$ MeV after applying the selection criteria for energy
interval I (solid), II (dashed), III (dotted), and IV (dash-dotted).
Right panel: The distribution of the open angle
between charged pions for data (solid histogram), simulated signal
(dashed histogram), and $K_S K_L$ (dotted histogram) events
at $E=1019.1$ MeV, selected with the condition $\chi^2_{3\pi}<30$.
The arrow indicates the condition $\psi<140^\circ$.
\label{fig5}}
\end{figure}

After applying the conditions on $\Delta\varphi$ and $\psi$, the main
background source in interval IV becomes the process $e^+e^- \to e^+e^-\gamma
(\gamma)$. This is demonstrated in Fig.~\ref{fig4} (right), which shows
the distribution of the normalized total energy deposition in the
calorimeter ($E_{\rm EMC}/E$) for data events at $E=679.78$ MeV in
comparison with the signal distribution. The peak near 0.9 is due to
$e^+e^-\gamma(\gamma)$ events. The condition $E_{\rm EMC}/E<0.75$ 
suppresses this background by a factor of 6. 
The two additional conditions in interval IV, $\mu{\rm veto}=0$ and
$n_{\rm ch}=2$, remove the cosmic-ray and beam-induced backgrounds.
The result of sequential application of selection conditions for
intervals I-IV to data events at $E=719.93$ MeV is demonstrated in
Fig.~\ref{fig5} (left), where the two-photon mass ($M_{\gamma\gamma}$)
spectrum is shown. The improvement in the signal to background ratio
is clearly seen.

In the $\phi$-resonance energy range (V), additional background
processes $e^+e^- \to K^+ K^-$ and $e^+e^- \to K_S K_L$ appear,
exhibiting a resonant energy dependence.
Figure~\ref{fig5} (right) shows the $\psi$ distribution at
the $\phi$-resonance maximum ($E=1019.1$ MeV) for data and
simulated signal and $K_S K_L$ events selected with the condition
$\chi^2_{3\pi}<30$. The data distribution contains two peaks: near
$180^\circ$ due to $K^+ K^-$ events and near
$150^\circ$ due to $K_S K_L$ events. The condition $\psi<140^\circ$ is
used to suppress these background sources. It is seen that below $140^\circ$
the data distribution is very close to the simulated $\pi^+\pi^-\pi^0$
distribution.

The additional nonresonant background process in the $\phi$ region is
$e^+e^-\to \pi^+\pi^-\pi^0\pi^0$. In the energy region under study it is
dominated by the $\omega\pi^0$ intermediate state. Its cross section
grows almost linearly from 0.4 nb at 920 MeV to 11 nb at 1100 MeV.
The fraction of $\pi^+\pi^-\pi^0\pi^0$ events in the total background
for events selected with conditions III is about 40\% at  $E=1000$ MeV
and about 50\% at  $E=1100$ MeV. Background $\pi^+\pi^-\pi^0\pi^0$ events
have a broad bump near 150 MeV in the $M_{\gamma\gamma}$ spectrum,
so it is desirable to suppress them. To do this we perform a kinematic
fit for events with four or more detected photons to the 
$e^+e^-\to \pi^+\pi^-\pi^0\pi^0$ hypothesis and reject events with
$\chi^2_{4\pi}<500$. This condition suppresses the $\pi^+\pi^-\pi^0\pi^0$
background by a factor of 2.3,  while the signal loss is 4\%.

\section{Fitting the $M_{\gamma\gamma}$ distribution\label{mggfit}}
\begin{figure}
\centering
\includegraphics[width=0.40\linewidth]{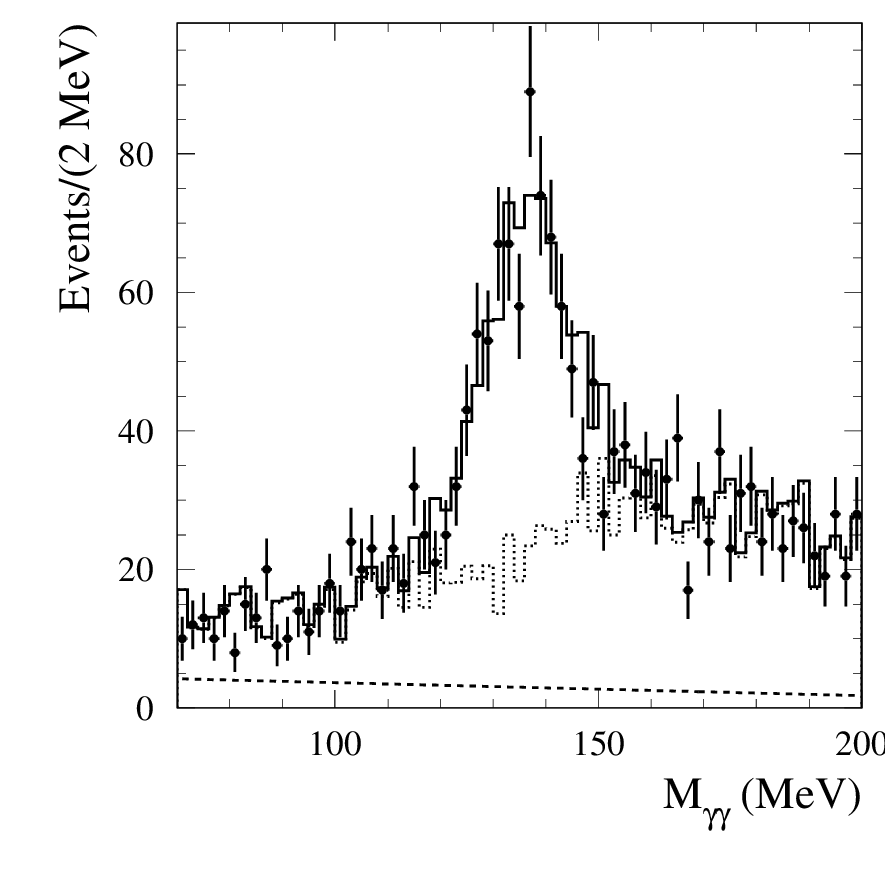}
\includegraphics[width=0.40\linewidth]{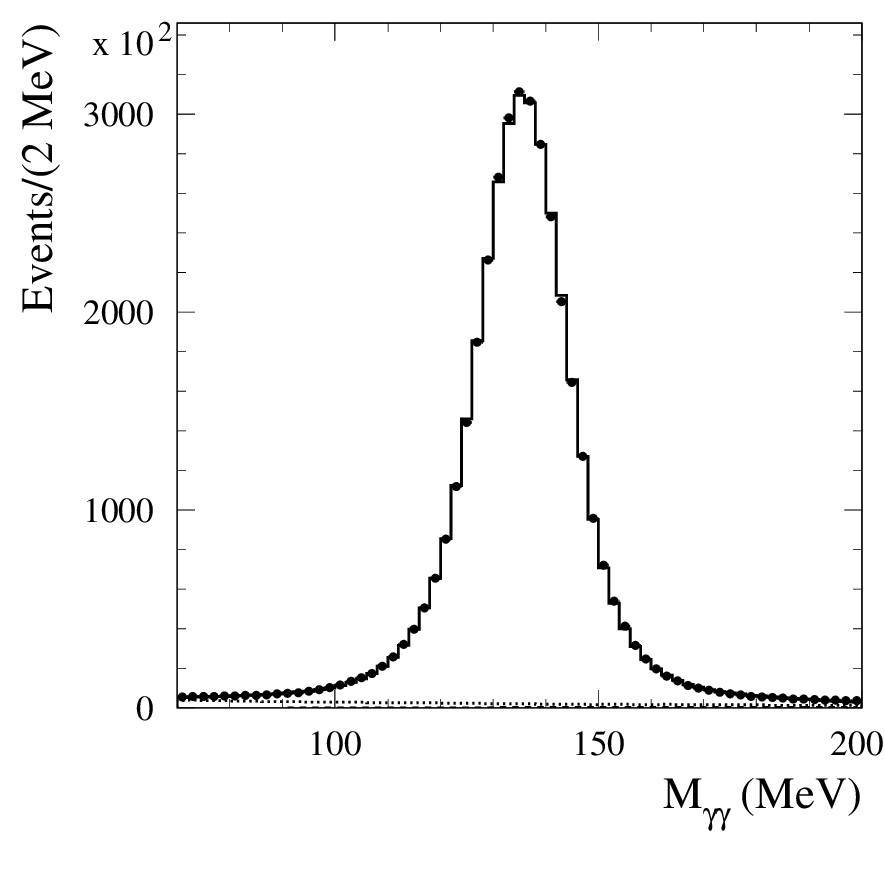}
\includegraphics[width=0.40\linewidth]{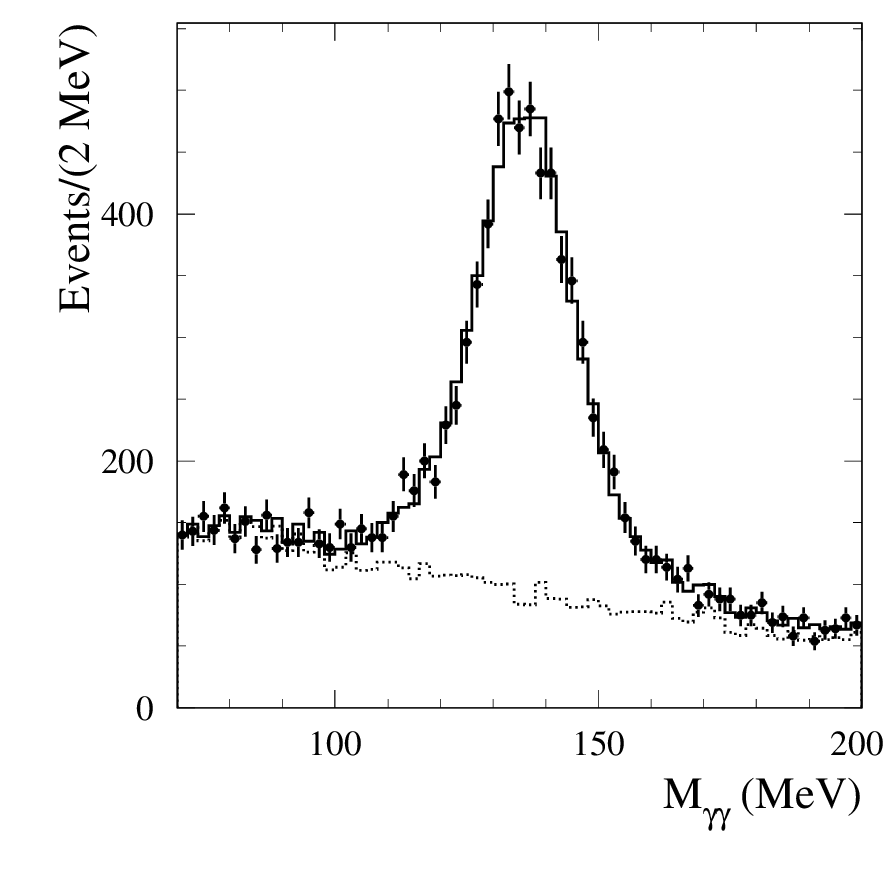}
\includegraphics[width=0.40\linewidth]{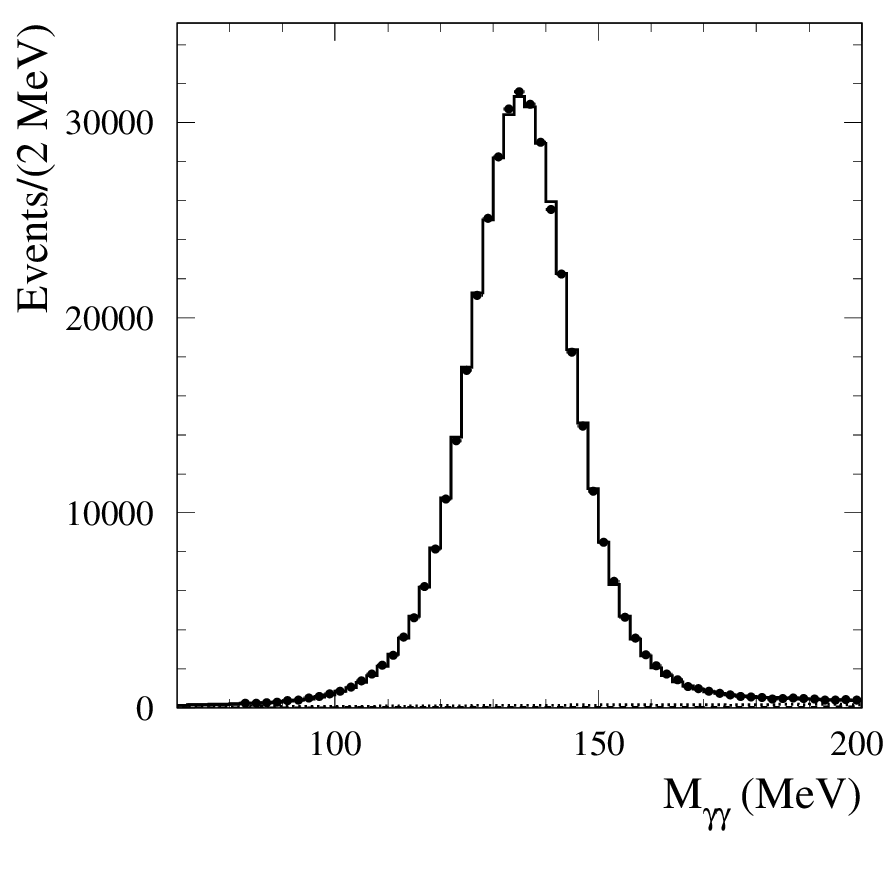}
\caption{The $M_{\gamma\gamma}$ spectra for selected data events
(points with error bars) at $E=679.78$ MeV (top,~left), $E=782.76$ MeV
(top,~right), $E=920.28$ MeV (bottom,~left), and $E=1019.10$ MeV
(bottom,~right). The solid histogram represents the result of the
fit described in the text. The dotted histogram is the fitted
background distribution. The dashed line represents the linear component
of the background.
\label{fig6}}
\end{figure}
The $M_{\gamma\gamma}$ spectra for data events are shown in
Fig.~\ref{fig6} at four energy points, below the $\omega$ resonance 
($E=679.78$ MeV), at the $\omega$ maximum ($E=782.76$ MeV), between the 
$\omega$ and $\phi$ resonances ($E=920.28$ MeV), and at the $\phi$ maximum
($E=1019.10$ MeV).
To determine the number of signal events ($N_{3\pi}$) at each energy point, 
we perform a binned likelihood fit to the $M_{\gamma\gamma}$ spectrum
with a sum of signal ($F_{sig}$) and background ($F_{bkg}$) distributions. 

The signal distribution is obtained by fitting the mass spectra for
simulated signal events by a sum of three Gaussian functions
$\epsilon_1G_1(a_1,\sigma_1)+\epsilon_2G_2(a_2,\sigma_2)+
(1-\epsilon_1-\epsilon_2)G_3(a_3,\sigma_3)$.
The fitted parameters are $a_1=136.1 (136.0) $ MeV, 
$a_2-a_1=1.6 (0.8) $ MeV, $a_3-a_1=9 (12) $ MeV, $\sigma_1=7.0 (8.2)$ MeV, 
$\sigma_2=12 (14)$ MeV, $\sigma_3=30 (40)$ MeV, $\epsilon_1\approx 0.6$,
$\epsilon_2\approx 0.3$ at the maximum of the $\omega$ ($\phi$) resonance. 
To take into account a diﬀerence between data and MC simulation in the
$\pi^0$ line shape, the signal distribution obtained in simulation is 
modiﬁed in the following way:
$a_1^{\rm data} = a_1^{\rm MC}+\Delta a_1$, 
$a_{2,3}^{\rm data} = a_{2,3}^{\rm MC}+\Delta a_2$,
$\sigma_{1,2,3}^{2,\rm data} = \sigma_{1,2,3}^{2,\rm MC}+\Delta \sigma^2$.
For data points where the $N_{3\pi}$ statistical uncertainty is less
than 2.5\%, $\Delta a_1$, $\Delta a_2$, and $\Delta \sigma^2$ are
free fit parameters. For the remaining points, they are fixed at the
values $\Delta a_1=-0.2\pm0.2$ MeV, $\Delta a_2=-2.0\pm 0.5$ MeV,
$\Delta \sigma^2=3.9\pm 1.0$ MeV$^2$. These values are obtained as 
averages over energy points near the $\omega$ and $\phi$ resonances. 
The uncertainties reflect their nonstatistical fluctuations. 
The variation of the parameters $\Delta a_1$, $\Delta a_2$, and
$\Delta \sigma^2$ within their errors is used to 
estimate the associated systematic uncertainty.

\subsection{Background distribution}
The background distribution is obtained using simulation of the background
processes $e^+e^-\to e^+e^-(\gamma\gamma)$, $\mu^+\mu^-(\gamma\gamma)$, 
$\pi^+\pi^-(\gamma)$, $\eta\gamma$, $\pi^0\gamma$, $\pi^0 e^+e^-$, 
$\pi^+\pi^-\pi^0\pi^0$, $K^+ K^-$, and $K_S K_L$. 
Analysis of the $M_{\gamma\gamma}$ distribution for events selected
with the additional condition $(\mu{\rm veto}=1)$ shows that the
cosmic ray background is well described by a linear function.
Therefore, a linear function with free parameters is added to the
simulated background distribution. This linear function also serves to
account for possible inaccuracies in the simulated background
distribution.

We observe that the simulation underestimates the number of background
events in data. Therefore, a scale factor is introduced for the expected 
number of events for each background process.

To determine scale factors for the main background processes $e^+e^-\to
e^+e^-(\gamma\gamma)$ ($S_{ee}$) and $\pi^+\pi^-(\gamma)$ ($S_{\pi\pi}$), 
we analyze data in the energy regions 680--750 MeV and 800--920 MeV, where 
background contribution is relatively large. To suppress the cosmic-ray 
background, the condition $\mu{\rm veto}=0$ is applied.
To study the dependence of the scale factors on the selection used,
three sets of selection criteria are analyzed. Two of them,
$\chi^2_{3\pi}<100$ and $\chi^2_{3\pi}<30$, are applied in energy
intervals I and II. In the third set ($\chi^2_{3\pi}<30$ and
$\psi<160^\circ$), the condition on $\psi$ is weakened compared to the
standard one for intervals III-V to increase sensitivity to $S_{\pi\pi}$.

Two subsamples with $E_{\rm EMC}/E < 0.75$ and $E_{\rm EMC}/E > 0.75$ are
analyzed to separate the $e^+e^-(\gamma\gamma)$ and $\pi^+\pi^-(\gamma)$
contributions. In both subsamples, the fit to the $M_{\gamma\gamma}$ 
distribution is performed, and the numbers of background events $N^{\rm data}
(E_{\rm EMC}/E < 0.75)$ and $N^{\rm data}(E_{\rm EMC}/E > 0.75)$
are determined. The scale factors are obtained by solving the system of linear 
equations:
\begin{eqnarray} 
N^{\rm data} (E_{\rm EMC}/E < 0.75) &=&
S_{ee} N^{\rm MC}_{ee} (E_{\rm EMC}/E < 0.75)+
S_{\pi\pi} N^{\rm MC}_{\pi\pi} (E_{\rm EMC}/E < 0.75)\nonumber \\
&+& N^{\rm MC}_{\mu\mu} (E_{\rm EMC}/E < 0.75),\nonumber \\
N^{\rm data} (E_{\rm EMC}/E > 0.75) &=&
S_{ee} N^{\rm MC}_{ee} (E_{\rm EMC}/E > 0.75)+
S_{\pi\pi} N^{\rm MC}_{\pi\pi} (E_{\rm EMC}/E > 0.75)\nonumber \\
&+& N^{\rm MC}_{\mu\mu} (E_{\rm EMC}/E > 0.75),
\end{eqnarray}
where $N^{\rm MC}_{ee}$, $N^{\rm MC}_{\pi\pi}$, and $N^{\rm MC}_{\mu\mu}$ are
the expected numbers of $e^+e^-(\gamma\gamma)$, $\pi^+\pi^-(\gamma)$
and $\mu^+\mu^-(\gamma\gamma)$ events obtained in simulation, respectively.
The fraction of the $e^+e^-\to \mu^+\mu^-(\gamma\gamma)$ process in the
overall background, even in energy interval IV, does not exceed 12\% .
Therefore, its scale factor is taken to be equal to unity.
We obtain $S_{ee}=1.25\pm0.10$ and $S_{\pi\pi}=1.55\pm0.05$.  
The quoted errors are systematic and reflect the dependence of
the scale factors on energy and selection conditions.

\begin{figure}
\centering
\includegraphics[width=0.40\linewidth]{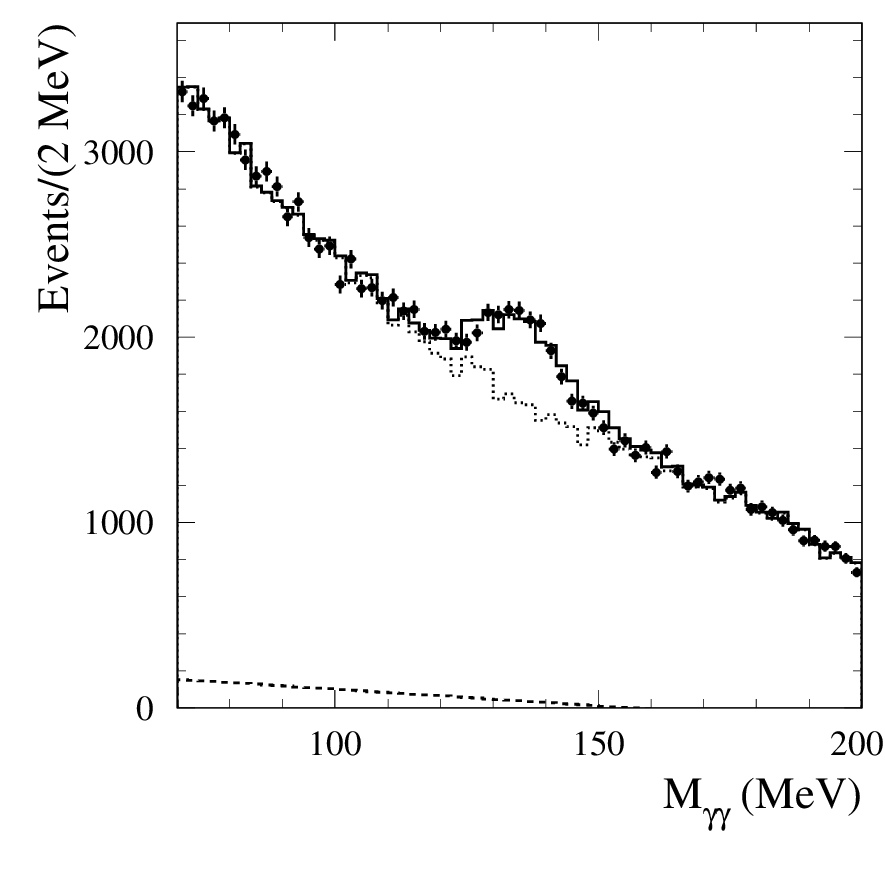}
\includegraphics[width=0.40\linewidth]{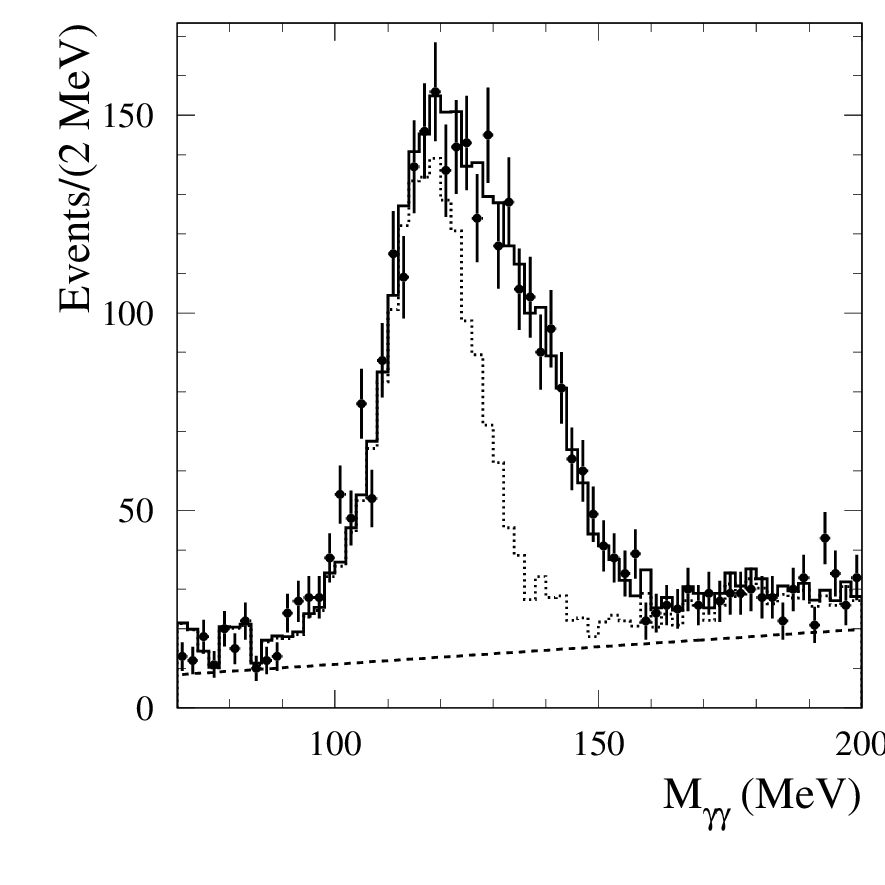}
\caption{Left panel: The $M_{\gamma\gamma}$ spectrum for data events (points
with error bars) from the energy range 679--720 MeV selected with conditions
$\chi^2_{3\pi}<100$ and $\mu{\rm veto}=0$. The solid histogram represents the
result of the fit described in the text. The dotted histogram is the fitted
background distribution. The dashed line represents the linear component
of the background.
Right panel: The $M_{\gamma\gamma}$ spectrum for data events
at $E=782.76$ MeV (points with error bars) selected with
the conditions $\chi^2_{3\pi}<100$, $\psi<40^\circ$, $E_{\rm EMC}/E >
0.75$. The solid histogram represents the result of the fit described in
the text. The dotted histogram is the fitted background distribution.
The peak in this distribution is due to $\pi^0 e^+e^-$ events.
The dashed line represents the linear component of the background.
\label{fig7}}
\end{figure}
Figure~\ref{fig7} (left) show the $M_{\gamma\gamma}$ spectrum for data
events from the energy range 679--720 MeV selected with conditions 
$\chi^2_{3\pi}<100$ and $\mu{\rm veto}=0$. The obtained above scale
factors are used to calculate the expected background distribution.
It is seen that the background shape is not very different from linear, and
the nonlinearity is well reproduced by the scaled simulation.
The relatively small background nonlinearity is characteristic of
energy intervals I-III and V. In interval IV the background has a more 
complex shape, which is also well described by the simulation,
as shown in Fig.~\ref{fig6} (top, left). To quantify how a possible 
incorrectness of the
background shape in simulation affects the signal yield, we
compare the signal data-to-MC ratio for the conditions $\psi<140^\circ$ and
$\psi<160^\circ$. Weakening the condition increases the background
by a factor of about two, while the signal detection efficiency increases
by only 30\%. The change in the ratio is $(2.6\pm4.2)\%$ at $E=679.78$
MeV and $-(1.3\pm1.3)\%$ at $E=719.93$ MeV. We conclude that the scaled
simulation describes the shape of the data background
$M_{\gamma\gamma}$ distribution reasonably well.

Near the $\omega$ resonance maximum, there is background from
the processes $e^+e^-\to \pi^0 e^+e^-$ and $e^+e^-\to \pi^0\gamma$.
In the latter process the $e^+e^-$ pair arises from photon conversion in
material before the tracking system. This
background has a peak in the $M_{\gamma\gamma}$ distribution shifted
by about 15 MeV relative to the $\pi^0$ mass. About 60\% of the $\pi^0
e^+e^-$ events with $\chi^2_{3\pi}<100$ have $\psi<40^\circ$ and  
$E_{\rm EMC}/E > 0.75$. The $M_{\gamma\gamma}$ spectrum for these
events at $E=782.76$ MeV is shown in Fig.~\ref{fig7} (right). 
In the fit to this spectrum, the additional parameter is introduced,
the scale factor for $\pi^0 e^+e^-$ events. It is found to be
$1.11\pm0.04$. With this scale factor the ratio of the $\pi^0 e^+e^-$ 
background to the signal events selected with the condition 
$\chi^2_{3\pi}<100$ 
is $7\times 10^{-4}$. The $\eta\gamma$ background is about four times 
smaller and have $M_{\gamma\gamma}$ distribution close to linear.
\begin{figure}
\centering
\includegraphics[width=0.40\linewidth]{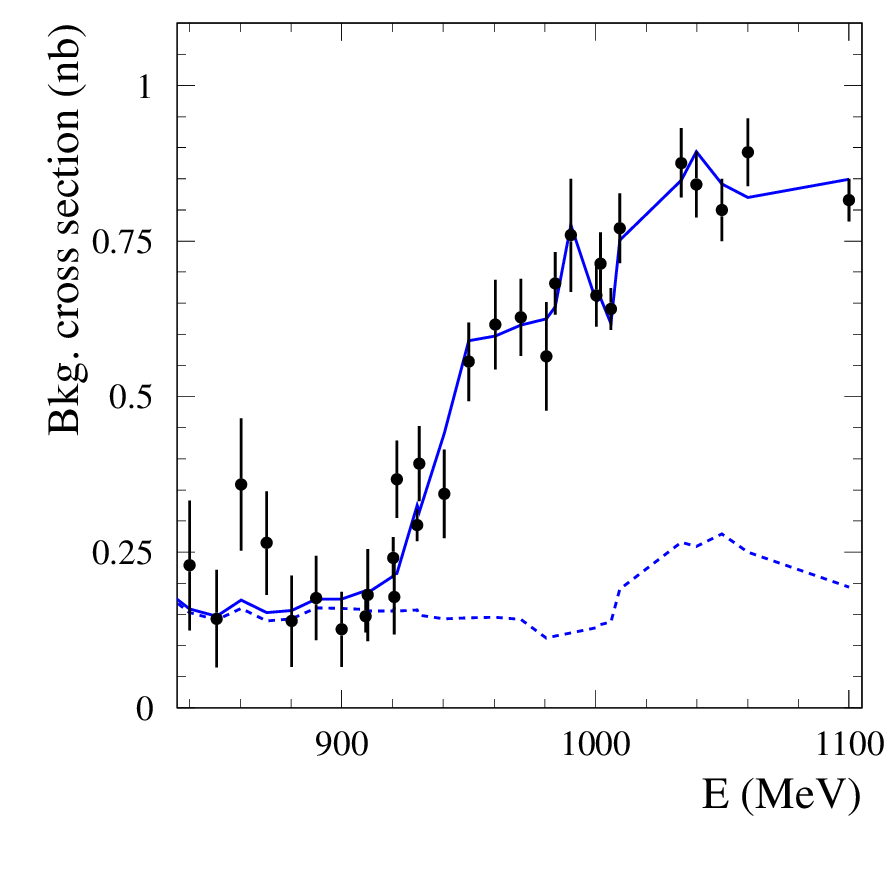}
\includegraphics[width=0.40\linewidth]{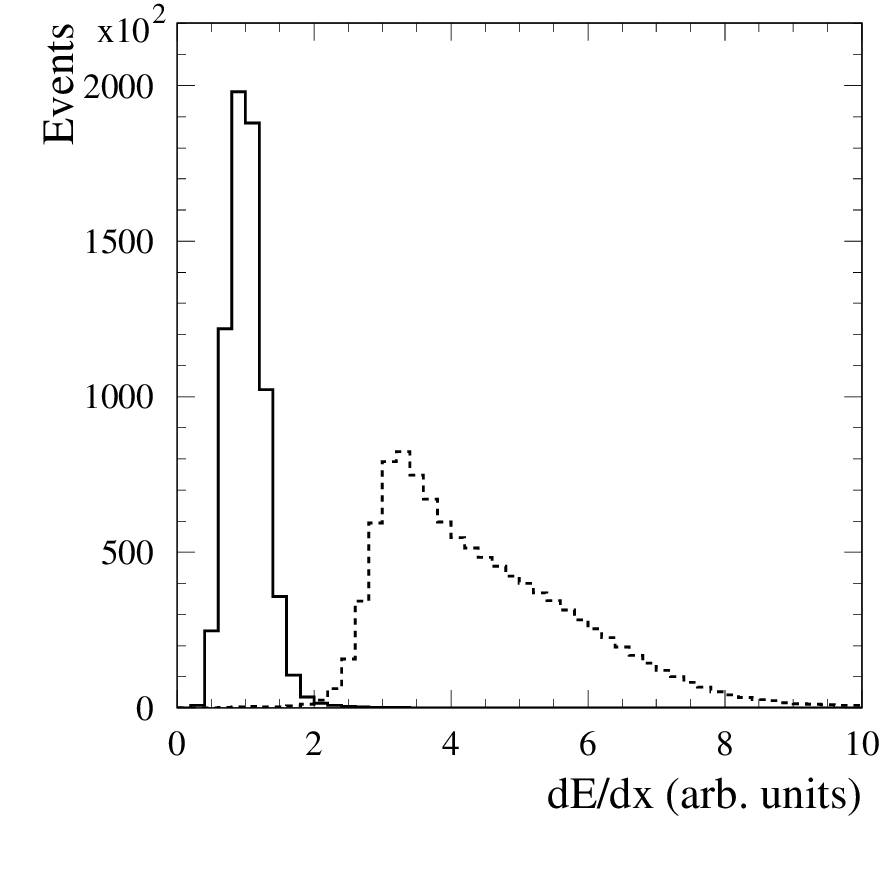}
\caption{Left panel: The energy dependence of background cross section for 
events selected with conditions $n_\gamma>2$, $\chi^2_{3\pi}<30$, 
$\psi<140^\circ$, and $\mu{\rm veto}=0$. The solid curve is the result
of the fit described in the text. The dashed curve is non-$4\pi$
background. Right panel: The distribution of ionization loss ($dE/dx$)
in the drift chamber for charged pions from the $e^+e^-\to \pi^+\pi^-\pi^0$
reaction (solid histogram) and charged kaons from the $e^+e^-\to K^+K^-$
reaction (dashed histogram) at $E=1019.10$ MeV.
\label{fig8}}
\end{figure}

Above 920 MeV the background from the process $e^+e^-\to \pi^+\pi^-\pi^0\pi^0$
must be taken into account. To find the scale factor for this process
($S_{4\pi}$), we fit to the $M_{\gamma\gamma}$ spectra for events in the energy
region above 840 MeV. To increase fraction of $\pi^+\pi^-\pi^0\pi^0$
background, the requirement $n_\gamma>2$ is used together with the  
conditions $\chi^2_{3\pi}<30$, $\psi<140^\circ$, and $\mu{\rm veto}=0$.
In the fit to the $M_{\gamma\gamma}$ spectrum, the scale factor $S_{4\pi}$
is set to unity. The energy dependence of the cross section for the fitted
background ($\sigma_{\rm{bkg}}$) is shown in Fig.~\ref{fig8} (left). 
The energy points near the $\phi$ resonance, where there are large $K\bar{K}$ 
background, are removed from this plot.
The background energy dependence is fitted by the function
$\sigma_{\rm{bkg},i}^{\rm exp}=\sigma_{\rm{non-}4\pi,i}^{\rm MC}+
S_{4\pi}\sigma_{4\pi,i}^{\rm MC}+C$, where $\sigma_{4\pi,i}^{\rm MC}$ and
$\sigma_{\rm{non-}4\pi}^{\rm MC}$ are the expected cross sections
for $4\pi$ and non-$4\pi$ backgrounds, respectively, and $C$  is a free
constant. The scale factor is found to be $S_{4\pi}=1.03\pm0.04$. The
consistent scale-factor value is obtained for events selected without the
condition $\psi<140^\circ$. 

Near the maximum of the $\phi$ resonance, about 70\% of background
events arise from the decays $\phi\to \eta\gamma$ with $\eta\to
\pi^+\pi^-\pi^0$, $\phi\to K^+ K^-$, and $\phi\to K_S K_L$. To determine
the scale factor for the process $e^+e^-\to \eta\gamma$
($S_{\eta\gamma}$), we analyze a sample
of events selected with the conditions for energy interval V and
specific conditions ($n_\gamma>2$, $E_{\gamma,max}>330$ MeV, and 
$\mu{\rm veto}=0$) that increase the fraction of $\eta\gamma$ events.
The $M_{\gamma\gamma}$ spectra for events with $E=1019.10$ MeV and
$E=1019.96$ MeV are fitted. A background distribution without a linear
component, but with a free parameter $S_{\eta\gamma}$, is used in the
fit. The scale factor is found to be $S_{\eta\gamma}=1.02\pm0.07$.
\begin{figure}
\centering
\includegraphics[width=0.40\linewidth]{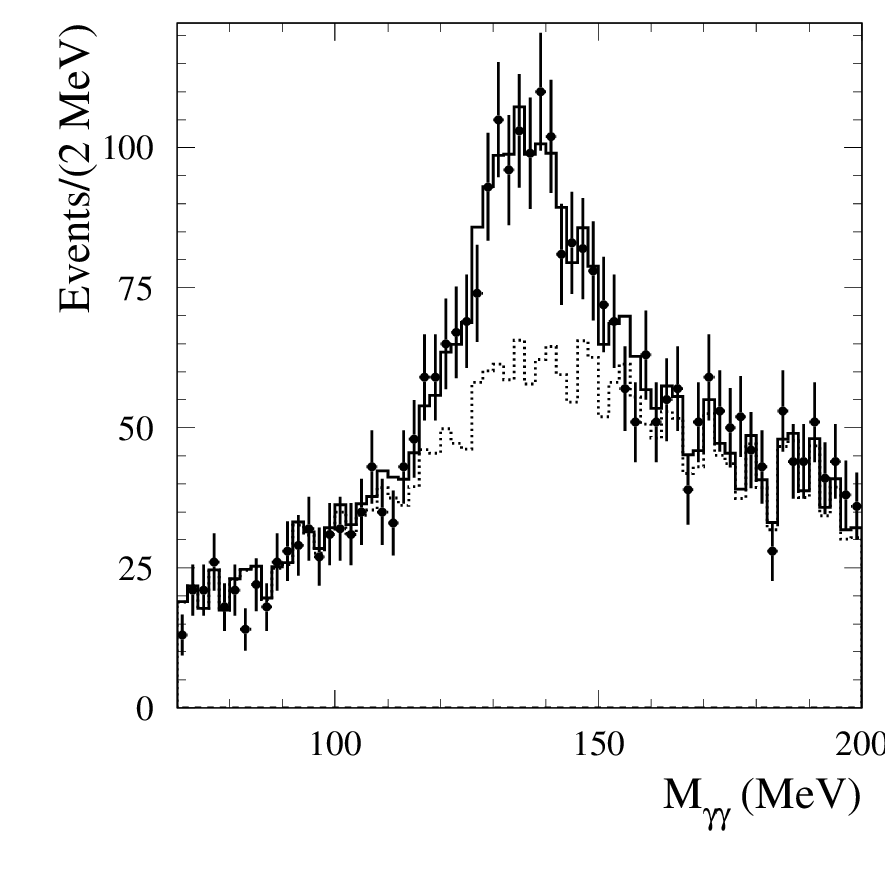}
\includegraphics[width=0.40\linewidth]{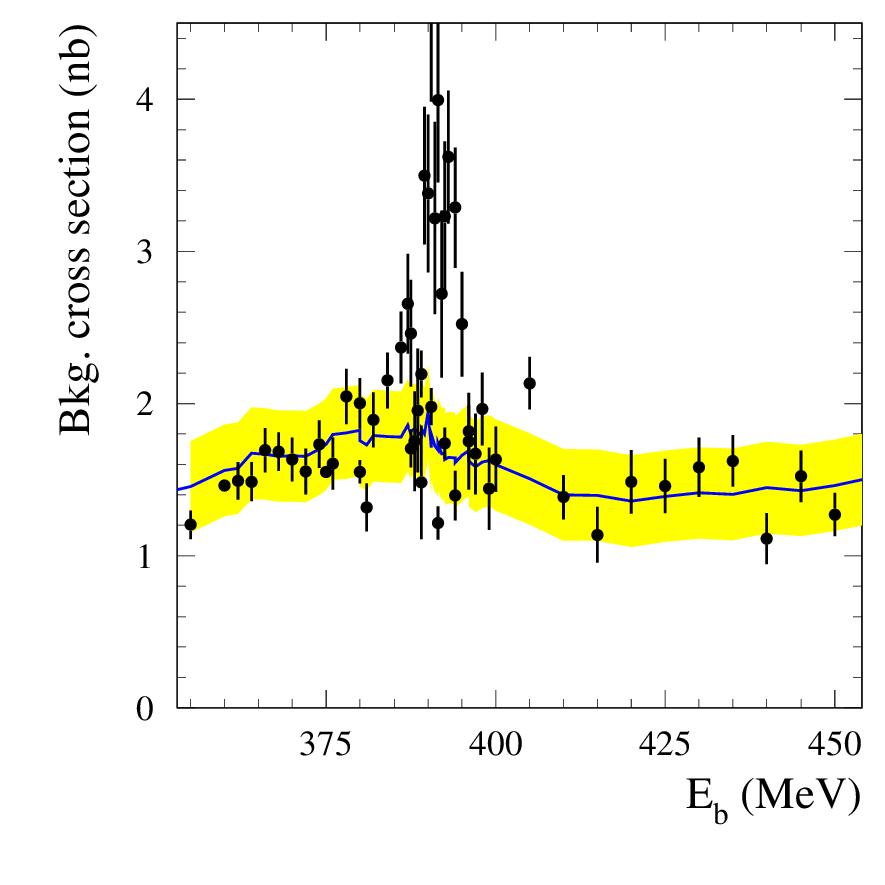}
\caption{Left panel: The $M_{\gamma\gamma}$ spectrum for data events
at $E=1019.10$ MeV (points with error bars), selected using the criteria for 
energy interval V and the condition $(dE/dx)_{\rm max}>3$. The solid histogram
represents the result of the fit described in the text. The dotted histogram
is the fitted background distribution. Right panel: The energy
dependence of the background cross section obtained from the fit to
the $M_{\gamma\gamma}$ spectrum for data events selected with
the conditions~(\ref{condsub}). The curve is the background
cross section calculated using Eq.~(\ref{csbkg}). The shaded band
represents the systematic uncertainty of Eq.~(\ref{csbkg}).
\label{fig9}}
\end{figure}

Most $e^+e^-\to K^+ K^-$ events are rejected by the condition
$\psi<140^\circ$. The remaining events typically contain a single charged
kaon, while the second charged particle is a pion, muon, or electron
from the second kaon decay or its nuclear interaction in the material before
the drift chamber. To obtain a data sample enriched in 
$e^+e^-\to K^+ K^-$ events, we use the difference 
in ionization losses ($dE/dx$) measured in the drift chamber between
kaons and pions. The $dE/dx$ distributions for charged pions from the
$e^+e^-\to \pi^+\pi^-\pi^0$ reaction and charged kaons from 
the $e^+e^-\to K^+K^-$ reaction  at $E=1019.10$ MeV are shown
Fig.~\ref{fig8} (right). The $M_{\gamma\gamma}$ spectrum for data events
at $E=1019.10$ MeV, selected using the criteria for energy interval V and
the condition on the the maximum $dE/dx$ for two charged tracks
$(dE/dx)_{\rm max}>3$, is shown in Fig.~\ref{fig9} (left). In the fitted
background distribution, the $K^+K^-$ background dominates. It is seen
that the background distribution differs from linear. A difference in
this distribution between the data and the simulation is also observed.
Therefore, for the fit shown in Fig.~\ref{fig9} (left),
the $K^+K^-$ distribution for data events from the four energy points
near the maximum of the $\phi$ resonance, selected with the condition
$(dE/dx)_{\rm max}>4.5$, is used. From the fits to the $M_{\gamma\gamma}$ 
spectra at $E=1019.10$ MeV and $E=1019.96$ MeV, the scale factor for $K^+K^-$
events is found to be $2.00\pm0.06$. 

The scale factor for the $K_SK_L$
events ($S_{K_SK_L}$) is then found from the fit to the
$M_{\gamma\gamma}$ spectrum for data events selected using the
criteria for energy interval V and the condition $\mu{\rm veto}=0$.
A background distribution without a linear component, but with a free 
parameter $S_{K_SK_L}$, is used in the fit. All other scale factors are fixed
at the values obtained above. From the fits to mass spectra at
$E=1019.10$ MeV and $E=1019.96$ MeV, the scale factor is found to be 
$S_{K_SK_L}=1.45\pm 0.20$.

\subsection{Systematic uncertainty in $N_{3\pi}$}
The quality of the fit to the $M_{\gamma\gamma}$ spectrum at energy points
with large statistics, for example, at $E=782.76$ MeV, is poor,
$\chi^2/\nu\approx 15$, where $\nu$ is the number degrees of freedom.
The fractions of the linear background at two close energy points,
782.76 and 782.81 MeV, collected at different running periods, differ
significantly. At these points, the gas gain in the tracking system
differed by approximately a factor of 2. It can be
concluded that the $\pi^0$ line shape used in the fit is not entirely
correct, and the line shape in the data can vary from point to point.
\begin{figure}
\centering
\includegraphics[width=0.40\linewidth]{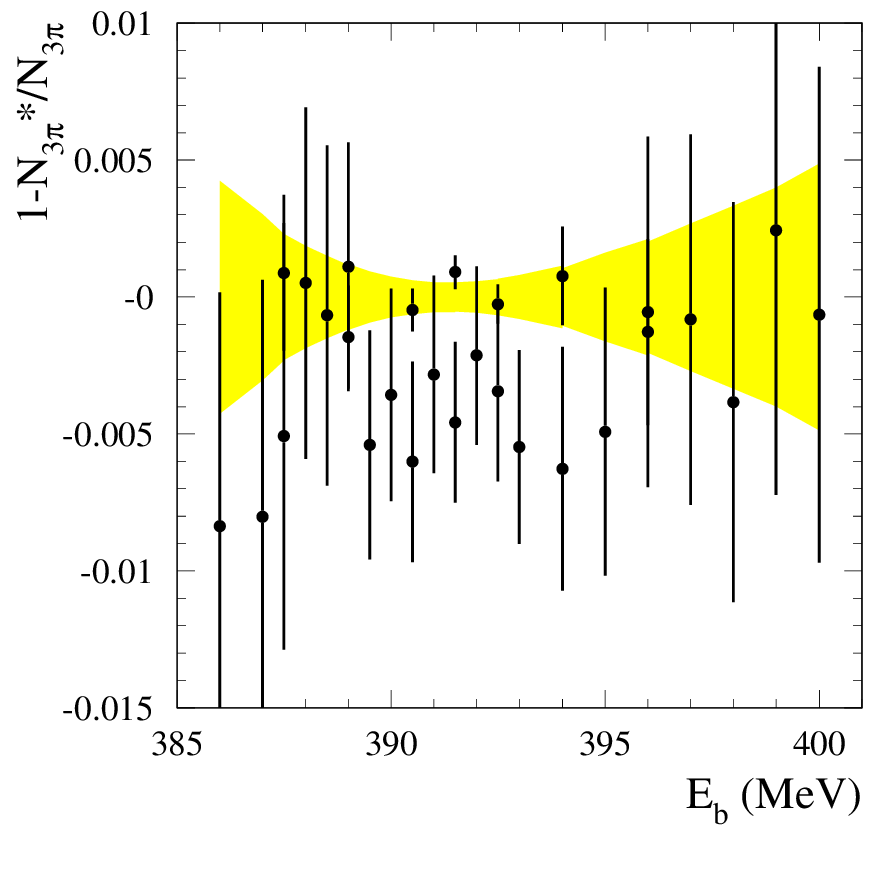}
\caption{Left panel: The relative difference between the number of
signal events obtained by fitting the $M_{\gamma\gamma}$ spectrum and 
the number of signal events obtained by subtracting the background. 
The error bars show the statistical uncertainties, and the shaded band 
shows the systematic uncertainty associated with background subtraction.
\label{fig10}}
\end{figure}

To study the systematic error in $N_{3\pi}$ arising from the imperfect
shape of the signal distribution, we employ an alternative background
subtraction method. Events selected with the following criteria are
analyzed:
\begin{equation}
\chi^2_{3\pi}<100,\,\psi<160^\circ,\,\mu{\rm veto}=0,\,E_{\rm EMC}/E <
0.75,\,n_{\rm ch}=2.
\label{condsub} 
\end{equation}
With these conditions, the background to signal ratio near the
$\omega$ resonance maximum becomes less than 0.01. We perform the fit
to the $M_{\gamma\gamma}$ spectra and determine the numbers of
background events. Figure~\ref{fig9} (right) shows the energy
dependence of the background cross section. Large fluctuations in the
cross section near the $\omega$ resonance maximum are associated with
the event transition from the signal to the background and vice versa.
Away from the resonance maximum, the signal to background ratio
decreases and the influence of this effect diminishes. The background
cross section far from the resonance is well described by the
following formula
\begin{equation}
\sigma_{\rm bkg}=\sigma_{\rm bkg,expec}+0.2\mbox{ nb},\label{csbkg}
\end{equation}
where $\sigma_{\rm bkg,expec}$ is the background cross section
expected from the simulation, and the constant (0.2 nb) is the contribution of
the linear background. The cross section calculated using
Eq.~(\ref{csbkg}) is shown in Fig.~\ref{fig9} (right). The shaded area around
the curve shows the systematic uncertainty of Eq.~(\ref{csbkg}), 
$\Delta \sigma_{\rm bkg}=0.3$ nb, conservatively estimated from the
deviations of the data points from the curve at beam energies $E_b < 385$ MeV
and $E_b>415$ MeV.
The number of signal events can be obtained by subtracting the
background calculated using Eq.~(\ref{csbkg}) from the number of
selected events with $70<M_{\gamma\gamma}<200$ MeV. Fig.~\ref{fig10}
shows the relative difference between the number of signal events
obtained by fitting the $M_{\gamma\gamma}$ spectrum and the number of
signal events obtained by subtracting the background. The error bars show
the statistical uncertainties, and the shaded band shows the systematic
uncertainty associated with background subtraction. The nonstatistical
deviations from zero do not exceed 0.6\%. This value is used as an
estimate of the systematic error associated with the uncertainty of
the signal shape. 
For energy points with small statistics, systematic uncertainties due to
fixing the parameters $\Delta a_1$, $\Delta a_2$, and $\Delta \sigma^2$
are quadratically added to this error, which are
estimated by variation of these parameters within their errors. 

Figure.~\ref{fig10} shows that for many energy points, the fit yields 
a shifted value of $N_{3\pi}$. This shift can be reduced by determining the
number of events ($N_{3\pi,1}$) selected with the conditions
(\ref{condsub}) by subtracting the background, as described above, and
by fitting to the  $M_{\gamma\gamma}$ spectrum  for the remaining
events with $\chi^2_{3\pi}<100$ ($N_{3\pi,2}$). The total number of
signal events is calculated as $N_{3\pi}=N_{3\pi,1}+N_{3\pi,2}$, with
the ratio
$N_{3\pi,2}/N_{3\pi}$ being approximately 0.2. This method is used in
the energy range 763.94--810.58 MeV, where its systematic uncertainty
$\sqrt{(\Delta \sigma_{\rm bkg}L)^2+(0.006N_{3\pi,2})^2}$ does not
exceed $0.006N_{3\pi}$.

To estimate the systematic error arising from uncertainty of the
background shape, we modified the function describing background.
Instead of a linear contribution, we introduced a common scale factor
for the background obtained from the simulation. The difference
between the results of fits with the two background functions is
calculated. To eliminate the contribution of the cosmic-ray
background, the condition $\mu{\rm veto}=0$ is applied. Systematic
errors due to uncertainties in the scale factors for background
processes are determined by their variation within the errors. All of
these contributions to the systematic uncertainty are added
quadratically to the uncertainty associated with the signal line shape.

The numbers of signal events $N_{3\pi}$ obtained for different energy
points with statistical and systematic errors are listed in
Tables~\ref{tab4a}, \ref{tab4b}, and \ref{tab4c}. Below 720 MeV, the systematic
uncertainty in $N_{3\pi}$ is dominated by the uncertainty associated with the
background shape. In the range of 760--1050 MeV, it is dominated
by the uncertainty associated the signal shape.

\section{Detection efficiency\label{deteff}}
\begin{figure}
\centering
\includegraphics[width=0.60\linewidth]{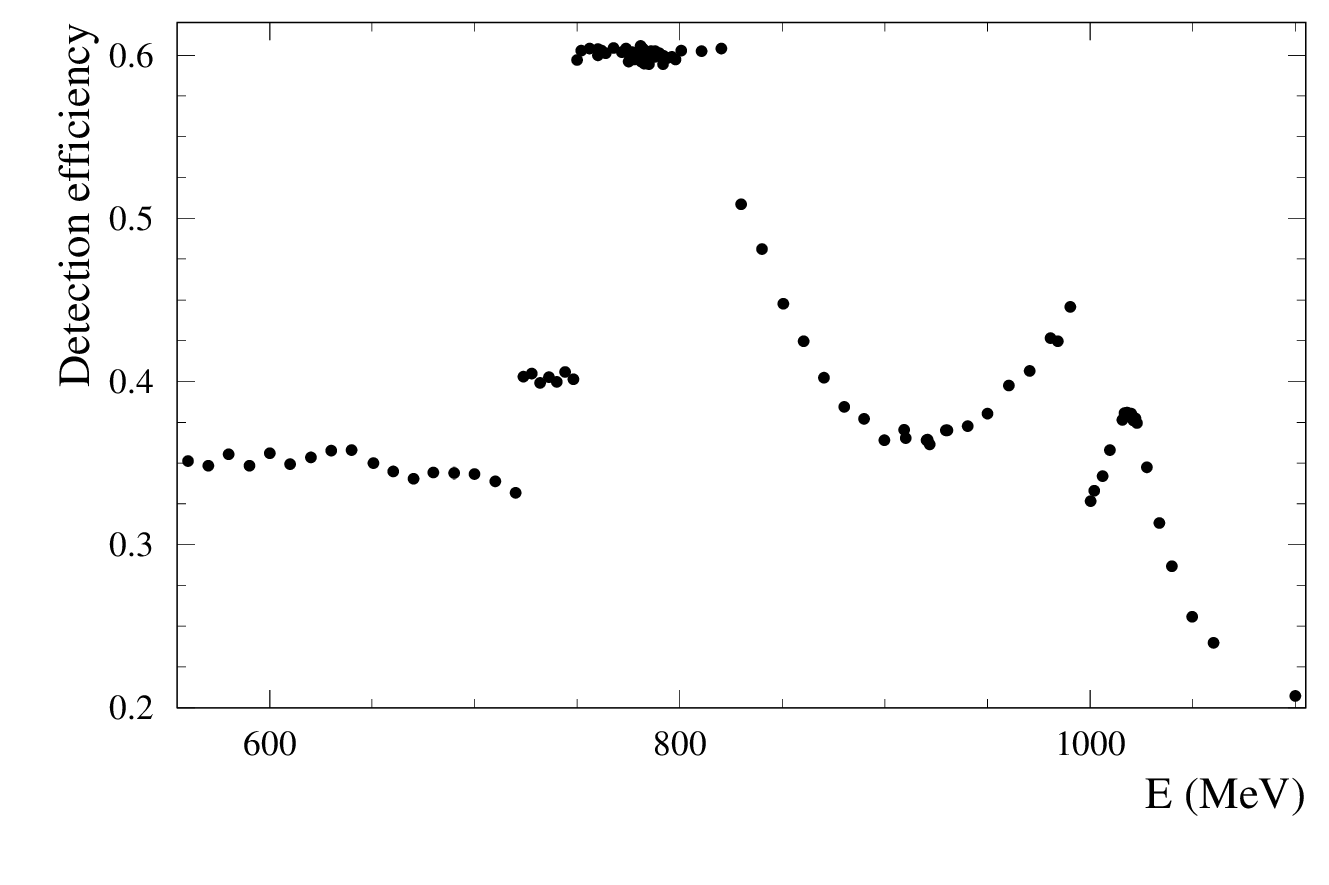}
\caption{The detection efficiency for
$e^+e^-\to\pi^+\pi^-\pi^0$ events obtained using MC simulation.
\label{fig11}}
\end{figure}
The energy dependence of the detection efficiency for 
$e^+e^-\to\pi^+\pi^-\pi^0$ events obtained using MC simulation is
shown in Fig.~\ref{fig11}. Five energy intervals with different
selection criteria (see Table~\ref{tab2}) are clearly visible. In the
region of the $\omega$ resonance (interval I), the efficiency reaches 60\%.
Nonstatistical fluctuations in the efficiency observed in this
interval are due to various experimental conditions (dead detector channels,
beam-induced background, etc.). In intervals IV, III, and I, the efficiency
exhibits a steplike behavior. A more complex energy dependence in intervals II
and V is due to radiative return to the resonances $\omega$ and
$\phi$. Events with a hard ISR photon are 
rejected by the condition $\chi^2_{3\pi}<30$.

The detection efficiency depends on the Born cross section used in simulation. 
Therefore an iterative procedure is performed to determine the
efficiency. The VMD model (see
Sec.~\ref{xsfit}) with the the $\rho$, $\omega$, and $\phi$ parameters
taken from Particle Data Group (PDG) tables~\cite{pdg}, and the
parameters of excited $\omega$-like resonances from
Ref.~\cite{BABAR-3pi-0} is used as
the initial Born cross section. Then we
determinate the detection efficiency and measure the Born cross section
as it is described in Sec.~\ref{xsfit}. The obtained cross section is used in the
second-iteration simulation. We do not perform full simulation, but instead reweight 
simulated events using the new Born cross section. Six iterations are
enough to reach the precision in the efficiency better 0.025\% for all
energy points.

The imperfect MC simulation of the detector response for charged pions
and photons leads to an inaccuracy in the simulation-derived detection 
efficiency. Therefore, we introduce efficiency corrections. 
As a first step, we determine the corrections due to
conditions in intervals II–V, which are additional to those in
interval I.
To do this, we remove one of the additional conditions used in these
intervals and estimate the efficiency correction factor as
\begin{equation}
C_i= \frac{N^{\rm data}/N_{\rm MC}}{N^{\rm data}_{0}/N^{\rm MC}_{0}}, 
\label{eqcorr}
\end{equation}
where
$N^{\rm data}$ ($N^{\rm data}_{0}$) and $N^{\rm MC}$ ($N^{\rm MC}_{0}$) are 
the numbers of signal events in the data and in the signal simulation,
respectively, for the modified (nominal) selection criteria.

To determine the correction for the condition $\chi^2_{3\pi}<30$
($C_{\chi^2_{3\pi}}$), we analyze three energy regions, where the
radiation from the initial state is small. The correction is
calculated as a weighted average over the energy points. 
It is found to be $C_{\chi^2_{3\pi}}=1.004\pm0.004$ and
$1.009\pm0.007$ near the maxima of the $\phi$ ($E=1015$--1023 MeV) 
and $\omega$ ($E=776$--797 MeV) resonances, respectively, and 
$1.013\pm0.005$ in the range $E=719$--751 MeV. 
Due to nonstatistical fluctuations of $C_{\chi^2_{3\pi}}$ values,
the uncertainties in the $\phi$ and $\omega$ regions are calculated as
the standard deviation of the corrections at different energy points.
In the latter range, the uncertainty is statistical.
The corrections in different energy regions are consistent with each
other. However, a slight decrease in $C_{\chi^2_{3\pi}}$ with
increasing energy cannot be ruled out. To account for this
possible decrease, the correction with the highest uncertainty, obtained
near the $\omega$ resonance, $1.009\pm0.007$, is used in energy 
interval III. 

\begin{figure}
\centering
\includegraphics[width=0.70\linewidth]{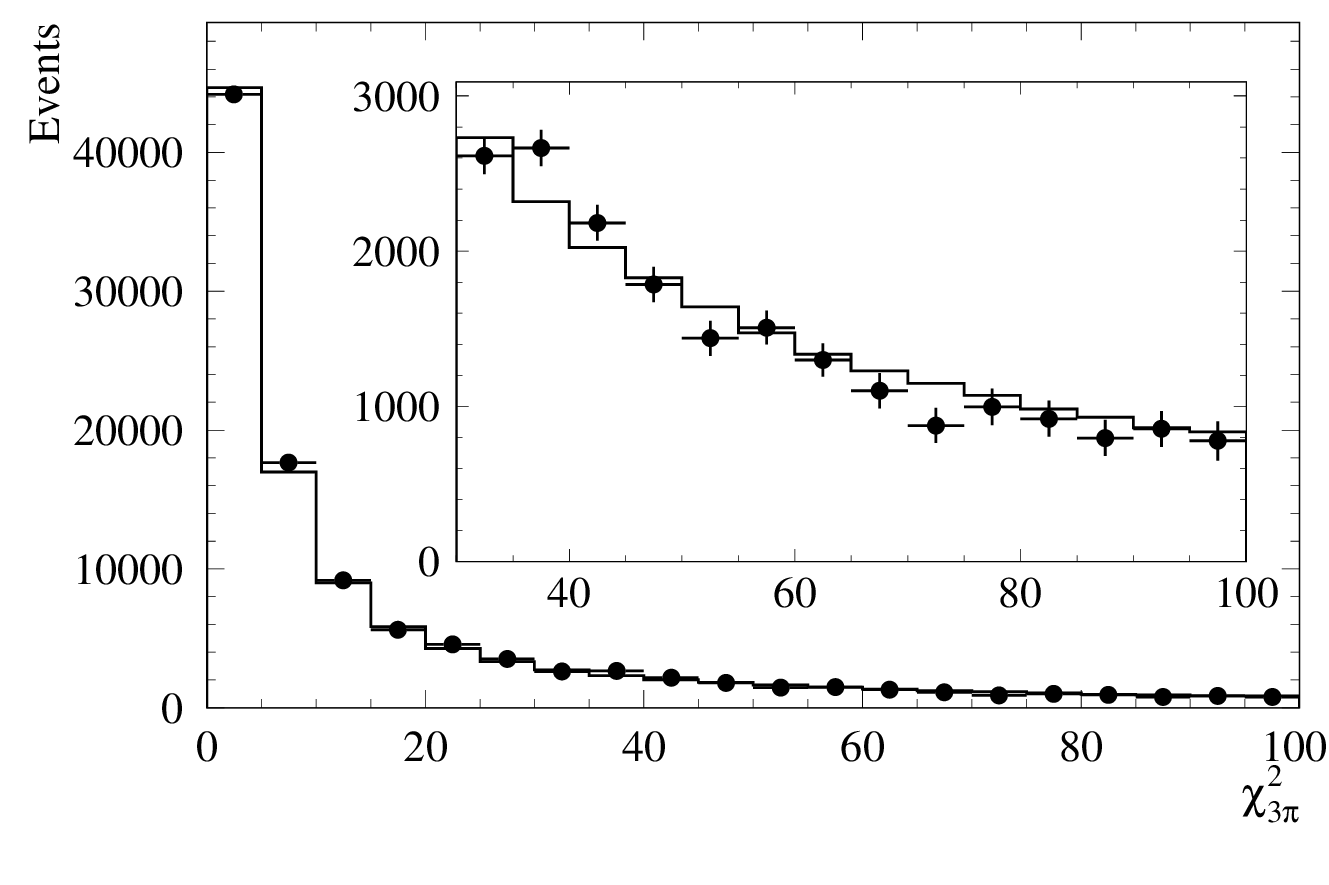}
\caption{The $\chi^2_{3\pi}$ distribution for data signal events
(points with error bars) and simulation (histogram) in the energy range
810--990 MeV.
\label{chi810-990}}
\end{figure}
The data-MC difference in the detector response for events with
a hard ISR photon and the ISR photon angular and energy distributions
(in particular, we simulate emission only one ISR photon) may be
additional sources of the systematic uncertainty. To estimate 
effect of these sources, we calculate the efficiency correction
$C_{\chi^2_{3\pi}}$ for the energy range 810--990 MeV, where
the $\chi^2_{3\pi}$ distribution and the $\pi^0$ line shape are 
strongly distorted by the radiative return to the $\omega$ resonance.
The obtained value $C_{\chi^2_{3\pi}}=0.998\pm0.005$ is approximately
1$\sigma$ lower than the the values obtained near the maxima of the 
$\omega$ and $\phi$ resonances. To test whether this deviation is
non-statistical, we construct a $\chi^2_{3\pi}$ distribution for signal
events in the 810--990 MeV range. This distribution is shown in Fig.~\ref{chi810-990}.
The number of $3\pi$ events in each histogram bin is determined from the fit
to the $M_{\gamma\gamma}$ distribution. It is seen that the data points
at $\chi^2_{3\pi}>50$ lie below the values expected from the 
$e^+e^-\to \pi^+\pi^-\pi^0$ simulation. Therefore, we consider the
difference between $C_{\chi^2_{3\pi}}$ values obtained in the energy
range of 810--990 MeV and at the $\omega$ resonance (1.1\%) as a 
non-statistical shift. The value of this shift is used as an estimate of the 
systematic uncertainty in $C_{\chi^2_{3\pi}}$ the for energy interval II. 
Since the measured Born cross section is determined from the number of 
events without ISR, the value of $C_{\chi^2_{3\pi}}$ is taken to be equal
to the value of 1.009 obtained in the $\omega$ region.
The same systematic uncertainty is introduced for $C_{\chi^2_{3\pi}}$
at $E>1045$ MeV, where a significant fraction of selected $3\pi$ events 
have high-energy ISR photons.

It should be noted that due to the strong distortion of the $\pi^0$
line shape, we do not expect the simulation to correctly reproduce the
detection efficiency at high values of $\chi^2_{3\pi}$. This is why
the strict condition $\chi^2_{3\pi}<30$ is used in interval II. 
At smaller values of $\chi^2_{3\pi}$, the agreement between the data and the
simulation is better. In particular, the correction $C_{\chi^2_{3\pi}}$
for energy range 810--990 MeV, calculated with the condition 
$\chi^2_{3\pi}<50$ instead of $\chi^2_{3\pi}<100$ is 
$1.003\pm0.002$ and agrees well with the corresponding values
for the $\omega$ and $\phi$ energy regions, $1.006\pm0.003$ and
$1.004\pm0.003$, respectively. So, the above value of the systematic
uncertainty in $C_{\chi^2_{3\pi}}$ in interval II is rather conservative.

The correction $C_\psi$, associated
with the angular conditions $|\Delta\varphi| >2.3^\circ$ and
$\psi<140^\circ$, used in intervals III, IV, V, is determined in 
the energy range near the $\omega$ resonance
($E=750$--815 MeV). It is found to be $C_\psi=0.9987\pm0.0003$.
Three additional corrections in energy interval IV, associated with the
conditions  
$E_{\rm EMC}/E < 0.75$, $\mu{\rm veto}=0$, and $n_{\rm ch}=2$, are
$C_{\rm EMC}=0.981\pm 0.003$, $C_{\mu{\rm veto}}=0.998\pm 0.002$,
and $C_{n_{\rm ch}}=0.983\pm 0.003$, respectively. They are obtained
in the energy range $E=695$--751 MeV.
The additional correction in energy interval V, 
associated with the condition  $\chi^2_{4\pi}>500$, is equal to 
$C_{n_{\chi^2_{4\pi}}}=1.0036\pm 0.0005$.

Above, we considered efficiency corrections calculated relative to events 
selected with the criteria for $\omega$ meson energy range (interval I). 
To determine the efficiency corrections for interval I, we analyze events 
that do not satisfy these criteria.

A sample of events (C1) with at least two charged tracks originating from the 
collider interaction region, one or more photon, and
$\chi^2_{3\pi}>100$ for events with $n_\gamma>1$ is
used to study the data-simulation difference in photon loss and the
fraction of signal events rejected by the condition $\chi^2_{3\pi}<100$.
These two effects cannot be studied separately due to the high
probability of detecting a spurious photon in an event.

To suppress the cosmic-ray background and physical background mainly from 
$e^+e^-\to e^+e^-\gamma$ and $e^+e^-\to \pi^+\pi^-\gamma$ events, the 
conditions $\mu{\rm veto}=0$ and $\psi<140^\circ$ are applied. The $e^+e^-\to
e^+e^-\gamma$ background is additionally suppressed by the condition
$E_{\rm EMC}/E < 0.75$. 
The beam-induced background is subtracted by fitting the
$z$-distribution of the charged track with the smallest $d_i$.
\begin{figure}
\centering
\includegraphics[width=0.40\linewidth]{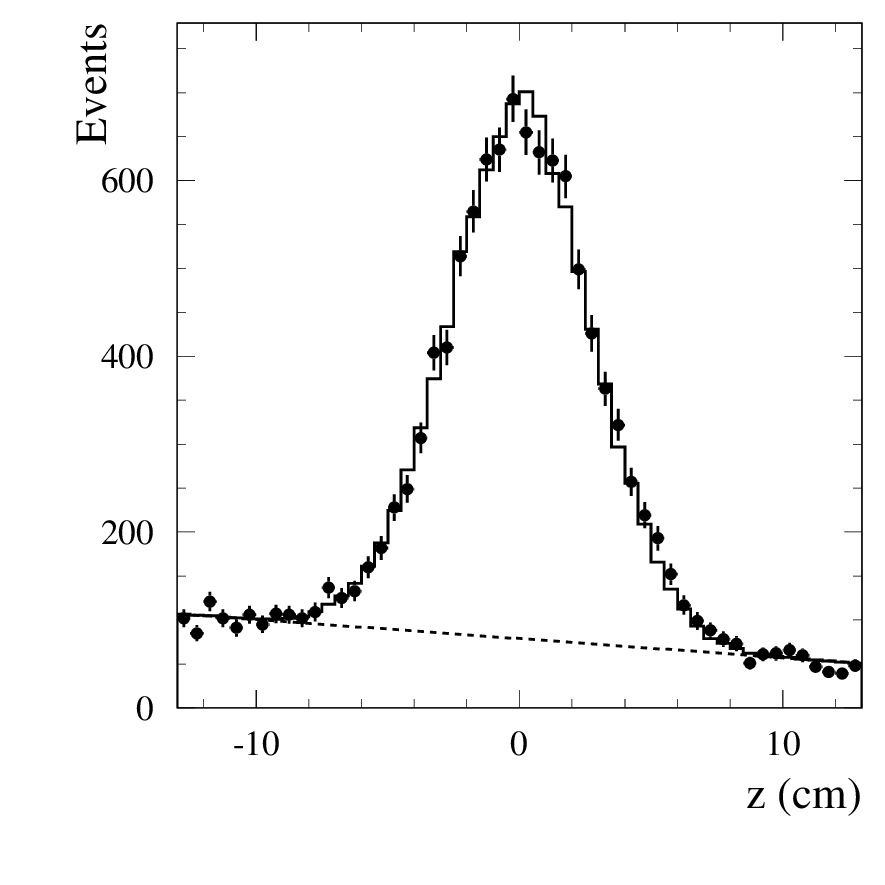}
\includegraphics[width=0.40\linewidth]{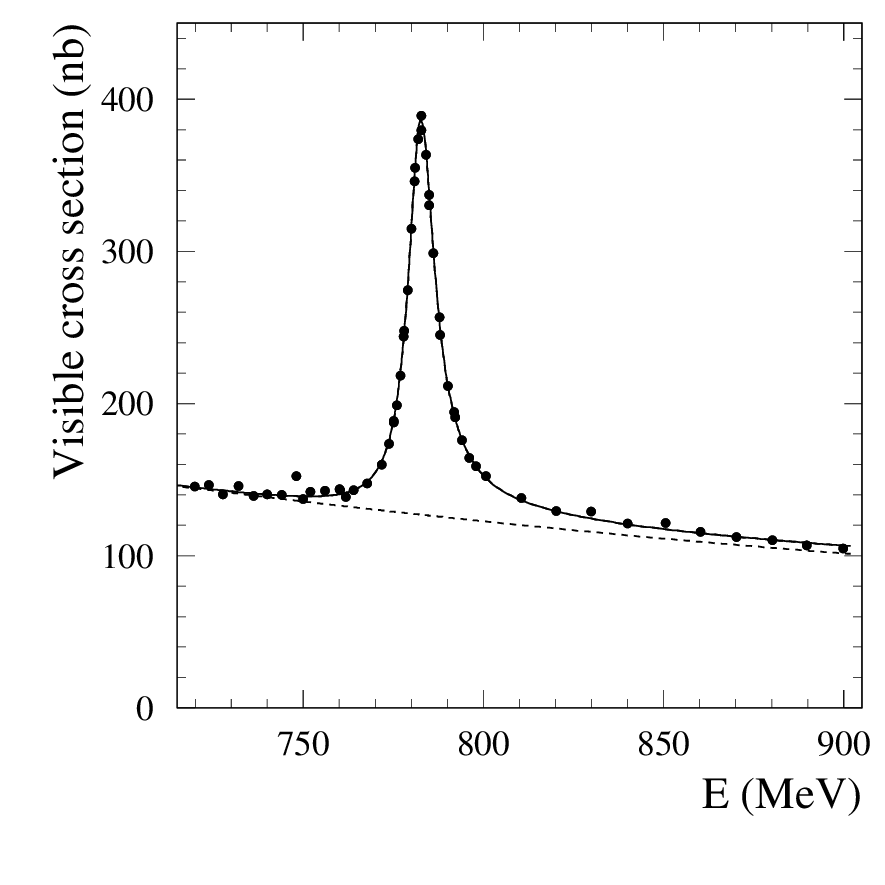}
\caption{Left panel: The $z$-distribution of the charged track with
the smallest $d$. The solid histogram is the result of the fit
described in the text. The dashed line represents the contribution of
the beam-induced background.
Right panel: The energy dependence of the visible cross section for
selected events from C1 sample. The solid curve is the result of the
fit by the sum of the $\omega$ line shape and the quadratic background.
The dashed curve is the fitted background cross section. 
\label{fig12}}
\end{figure}
This distribution is shown in Fig.~\ref{fig12} (left). It is fitted
by the sum of the simulated signal distribution and a linear function
for the beam-induced background. From the fit, the number of
beam-induced background events ($N_{\rm beam}$) is obtained at each
energy point. 

\begin{figure}
\centering
\includegraphics[width=0.40\linewidth]{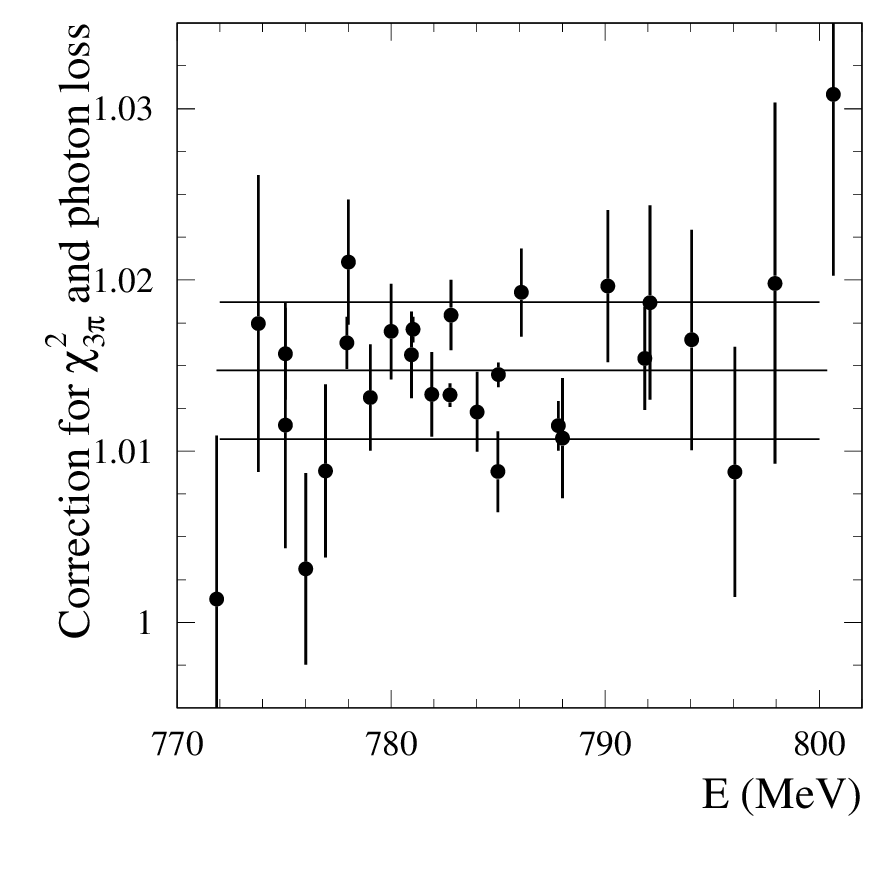}
\includegraphics[width=0.40\linewidth]{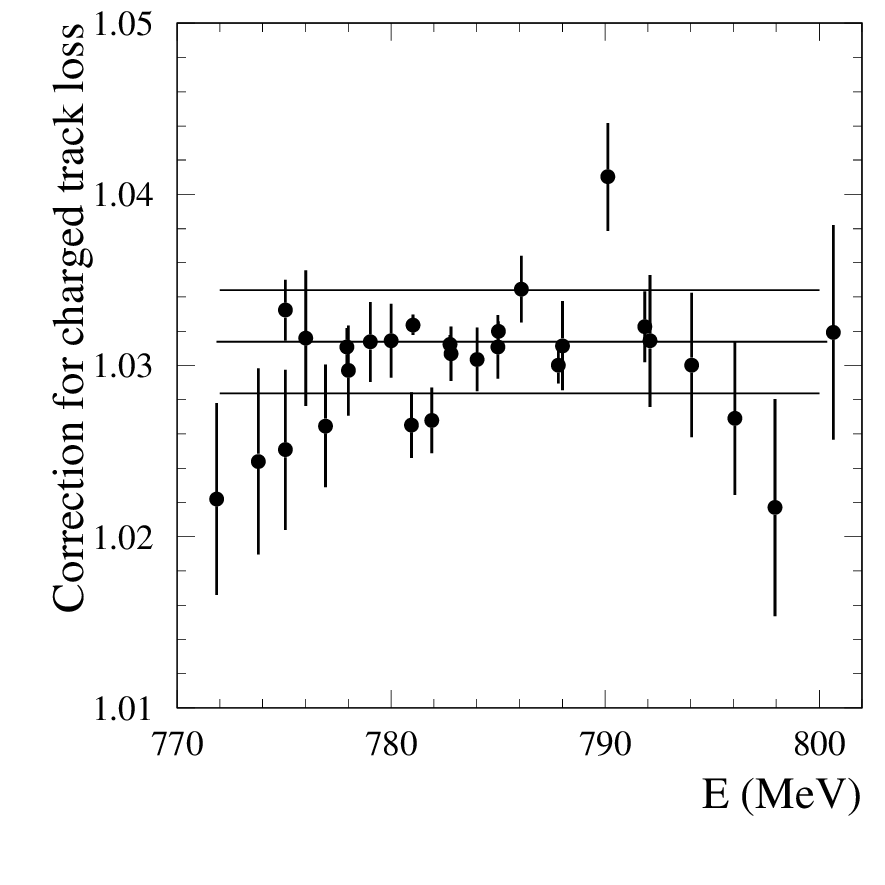}
\caption{The efficiency corrections for photon loss and the condition
$\chi^2_{3\pi}<100$ (left) and charged track loss (right). The middle line
shows the result of the fit with a constant. The upper and lower lines show
the interval $\pm 1\sigma_{\rm sys}$. 
\label{fig13}}
\end{figure}

To determine the remaining background, we use the fact that the process 
$e^+e^-\to \pi^+\pi^-\pi^0$ dominates in the $\omega$-resonance region.
The energy dependence of the visible cross section for selected events
after subtracting the expected $e^+e^-\to \pi^+\pi^-\gamma$
($N_{\pi^+\pi^-\gamma}$) and $\eta\gamma$ ($N_{\eta\gamma} $) backgrounds,
and the beam-induced background is shown in Fig.~\ref{fig12} (right). The
peak of the $\omega$-resonance from $e^+e^-\to \pi^+\pi^-\pi^0$ events is
clearly seen above the almost linear background. The cross section
is fitted by the sum of the $\omega$ line shape obtained in Sec.~\ref{xsfit}
and the quadratic background contribution.
Finally, the number of $\pi^+\pi^-\pi^0$ events at $i$th energy point
of C1 sample is determined as
\begin{equation}
N_{C1,3\pi}=N_{C1}-N_{\rm beam}- N_{\pi^+\pi^-\gamma}-N_{\eta\gamma}-
\sigma_{\rm bkg}(E_i)L_i,
\end{equation}
where $N_{C1}$ is the number of selected events at $i$th point of C1 sample,
and $\sigma_{\rm bkg}(E_i)$ is the background cross section obtained
as result of the fit shown in Fig.~\ref{fig12} (right).
For normalization, events with $\chi^2_{3\pi}<100$, satisfying the background
suppression criteria described above, are used. The number of
$\pi^+\pi^-\pi^0$ events at $i$th energy point is determined 
by fitting the $M_{\gamma\gamma}$ distribution. The energy dependence
of the efficiency correction for photon loss and the condition 
$\chi^2_{3\pi}<100$ is shown in Fig.~\ref{fig13} (left). The average
value is $C_\gamma=1.015\pm0.004$,
where the uncertainty is calculated as the standard deviation of the 
corrections at different points in the energy range $E=777$--795 MeV.

A sample of events (C2) with exactly one charged track originating from the 
collider interaction region, and two or more photon is
used to study the data-simulation difference in charged track loss.
Any number of charged tracks with $d>1$ cm or $|z|>15$ cm is allowed.   

To suppress background, the conditions $\mu{\rm veto}=0$ and 
$100<m^\ast_{2\gamma}<170$ MeV are applied, where $m^\ast_{2\gamma}$
is the invariant mass of two photons with energy greater than 50 MeV,
closest to the $\pi^0$ mass. To suppress the background from the 
$e^+e^-\to \pi^0\gamma$ and $e^+e^-\to \pi^0 e^+e^-$ events, 
we reject events
with $E_{\rm EMC}/E > 0.8$ and $E_{\gamma, \rm max}/E > 0.8$, where
$E_{\gamma, \rm max}$ is the energy of the most energetic photon in the
event.

The beam-induced background is determined from the fit to the
$z$-distribution for the charged track originating from the 
interaction region. Similar to C1 sample, we subtract the beam-induced
background, and expected backgrounds from the 
$\pi^+\pi^-\gamma$, $\pi^0 e^+e^-$, $\pi^0\gamma$, and $\eta\gamma$
events, and fit to the energy dependence of the visible cross section
to determine the remaining background cross section. 
The same procedure is used for the normalization sample containing 
events with at least two charged tracks originating from the
collider interaction region. The energy dependence of the efficiency 
correction for charged track loss is shown in Fig.~\ref{fig13} (right).
The average value is $C_{\rm ch. tr.}=1.032\pm0.003$,
where the uncertainty is calculated as the standard deviation of the
corrections at different points in the energy range $E=777$--795 MeV.

The correction $C_{\rm ch. tr.}$ is mainly due to the difference
between data and simulation in the efficiency of the tracking system 
at small polar angles. In particular, the simulation does not take into 
account the reduction in gas gain near the ends of the sensitive wires. 
We can estimate this effect by comparing the fractions of signal events
with ($\theta_\pi<36^\circ$ or $\theta_\pi>144^\circ$) in data and
simulation, where $\theta_\pi$ is the charged pion polar angle.
The tracks with $36^\circ<\theta_\pi<144^\circ$ contain more
than 4 hits in the drift chamber and, therefore, are expected to have a 
reconstruction efficiency close to unity. We study events near the
maximum of the $\omega$ resonance selected with the condition 
$\chi^2_{3\pi}<100$. Assuming unit reconstruction efficiency for tracks with
$36^\circ<\theta_\pi<144^\circ$, we obtain that the number of signal
events without any polar angle constraints in the data is
$(2.2\pm0.4)\%$ less than in the simulation.

The superimposing the charged track from the beam background on the 
signal event can also lead to charged track loss, which may differ
between data and simulation. We study events near
the maximum of the $\omega$ resonance selected with the condition
$\chi^2_{3\pi}<100$ and observe that the fraction of signal events with
$n_{\rm ch}>2$ in simulation ($\sim 12\%$) is larger than that in data
($\sim 11\%$). This difference of approximately 1\% can be partly
explained by the difference between the data and the simulation in track
loss for events with extra charged tracks.

The third effect leading to the charged track loss
is nuclear interaction in material before the tracking system. The
cross section of nuclear interaction strongly depends on charged pion
momentum $p_\pi$. This can lead to the energy dependence of the correction
$C_{\rm ch. tr.}$. To study the possible energy dependence, we determine
the corrections for events with $E_{\pi^0}>290$ MeV and $E_{\pi^0}<240$
MeV. The average charged pion momenta for signal events selected with these
conditions are 190 MeV/$c$ and 240 MeV/$c$, respectively. In this
momentum range, the nuclear interaction cross section grows more than
two times. Since we do not observe any significant difference between
the corrections with $E_{\pi^0}>290$ MeV and $E_{\pi^0}<240$
MeV, we conclude that the nuclear interaction is reproduced by
simulation correctly.

The total correction factor ($C$) for energy intervals I--V obtained as 
products of correction factors for individual conditions are listed in
Table~\ref{tab2}. It is assumed that the corrections $C_\gamma$ and
$C_{\rm ch. tr.}$ are independent of energy and are the same for all
intervals. This assumption was verified above for the $C_{\rm ch.tr.}$
correction. 
The energy dependence of $C_\gamma$ can arise from a change
in the shape of the $\chi^2_{3\pi}$ distribution. To estimate the
value of this effect, we use the energy dependence of the
$C_{\chi^2_{3\pi}}$ correction. This correction was measured at three
energy points. The deviation from the value measured near the $\omega$
resonance does not exceed 0.5\%. This deviation is taken as a measure
of the systematic uncertainty associated with the possible energy
dependence of $C_\gamma$ for intervals II-V.
The corrected efficiencies calculated as
$\varepsilon= \varepsilon^{\rm MC} /C$, where $\varepsilon^{\rm MC}$ is
the efficiency obtained from the MC simulation, are listed
in Tables~\ref{tab4a}, \ref{tab4b}, and \ref{tab4c}.
\begin{table}
\caption{\footnotesize The center-of-mass energy ($E$), detection
efficiency ($\varepsilon$), number of selected events ($N{3\pi}$),
radiative correction ($1+\delta$), correction for the energy spread
($1+\delta_E$), Born cross section for the process $e^+e^-\to
\pi^+\pi^-\pi^0$ ($\sigma$). 
The first error in the number of events and the cross section is statistical,
the second is systematic. 
\label{tab4a}}
\begin{ruledtabular}
\fontsize{9pt}{9pt}\selectfont
\begin{tabular}{cccccc}
$E$, MeV & $\varepsilon$ & $N{3\pi}$ & $1+\delta$ & $1+\delta_E$ & $\sigma$, nb \\
\hline
 560.237& 0.3440&$       0.1\pm     2.5\pm     4.2$&  $0.9055\pm0.0008$&
 1.0000&$    0.01\pm 0.15 \pm 0.26$\\		       	         
 570.215& 0.3410&$      11.1\pm     6.0\pm     9.3$&  $0.9046\pm0.0007$&
 1.0000&$    0.47\pm 0.26 \pm 0.40$\\		       	         
 580.145& 0.3482&$       4.5\pm     5.1\pm     6.3$&  $0.9035\pm0.0009$&
 1.0000&$    0.16\pm 0.19 \pm 0.23$\\		       	         
 590.206& 0.3410&$       0.0\pm     2.8\pm     5.0$&  $0.9023\pm0.0008$&
 1.0000&$    0.00\pm 0.15 \pm 0.26$\\		       	         
 600.151& 0.3486&$       0.0\pm     4.7\pm    12.3$&  $0.9010\pm0.0009$&
 1.0000&$    0.00\pm 0.16 \pm 0.43$\\		       	         
 610.068& 0.3421&$       1.2\pm     2.9\pm     2.3$&  $0.8996\pm0.0008$&
 1.0000&$    0.06\pm 0.15 \pm 0.12$\\		       	         
 620.167& 0.3461&$       0.3\pm     2.8\pm     4.1$&  $0.8979\pm0.0010$&
 1.0000&$    0.01\pm 0.11 \pm 0.16$\\		       	         
 630.085& 0.3502&$       1.4\pm     3.8\pm     5.3$&  $0.8962\pm0.0009$&
 1.0000&$    0.05\pm 0.15 \pm 0.20$\\		       	         
 640.019& 0.3504&$       4.7\pm     4.2\pm     6.2$&  $0.8942\pm0.0010$&
 1.0000&$    0.18\pm 0.16 \pm 0.24$\\		       	         
 650.612& 0.3426&$       2.8\pm     3.9\pm     5.7$&  $0.8919\pm0.0011$&
 1.0000&$    0.11\pm 0.15 \pm 0.23$\\		       	         
 660.328& 0.3377&$       1.5\pm     4.0\pm     0.7$&  $0.8896\pm0.0010$&
 1.0000&$    0.06\pm 0.16 \pm 0.03$\\		       	         
 670.201& 0.3333&$       5.8\pm     4.5\pm     1.1$&  $0.8869\pm0.0011$&
 1.0000&$    0.23\pm 0.18 \pm 0.04$\\		       	         
 679.781& 0.3370&$     557.5\pm    34.3\pm    13.1$&  $0.8841\pm0.0011$&
 1.0000&$    0.64\pm 0.04 \pm 0.02$\\		       	         
 689.880& 0.3368&$      69.3\pm    11.2\pm     3.0$&  $0.8807\pm0.0011$&
 1.0000&$    0.82\pm 0.13 \pm 0.04$\\		       	         
 699.860& 0.3362&$     110.5\pm    12.8\pm     2.8$&  $0.8770\pm0.0010$&
 1.0001&$    1.55\pm 0.18 \pm 0.04$\\		       	         
 709.969& 0.3319&$     118.7\pm    12.8\pm     4.6$&  $0.8726\pm0.0011$&
 1.0000&$    2.15\pm 0.23 \pm 0.09$\\		       	         
 719.930& 0.3250&$    2348.5\pm    56.2\pm    27.1$&  $0.8676\pm0.0009$&
 1.0001&$    3.52\pm 0.09 \pm 0.06$\\		       	         
 723.798& 0.3818&$     177.7\pm    18.8\pm     1.8$&  $0.8654\pm0.0009$&
 1.0001&$    3.71\pm 0.39 \pm 0.06$\\		       	         
 727.733& 0.3836&$     198.5\pm    18.2\pm     3.4$&  $0.8631\pm0.0008$&
 1.0001&$    4.83\pm 0.45 \pm 0.10$\\		       	         
 731.994& 0.3782&$     237.9\pm    19.7\pm     3.5$&  $0.8602\pm0.0008$&
 1.0002&$    6.25\pm 0.52 \pm 0.12$\\		       	         
 736.202& 0.3815&$     413.8\pm    25.1\pm     6.5$&  $0.8571\pm0.0007$&
 1.0002&$    7.54\pm 0.46 \pm 0.15$\\		       	         
 739.948& 0.3790&$     453.7\pm    25.3\pm     4.7$&  $0.8541\pm0.0006$&
 1.0001&$   10.08\pm 0.57 \pm 0.16$\\		       	         
 744.089& 0.3845&$     536.4\pm    27.0\pm     4.1$&  $0.8503\pm0.0005$&
 1.0001&$   12.71\pm 0.65 \pm 0.18$\\		       	         
 748.048& 0.3803&$     791.0\pm    33.1\pm    10.4$&  $0.8462\pm0.0005$&
 1.0001&$   16.55\pm 0.70 \pm 0.30$\\		       	         
 750.019& 0.5659&$   49865.0\pm   395.9\pm   371.4$&  $0.8440\pm0.0004$&
 1.0003&$   19.03\pm 0.18 \pm 0.27$\\		       	         
 751.989& 0.5758&$    1242.6\pm    56.2\pm    22.7$&  $0.8416\pm0.0004$& 
 1.0002&$   21.56\pm 0.98 \pm 0.44$\\		       	         
 756.093& 0.5768&$    1848.1\pm    63.8\pm    26.8$&  $0.8359\pm0.0003$&
 1.0002&$   30.25\pm 1.06 \pm 0.51$\\		       	         
 760.024& 0.5766&$    3834.0\pm    92.6\pm    45.7$&  $0.8295\pm0.0002$&
 1.0004&$   43.50\pm 1.07 \pm 0.64$\\		       	         
 760.188& 0.5731&$   14974.7\pm   179.3\pm    97.7$&  $0.8292\pm0.0003$&
 1.0005&$   44.30\pm 0.55 \pm 0.48$\\		       	         
 761.852& 0.5757&$    4367.9\pm    91.0\pm    29.7$&  $0.8261\pm0.0002$&
 1.0006&$   55.26\pm 1.19 \pm 0.61$\\		       	         
 763.938& 0.5743&$    5383.5\pm    93.3\pm    60.6$&  $0.8219\pm0.0003$&
 1.0004&$   65.56\pm 1.19 \pm 0.93$\\		       	         
 767.763& 0.5774&$   13025.7\pm   134.7\pm    85.5$&  $0.8128\pm0.0002$&
 1.0008&$  106.96\pm 1.19 \pm 1.16$\\		       	         
 771.846& 0.5749&$   20403.4\pm   163.8\pm    80.8$&  $0.8008\pm0.0002$&
 1.0011&$  197.54\pm 1.81 \pm 1.87$\\		       	         
 773.788& 0.5770&$   19891.0\pm   158.4\pm    55.4$&  $0.7941\pm0.0002$&
 1.0014&$  280.56\pm 2.64 \pm 2.54$\\		       	         
 775.057& 0.5737&$   23889.0\pm   169.1\pm    55.2$&  $0.7895\pm0.0002$&
 1.0017&$  370.25\pm 3.27 \pm 3.30$\\		       	         
 775.059& 0.5692&$  178207.2\pm   463.0\pm   434.9$&  $0.7895\pm0.0002$&
 1.0027&$  371.86\pm 1.54 \pm 3.33$\\		       	         
 776.019& 0.5742&$   34459.9\pm   201.7\pm    69.4$&  $0.7860\pm0.0002$&
 1.0017&$  456.18\pm 3.38 \pm 4.03$\\		       	         
 776.944& 0.5747&$   36835.0\pm   206.1\pm    68.0$&  $0.7828\pm0.0002$&
 1.0022&$  568.16\pm 4.31 \pm 5.00$\\		       	         
 777.925& 0.5705&$  412833.6\pm   684.2\pm   755.7$&  $0.7798\pm0.0002$&
 1.0031&$  728.24\pm 2.51 \pm 6.41$\\		       	         
 778.017& 0.5735&$   69411.5\pm   280.1\pm   120.8$&  $0.7796\pm0.0002$&
 1.0018&$  737.60\pm 4.47 \pm 6.47$\\			         
 779.023& 0.5729&$   83795.9\pm   306.5\pm   139.5$&  $0.7778\pm0.0002$&
 1.0013&$  944.72\pm 5.17 \pm 8.28$\\			         
 779.994& 0.5736&$   96689.0\pm   327.0\pm   171.5$&  $0.7782\pm0.0002$&
 1.0002&$ 1178.99\pm 6.32 \pm 10.36$\\
\end{tabular}
\end{ruledtabular}
\end{table}
\begin{table}
\caption{\footnotesize The center-of-mass energy ($E$), detection
efficiency ($\varepsilon$), number of selected events ($N{3\pi}$),
radiative correction ($1+\delta$), correction for the energy spread
($1+\delta_E$), Born cross section for the process $e^+e^-\to
\pi^+\pi^-\pi^0$ ($\sigma$). 
The first error in the number of events and the cross section is statistical,
the second is systematic. 
\label{tab4b}}
\begin{ruledtabular}
\fontsize{9pt}{9pt}\selectfont
\begin{tabular}{cccccc}
$E$, MeV & $\varepsilon$ & $N{3\pi}$ & $1+\delta$ & $1+\delta_E$ & $\sigma$, nb \\
\hline
 780.960& 0.5786&$  108314.5\pm   343.3\pm   162.9$&  $0.7821\pm0.0002$&
 0.9990&$ 1396.92\pm 6.95 \pm 12.20$\\			         
 781.023& 0.5692&$ 2331460.0\pm  1596.6\pm  3440.8$&  $0.7825\pm0.0002$&
 0.9977&$ 1415.55\pm 2.87 \pm 12.36$\\			         
 781.898& 0.5771&$  110880.5\pm   346.3\pm   161.3$&  $0.7910\pm0.0002$&
 0.9982&$ 1564.83\pm 7.77 \pm 13.66$\\		       	         
 782.760& 0.5681&$ 3678420.0\pm  1997.2\pm  5338.3$&  $0.8047\pm0.0003$&
 0.9949&$ 1603.83\pm 3.08 \pm 14.01$\\		       	         
 782.807& 0.5757&$  172385.6\pm   433.4\pm   285.9$&  $0.8056\pm0.0003$&
 0.9969&$ 1587.08\pm 6.56 \pm 13.92$\\		       	         
 784.019& 0.5746&$  136124.9\pm   384.6\pm   209.9$&  $0.8345\pm0.0003$& 
 0.9984&$ 1430.95\pm 7.97 \pm 12.52$\\		       	         
 784.997& 0.5729&$  134076.0\pm   381.8\pm   208.8$&  $0.8641\pm0.0003$&
 0.9997&$ 1207.90\pm 5.97 \pm 10.57$\\		       	         
 785.023& 0.5679&$ 2737805.0\pm  1728.1\pm  4027.1$&  $0.8649\pm0.0003$&
 0.9996&$ 1188.83\pm 2.45 \pm 10.39$\\		       	         
 786.090& 0.5753&$  117118.2\pm   358.2\pm   191.4$&  $0.9013\pm0.0005$&
 1.0007&$  943.54\pm 4.74 \pm  8.28$\\		       	         
 787.811& 0.5721&$  441148.3\pm   699.0\pm   691.7$&  $0.9640\pm0.0006$&
 1.0016&$  662.77\pm 1.95 \pm  5.81$\\		       	         
 788.023& 0.5754&$   74116.1\pm   287.7\pm   130.4$&  $0.9719\pm0.0007$&
 1.0014&$  618.71\pm 3.51 \pm  5.45$\\		       	         
 790.133& 0.5742&$   53766.7\pm   249.8\pm   124.9$&  $1.0499\pm0.0014$&
 1.0011&$  404.24\pm 2.64 \pm  3.64$\\		       	         
 791.856& 0.5677&$  124984.0\pm   381.5\pm   266.0$&  $1.1119\pm0.0018$&
 1.0014&$  301.75\pm 1.34 \pm  2.72$\\		       	         
 792.108& 0.5728&$   36237.8\pm   206.4\pm    79.8$&  $1.1208\pm0.0017$&
 1.0009&$  289.36\pm 2.21 \pm  2.61$\\		       	         
 794.047& 0.5715&$   32269.8\pm   197.0\pm    87.9$&  $1.1876\pm0.0030$&
 1.0009&$  214.76\pm 1.80 \pm  2.01$\\		       	         
 796.067& 0.5719&$   28212.8\pm   186.6\pm    86.6$&  $1.2538\pm0.0027$&
 1.0006&$  165.92\pm 1.53 \pm  1.56$\\		       	         
 797.955& 0.5706&$   16295.7\pm   144.7\pm    72.5$&  $1.3125\pm0.0041$&
 1.0003&$  132.80\pm 1.67 \pm  1.35$\\		       	         
 800.668& 0.5757&$   18382.6\pm   155.2\pm    75.1$&  $1.3914\pm0.0042$& 
 1.0002&$  101.38\pm 1.26 \pm  1.01$\\		       	         
 810.583& 0.5755&$   15882.5\pm   147.3\pm   105.1$&  $1.6283\pm0.0066$&
 1.0001&$   50.09\pm 0.78 \pm  0.58$\\		       	         
 820.284& 0.5768&$   11700.6\pm   148.9\pm    74.9$&  $1.7901\pm0.0107$& 
 1.0001&$   32.09\pm 0.74 \pm  0.39$\\		       	         
 829.859& 0.4815&$    4792.0\pm    94.2\pm    35.4$&  $1.8930\pm0.0124$& 
 1.0000&$   23.24\pm 0.87 \pm  0.42$\\		       	         
 840.025& 0.4555&$    3190.0\pm    78.6\pm    22.1$&  $1.9525\pm0.0131$&
 1.0000&$   18.29\pm 0.89 \pm  0.33$\\		       	         
 850.553& 0.4240&$    3475.3\pm    84.9\pm    23.9$&  $1.9721\pm0.0126$& 
 1.0000&$   15.49\pm 0.75 \pm  0.28$\\		       	         
 860.252& 0.4022&$    1790.8\pm    59.2\pm    13.4$&  $1.9618\pm0.0118$& 
 1.0000&$   13.01\pm 0.85 \pm  0.24$\\		       	         
 870.291& 0.3811&$    2331.8\pm    70.0\pm    23.9$&  $1.9303\pm0.0101$& 
 1.0000&$   11.83\pm 0.69 \pm  0.23$\\		       	         
 880.194& 0.3640&$    1684.1\pm    58.9\pm    12.7$&  $1.8848\pm0.0092$&
 1.0000&$   11.30\pm 0.75 \pm  0.20$\\		       	         
 889.788& 0.3570&$    1510.3\pm    56.7\pm    11.4$&  $1.8316\pm0.0084$&
 1.0000&$    9.65\pm 0.67 \pm  0.17$\\		       	         
 899.816& 0.3448&$    1815.6\pm    62.0\pm    12.6$&  $1.7703\pm0.0073$&
 1.0000&$    9.98\pm 0.61 \pm  0.17$\\		       	         
 909.399& 0.3508&$    8084.0\pm   144.1\pm    70.9$&  $1.7088\pm0.0069$&
 1.0000&$    9.80\pm 0.30 \pm  0.18$\\		       	         
 910.112& 0.3459&$    1069.7\pm    45.9\pm     7.7$&  $1.7042\pm0.0073$&
 1.0000&$   10.47\pm 0.77 \pm  0.18$\\		       	         
 920.282& 0.3447&$    5394.0\pm   112.7\pm    39.8$&  $1.6377\pm0.0067$&
 1.0000&$   10.42\pm 0.36 \pm  0.18$\\		       	         
 920.685& 0.3451&$    1343.8\pm    51.0\pm     9.9$&  $1.6350\pm0.0077$&
 1.0000&$   10.64\pm 0.66 \pm  0.18$\\		       	         
 921.726& 0.3421&$    1603.1\pm    56.9\pm    12.6$&  $1.6282\pm0.0070$&
 1.0000&$   10.17\pm 0.59 \pm  0.18$\\		       	         
 929.637& 0.3505&$    8247.2\pm   141.5\pm    60.5$&  $1.5763\pm0.0064$&
 1.0000&$   10.21\pm 0.28 \pm  0.17$\\		       	         
 930.427& 0.3504&$    1432.1\pm    54.4\pm    13.8$&  $1.5711\pm0.0062$&
 1.0000&$    9.60\pm 0.58 \pm  0.17$\\		       	         
 940.342& 0.3529&$    1102.3\pm    47.6\pm     7.4$&  $1.5062\pm0.0071$&
 1.0000&$   10.66\pm 0.70 \pm  0.18$\\		       	         
 950.000& 0.3602&$    1571.9\pm    55.9\pm    13.0$&  $1.4423\pm0.0058$&
 1.0000&$   10.58\pm 0.55 \pm  0.18$\\		       	         
 960.424& 0.3763&$    1275.9\pm    48.5\pm    11.8$&  $1.3716\pm0.0053$&
 1.0000&$   11.25\pm 0.59 \pm  0.20$\\		       	         
 970.651& 0.3850&$    1989.7\pm    60.9\pm    16.2$&  $1.2979\pm0.0050$&
 1.0000&$   12.27\pm 0.49 \pm  0.21$\\		       	         
 980.618& 0.4041&$    1090.6\pm    42.7\pm     9.1$&  $1.2183\pm0.0051$&
 1.0000&$   15.26\pm 0.74 \pm  0.26$\\		       	         
 984.201& 0.4021&$    3540.0\pm    81.2\pm    22.3$&  $1.1869\pm0.0037$&
 1.0001&$   15.36\pm 0.43 \pm  0.25$\\		       	         
 990.308& 0.4221&$    1508.8\pm    50.1\pm     9.9$&  $1.1285\pm0.0046$&
 1.0001&$   17.05\pm 0.65 \pm  0.28$\\
\end{tabular}
\end{ruledtabular}
\end{table}
\begin{table}
\caption{\footnotesize The center-of-mass energy ($E$), detection
efficiency ($\varepsilon$), number of selected events ($N{3\pi}$),
radiative correction ($1+\delta$), correction for the energy spread
($1+\delta_E$), Born cross section for the process $e^+e^-\to
\pi^+\pi^-\pi^0$ ($\sigma$).
The first error in the number of events and the cross section is statistical,
the second is systematic. 
\label{tab4c}}
\begin{ruledtabular}
\fontsize{9pt}{9pt}\selectfont
\begin{tabular}{cccccc}
$E$, MeV & $\varepsilon$ & $N{3\pi}$ & $1+\delta$ & $1+\delta_E$ & $\sigma$, nb \\
\hline
1000.348& 0.3101&$    4944.2\pm    79.3\pm    32.4$&$  1.0139\pm0.0024$&
1.0002&$   26.47\pm 0.55 \pm 0.34$\\		     	       	 
1001.900& 0.3160&$    5657.0\pm    83.7\pm    41.4$&$  0.9934\pm0.0017$&
1.0004&$   28.52\pm 0.45 \pm 0.37$\\		     	       	 
1005.998& 0.3244&$   19858.6\pm   154.7\pm   128.8$&$  0.9351\pm0.0013$&
1.0009&$   39.30\pm 0.37 \pm 0.49$\\		     	       	 
1009.610& 0.3397&$   12404.9\pm   120.7\pm    75.2$&$  0.8779\pm0.0010$&
1.0019&$   57.49\pm 0.63 \pm 0.71$\\		     	       	 
1015.742& 0.3572&$   38647.4\pm   207.8\pm   232.1$&$  0.7702\pm0.0004$&
1.0118&$  221.32\pm 2.50 \pm 2.71$\\		     	       	 
1016.843& 0.3610&$  146106.0\pm   403.7\pm   884.2$&$  0.7528\pm0.0003$&
1.0103&$  322.18\pm 2.83 \pm 3.94$\\		     	       	 
1017.959& 0.3613&$  162634.0\pm   426.1\pm   976.7$&$  0.7427\pm0.0005$&
0.9996&$  483.82\pm 3.73 \pm 5.92$\\		     	       	 
1019.101& 0.3607&$  399890.0\pm   669.4\pm  2400.0$&$  0.7573\pm0.0008$&
0.9721&$  616.77\pm 1.73 \pm 7.55$\\		     	       	 
1019.955& 0.3609&$  386590.0\pm   656.9\pm  2321.3$&$  0.8004\pm0.0010$&
0.9822&$  520.51\pm 3.40 \pm 6.39$\\		     	       	 
1020.932& 0.3569&$  145488.0\pm   404.4\pm   873.5$&$  0.8871\pm0.0019$&
1.0118&$  325.29\pm 5.77 \pm 4.03$\\		     	       	 
1022.114& 0.3580&$   75822.0\pm   292.4\pm   457.7$&$  1.0378\pm0.0043$&
1.0158&$  164.23\pm 2.18 \pm 2.12$\\		     	       	 
1022.930& 0.3552&$   36257.0\pm   204.1\pm   217.9$&$  1.1708\pm0.0066$&
1.0143&$  105.42\pm 1.73 \pm 1.42$\\		     	       	 
1027.752& 0.3296&$    8375.4\pm   101.0\pm    55.2$&$  2.7558\pm0.0697$&
1.0039&$   13.95\pm 0.47 \pm 0.39$\\		     	       	 
1033.822& 0.2972&$    3037.1\pm    64.2\pm    19.6$&$ 11.6589\pm2.1719$&
1.0009&$    1.64\pm 0.40 \pm 0.27$\\		     	       	 
1039.826& 0.2721&$    2192.1\pm    55.3\pm    16.6$&$ 59^{+220}_{-38}$&
1.0005&$    0.23\pm 0.35 ^{+0.41}_{-0.18}$\\	     	       	 
1049.812& 0.2428&$    1650.2\pm    51.3\pm    11.9$&$ 18.4^{+8.8}_{-5.7}$&
1.0001&$    0.59\pm 0.34 ^{+0.26}_{-0.19}$\\    	       	 
1060.030& 0.2274&$    1233.9\pm    46.2\pm    15.9$&$  7.5355\pm1.4064$&
1.0000&$    1.20\pm 0.34 \pm 0.25$\\		     	       	 
1099.980& 0.1968&$    2313.6\pm    65.8\pm    25.5$&$  2.7744\pm0.2266$&
1.0000&$    3.01\pm 0.24 \pm 0.25$\\			        

\end{tabular}
\end{ruledtabular}
\end{table}

\section{Fitting the measured cross section\label{xsfit}}
Directly from the experiment we obtain the visible cross section
\begin{equation}
\label{viscrs00}
\sigma_{{\rm vis},i} = \frac{N_{3\pi,i}}{\varepsilon_i IL_i},
\end{equation}
where $i$ is the number of the energy point. It is related to the Born cross
section $\sigma(E)$ by Eq.~(\ref{viscrs}), which can be rewritten
in the conventional form:
\begin{equation}
\label{viscrs1}
\sigma_{vis}(E) = \int\limits_{0}^{x_{max}} F(x,E)
\sigma(E\sqrt{1-x})dx=\sigma(E)(1+\delta(E)),
\end{equation}
where $\delta(E)$ is the radiative correction.

To take into account the beam energy spread, it is necessary to perform a 
convolution of the visible cross section (\ref{viscrs1}) with a Gaussian 
describing the c.m. energy distribution of the electron-positron pair.
Since the energy spread is 
much smaller than the width of the $\omega$ and $\phi$ resonances, 
we use an approximate formula instead of convolution,
\begin{equation}
\sigma_{\rm vis}(E) \Longrightarrow
\sigma_{\rm vis}(E)+\frac{1}{2}\frac{d^2\sigma_{\rm
vis}}{dE^2}(E)\sigma_E^2=
\sigma_{\rm vis}(E)(1+\delta_E(E))
\label{conv}
\end{equation}

To obtain the experimental Born cross section, the visible cross
section data are fitted using a model for $\sigma(E)$ that
describe the data well. Using the fitted model parameters, the radiative and
energy spread corrections are calculated. The Born cross section at 
the energy point $i$ is then determined as
\begin{equation}
\label{viscrs0}
\sigma_{i} = \frac{N_{3\pi,i}}
{IL_i \varepsilon_i (1+\delta_i)(1+\delta_{E,i})},
\end{equation}

The uncertainty in the collider energy listed in Table~\ref{tab1} effectively
increases the uncertainty in the measured visible cross section. When fitting
the cross section energy dependence, the following term is added in 
quadrature to the statistical error of $\sigma_{{\rm vis},i}$:  
\begin{equation} 
\delta E_i \frac{d\sigma_{\rm vis}}{dE}(E_i),
\label{desys} 
\end{equation} 
where $\delta E_i$ is the uncertainty in the energy of $i$th point.

To describe the Born cross section, the VMD model is used, which
includes the
amplitudes of the $\rho$, $\omega$, and $\phi$ mesons and an amplitude that
takes into account the contribution of the higher vector
resonances:
\begin{eqnarray}
\sigma_{3\pi}(E)&=&\frac{12\pi}{E^3}|F(E)|^2P_{\rho\pi}(E)\nonumber\\
F(E)&=&\sum_{V=\rho,\omega,\phi,\omega^\prime}
\frac{\Gamma_V m_V^{3/2}\sqrt{{\cal B}(V\to e^+e^-){\cal B}(V\to
3\pi)}}{D_V(E)}
\frac{e^{i\varphi_V}}{\sqrt{P_{\rho\pi}(m_V)}},\label{born}
\end{eqnarray}
where $m_V$ and $\Gamma_V$ are the mass and width of the resonance $V$,
$\varphi_V$ is its phase, and
${\cal B}(V\to e^+e^-)$ and ${\cal B}(V\to 3\pi)$ are the branching
fractions of $V$ into $e^+e^-$ and $3\pi$,
\begin{eqnarray}
D_V(E)&=&m_V^2 - E^2 - i E\Gamma_V(E),\nonumber\\
\Gamma_V(E)&=&\sum_{f}\Gamma_f(E).
\end{eqnarray}
Here $\Gamma_f(E)$ is the mass-dependent partial width of the
resonance decay
into the final state $f$, and $\Gamma_f(m_V)=\Gamma_V {\cal B}(V\to f)$.
The mass-dependent width for the $\omega$ and $\phi$ mesons has been
calculated taking into account all significant decay modes.
The corresponding formulae can be found,
for example, in Ref.~\cite{snd-3pi-2}. We assume that the $V\to 3\pi$ decay
proceeds
via the $\rho\pi$ intermediate state, and $P_{\rho\pi}(E)$ is the $3\pi$
phase space volume calculated under this hypothesis.
The formula for the $P_{\rho\pi}$ calculation can be found in
Ref.~\cite{snd-3pi-2}.

The phase $\varphi_\omega$ is set to zero. Thus, all other phases
$\varphi_V$ represent relative phases between the amplitudes of the $V$ and
$\omega$ resonances. In general, they depend on energy, for example,
due to vector meson mixing. Since the value of the form factor $F(E)$ at 
$E=0$ is a real number determined by the $\gamma\to 3\pi$ chiral anomaly (see, 
for example, Ref.~\cite{HHK0}), all phases tend
to $0^\circ$ or $180^\circ$ when the energy goes to 0.
For the energy dependence of $\varphi_\phi$, there is a theoretical
prediction~\cite{phiomphase}. The phase has a minimum of about $160^\circ$ 
at the maximum of the $\phi$ resonance and increases with decreasing energy.
At the $\omega$ mass it is approximately $179^\circ$.

In our nominal fit, we use only the data obtained in this article, and the 
contribution of the excited vector resonances is
modeled by a single resonance $\omega^\prime$ with mass and width
equal to the Particle Data Group (PDG) values for 
the $\omega(1420)$~\cite{pdg}. The parameter $\varphi_{\omega^\prime}$
is set to its expected value of $180^\circ$~\cite{Clegg}.

The free fit parameters are 
${\cal B}(\omega\to e^+e^-){\cal B}(\omega\to 3\pi)$,
$m_\omega$, $\Gamma_\omega(m_\omega)$,
${\cal B}(\phi\to e^+e^-){\cal B}(\phi\to 3\pi)$,
$m_\phi$, $\Gamma_\phi(m_\phi)$, $\varphi_\phi$,
${\cal B}(\rho\to 3\pi)$, $\varphi_\rho$, 
${\cal B}(\omega^\prime\to e^+e^-){\cal B}(\omega^\prime\to 3\pi)$.
This model is close to those used to fit the data in previous 
measurements~\cite{cmd-3pi-1,cmd-3pi-2,snd-3pi-1,snd-3pi-2,BABAR-3pi}.
\begin{table}
\caption{The fitted VMD model parameters in comparison with the PDG
data~\cite{pdg} and results of Refs.~\cite{snd-3pi-2} and
\cite{BABAR-3pi} for $\varphi_\phi$ and $\varphi_\rho$, respectively. 
\label{tab3}}
\begin{ruledtabular}
\begin{tabular}{ccc}
Parameter & Fitted value & PDG value\\
\hline
${\cal B}(\omega\to e^+e^-){\cal B}(\omega\to 3\pi)\times10^5$ &
$6.648\pm 0.022\pm 0.058$ & $6.61\pm0.16$\\
$m_\omega$, MeV & $782.703\pm0.011\pm0.047$ & $782.66\pm0.13$ \\
$\Gamma_\omega$, MeV & $8.598\pm0.023\pm0.004$ & $8.68\pm0.13$\\
${\cal B}(\phi\to e^+e^-){\cal B}(\phi\to 3\pi)\times10^5$ &
$4.154^{+0.102}_{-0.066}\pm 0.066$ & $4.42\pm0.11$\\
$m_\phi$, MeV & $1019.488\pm0.017\pm0.061$ & $1019.460\pm0.016$ \\
$\Gamma_\phi$, MeV & $4.265\pm0.033\pm0.014$ & $4.249\pm0.013$ \\  
$\varphi_\phi$, deg & $156.8^{+6.5}_{-4.6}\pm1.9$ & $163\pm 7$ \\
${\cal B}(\rho\to 3\pi)\times10^5$ & $4.7\pm1.2\pm1.3$ & $9\pm4$ \\
$\varphi_\rho$, deg & $-95.3\pm8.3\pm4.0$ & $-99\pm17$ \\
\end{tabular}
\end{ruledtabular}
\end{table}

Fitting the data with purely statistical uncertainties yields
$\chi^2/\nu = 167/93$. The poor quality of the fit is explained by the
presence of uncorrelated systematic uncertainties. These uncertainties
manifest themselves, for example, as nonstatistical fluctuations between
energy points in the correction factors $C_\gamma$ and $C_{\rm ch.tr.}$ 
in Fig.~\ref{fig13}. We estimate the value uncorrelated systematic
uncertainty from the error of the product $C_\gamma C_{\rm ch.tr.}$ (0.5\%),
and the error associated with the signal shape
in the fit to the $M_{\gamma\gamma}$ spectrum (0.6\%). 
It is 0.52\% in the energy range 763.94--810.58 MeV, where the
alternative method of background subtraction is used,
and 0.78\% outside this range. 
Adding these uncertainties to the statistical uncertainties of
$N_{3\pi}$ in quadrature, we obtain a reasonable value of $\chi^2/\nu
= 96/93$.

\begin{figure}
\centering
\includegraphics[width=0.40\linewidth]{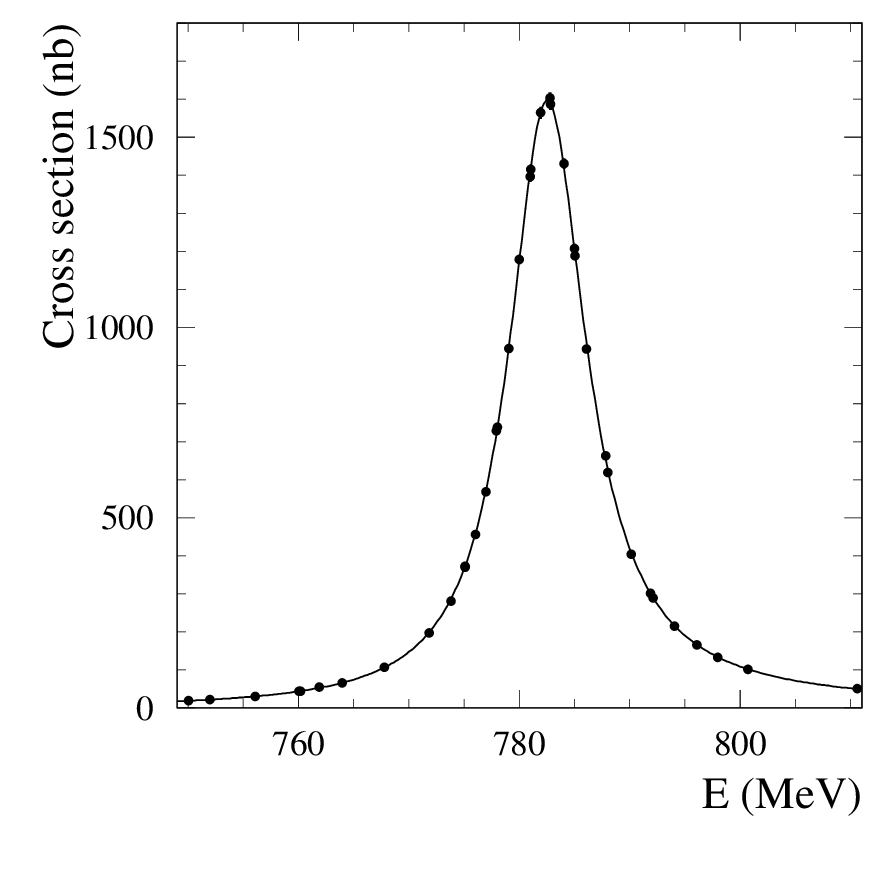}
\includegraphics[width=0.40\linewidth]{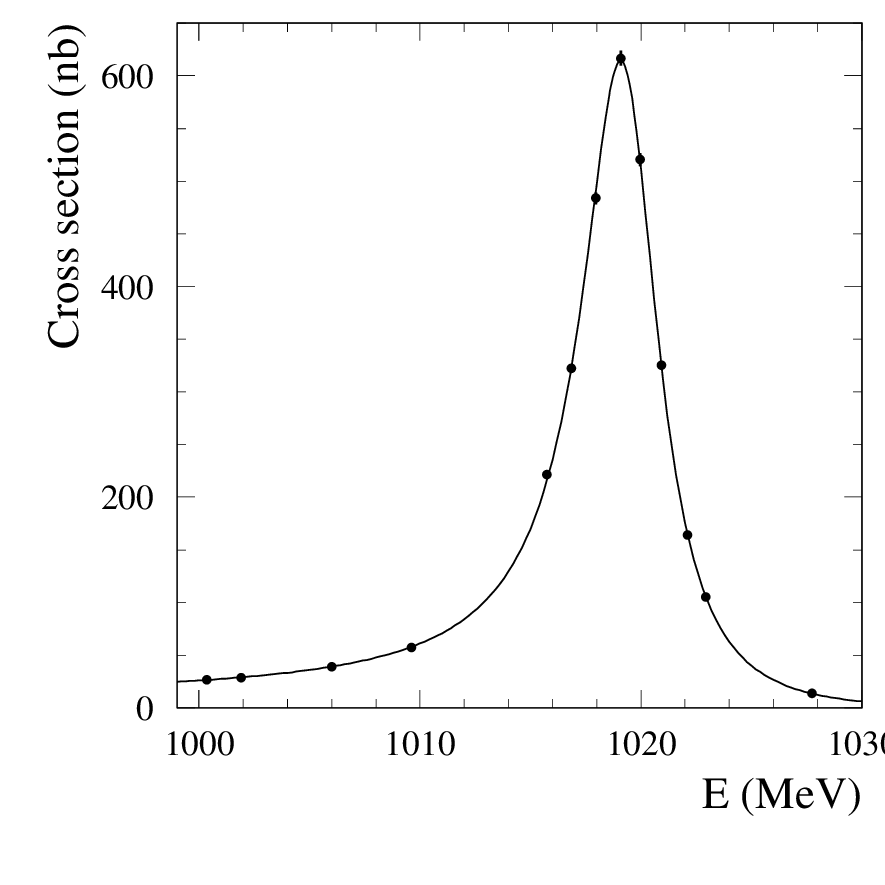}
\includegraphics[width=0.30\linewidth]{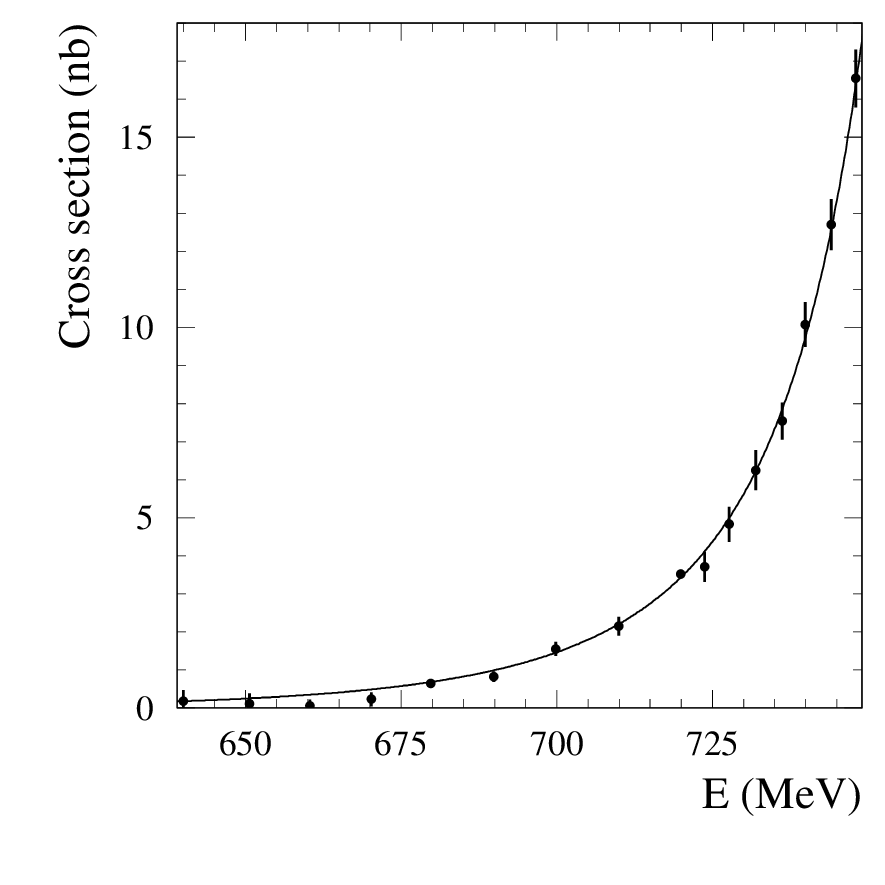}
\includegraphics[width=0.30\linewidth]{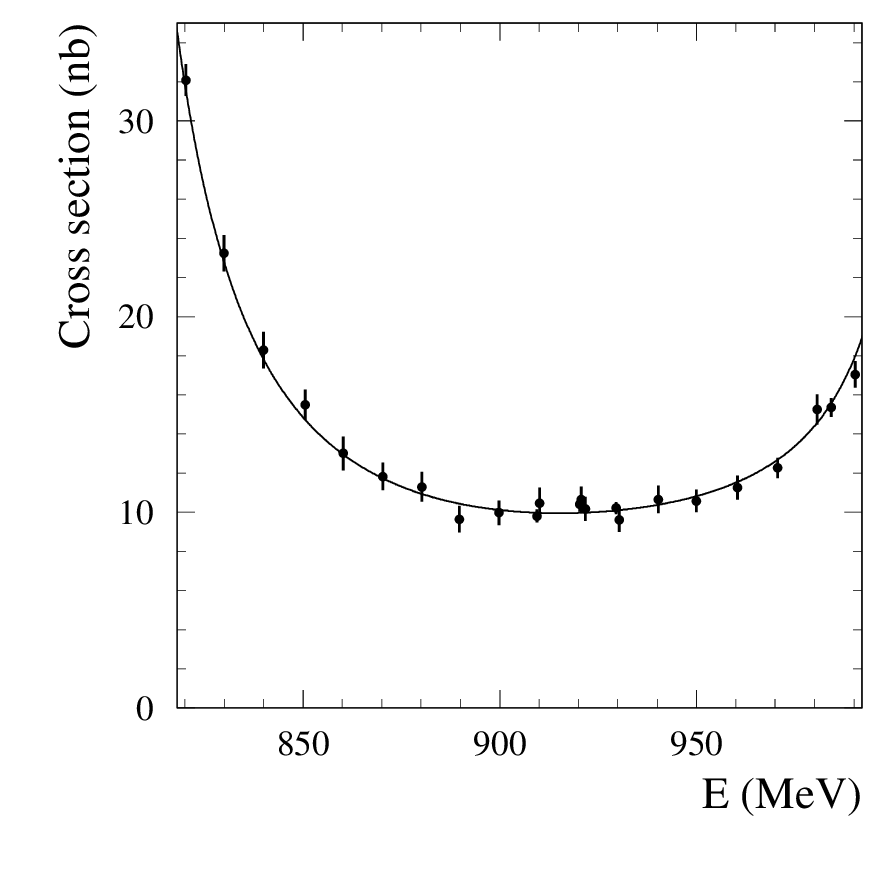}
\includegraphics[width=0.30\linewidth]{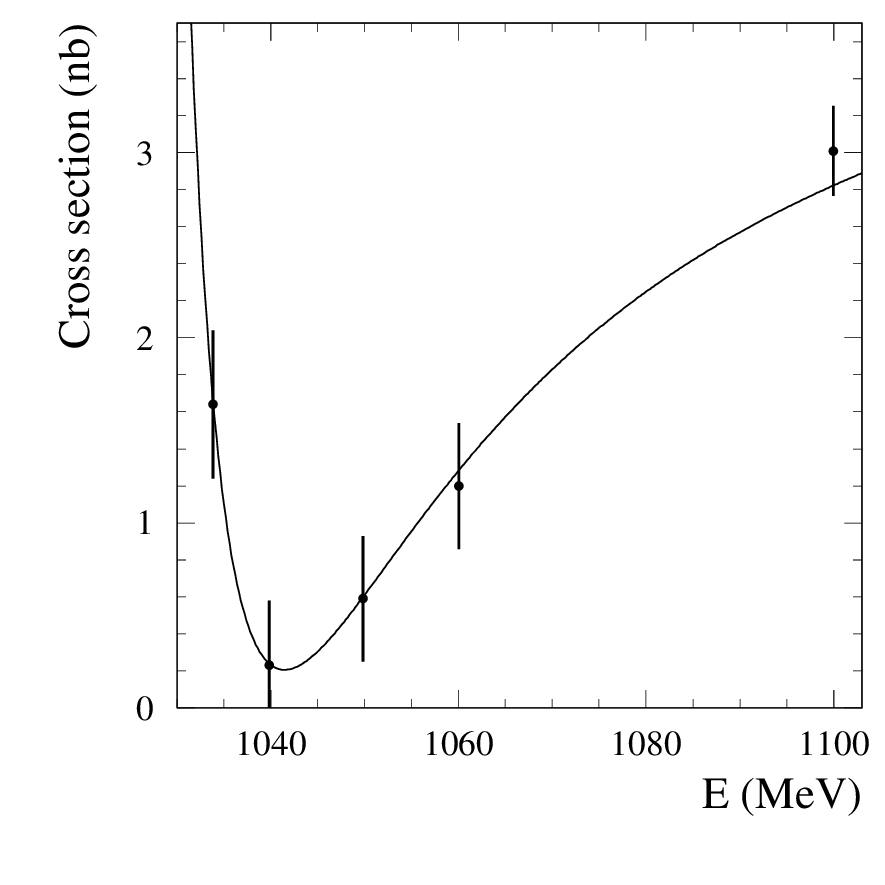}
\caption{The measured Born cross section for the process $e^+e^-\to
\pi^+\pi^-\pi^0$ in five energy regions. The error bars represent combined
statistical and systematic uncertainties.
The curve is the result of the fit described in the text.
\label{fig14}}
\end{figure}
The fit results are used to calculate the radiative correction and 
the energy spread correction. 
The measured energy dependence of the Born cross section
is shown in Fig.~\ref{fig14} together with the fitted curve.

The parameters of the $\rho$, $\omega$, and $\phi$ resonances
obtained from the fit to the visible cross
section data are listed in Table~\ref{tab3} in comparison with PDG
values~\cite{pdg} and results of Refs.~\cite{snd-3pi-2} and
\cite{BABAR-3pi} for $\varphi_\phi$ and $\varphi_\rho$, respectively.
The quoted errors are statistical and systematic.
The systematic error in the masses $m_\omega$ and $m_\phi$ is completely
determined by the systematic uncertainty in the collider energy measurement.
To determine the systematic uncertainties for other parameters, we shift all
$IL_i$, $\sigma_{E,i}$, $N_{3\pi,i}$, or efficiency corrections up and down by
the value of the systematic uncertainty. To determine the model uncertainties
of the parameters, three alternative fit models are tested. 

In model 1, instead of the constant $\varphi_\phi$, the
energy-dependent quantity $\varphi_\phi(E)$~\cite{phiomphase} is used,
multiplied by the free parameter $a_\phi$. The model describes the
cross section data well. The fitted value $a_\phi=0.96\pm0.03$.
In this model, the significant, 0.9\%, decrease of the fitted value of ${\cal
B}(\phi\to e^+e^-){\cal B}(\phi\to 3\pi)$ is observed.

In the second and third models we include to the fit the cross section
data obtained by SND above 1.1 GeV~\cite{snd-3pi-4}.
In this energy region, in addition to $\rho\pi$, the intermediate
mechanisms $\omega\pi^0$ and $\rho(1450)\pi$ are significant.
In Ref.~\cite{snd-3pi-4}, the total $e^+e^-\to \pi^+\pi^-\pi^0$ cross
section, and the cross section for the mechanisms $\rho\pi$ and
$\rho(1450)\pi$ were measured. 
It was shown that the process
$e^+e^-\to \omega(1650)\to \pi^+\pi^-\pi^0$ is dominated by the 
intermediate mechanism $\rho(1450)\pi$. 
It should be noted that the Dalitz distributions for the $\rho\pi$ and 
$\rho(1450)\pi$ intermediate states below 0.95 GeV are not
distinguishable, and the amplitudes of the $\omega$ and $\rho$ resonances
and the low-energy tail of the $\omega(1650)$ amplitude are coherent. 
The total contribution (including interference) of the isovector
intermediate state $\omega\pi^0$ to the $e^+e^- \to\pi^+\pi^-\pi^0$
cross section reaches 16\% at $E=1.4$ GeV. However, it decreases very
rapidly with decreasing energy and can be neglected below the
$\omega\pi^0$ threshold of 920 MeV.

\begin{figure}
\centering
\includegraphics[width=0.40\linewidth]{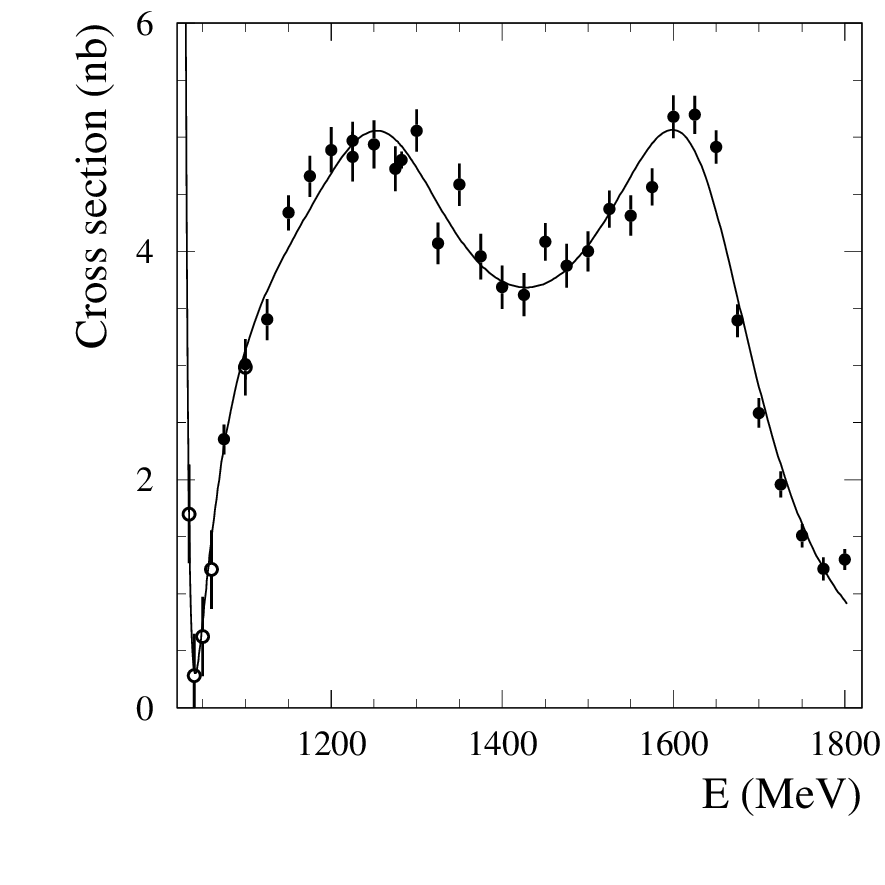}
\includegraphics[width=0.40\linewidth]{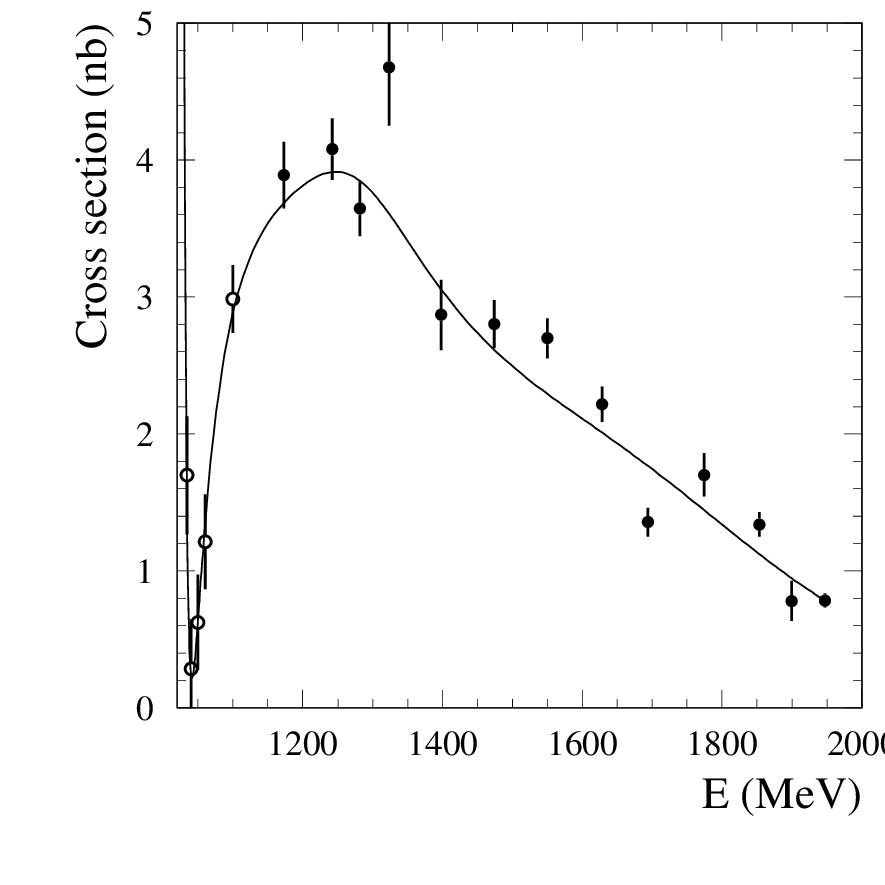}
\caption{The total $e^+e^-\to \pi^+\pi^-\pi^0$ 
cross section (left) and the cross section for $e^+e^-\to \rho\pi$
(right) measured in Ref.~\cite{snd-3pi-4}.
The curves are the results of the fits described in the text.
\label{fig15}}
\end{figure}

The total $e^+e^-\to \pi^+\pi^-\pi^0$
cross section and the cross section for $e^+e^-\to \rho\pi$ measured in
Ref.~\cite{snd-3pi-4} are shown in Fig.~\ref{fig15}.
To take into account intermediate states other than $\rho\pi$, the 
formula~(\ref{born}) must be modified. 
However, in our case, where data above 1.1 GeV are used only to
estimate the model uncertainties of the fitted parameters,
a simplified approach can be applied. Assuming the
intermediate $\rho\pi$ mechanism, we include in the fit either the
$e^+e^-\to \rho\pi$ data (model 2) or the $e^+e^-\to \pi^+\pi^-\pi^0$ data
(model 3) above 1.1 GeV. In model 2, the contribution of the 
$\omega(1650)\to \rho(1450)\pi$ amplitude to the $e^+e^-\to
\pi^+\pi^-\pi^0$ cross section is neglected.  
In contrast, in model 3, the contribution of the $\omega(1650)$ amplitude 
is overestimated. In both models, the cross section
above 1.1 GeV is described by contributions of two resonances with
phases $\varphi_{\omega^\prime}=180^\circ$ and
$\varphi_{\omega^{\prime\prime}}=0^\circ$. The fitting curves are 
shown in Fig.~\ref{fig15}. In model 3, the fitted resonance masses and
widths are close to the PDG values for $\omega(1420)$ and
$\omega(1650)$~\cite{pdg}.  
In model 2, the fitted parameters of the first resonance,
$m_{\omega^\prime}=1400\pm 25$ MeV and
$\Gamma_{\omega^\prime}=710\pm90$ MeV, do not contradict the PDG
values for $\omega(1420)$, the $\omega(1650)$ contribution
is not needed, instead a very broad structure with
$m_{\omega^{\prime\prime}}=2550\pm130$ MeV and
$\Gamma_{\omega^{\prime\prime}}=8000\pm3000$ MeV is additionally
required to describe the $e^+e^-\to \rho\pi$ data above 1.6 GeV. 
The differences between parameters of the $\rho$, $\omega$, and $\phi$
resonances obtained in the nominal model and models 1, 2 and 3 are
taken as estimates of the model uncertainties. The model uncertainties
determine the systematic errors of the parameters $\varphi_\phi$,
${\cal B}(\rho\to 3\pi)$, and $\varphi_\rho$, and contribute to the
errors of $\Gamma_\omega$ and ${\cal B}(\phi\to e^+e^-){\cal
B}(\phi\to 3\pi)$.

The obtained parameters of the $\omega$ meson listed in
Table~\ref{tab3} are consistent with the PDG values~\cite{pdg}, but
are much more accurate. The values of the $\phi$ mass and width agree 
with the PDG data~\cite{pdg}, but have worse accuracy. The value of
${\cal B}(\phi\to e^+e^-){\cal B}(\phi\to 3\pi)$ is smaller but 
consistent within errors with the PDG value~\cite{pdg}.
The phase $\varphi_\phi$ agrees
with the only previous measurement~\cite{snd-3pi-2}.
The accuracy of ${\cal B}(\rho\to 3\pi)$ and $\varphi_\rho$ are improved 
compared with the most accurate previous BABAR measurement~\cite{BABAR-3pi}.

\begin{figure}
\centering
\includegraphics[width=0.70\linewidth]{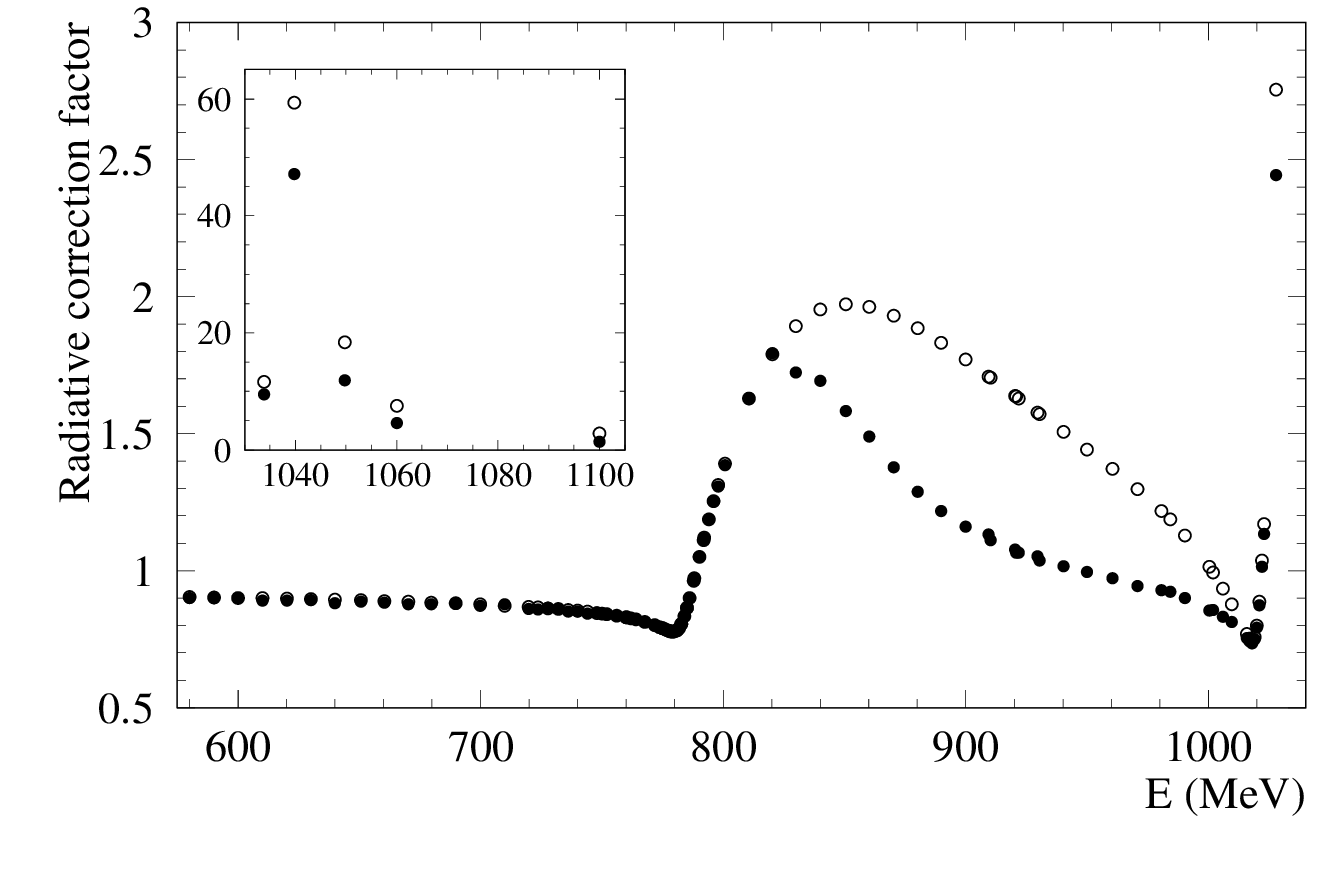}
\caption{The energy dependence of the radiative correction factor calculated 
using the nominal model for the Born cross section without any detector 
restriction [Eq.~(\ref{viscrs1})] (open circle) and with the condition 
$\chi^2_{3\pi}<30$ (filled circle).
\label{radcor}}
\end{figure}
The values of $(1+\delta)$ with errors and $(1+\delta_{E})$ obtained
in the nominal model are listed in Tables~\ref{tab4a}, \ref{tab4b}, and \ref{tab4c}.
The initial-state radiative correction factor $(1+\delta)$ calculated using
Eq.~(\ref{viscrs1}) depends only on the shape of the fitted Born cross
section. The limitation on the ISR photon energy due to 
condition on $\chi^2_{3\pi}$ manifests itself through the detection efficiency.
The measured Born cross section [(see Eq.~(\ref{viscrs0})] is
inversely proportional to the product $P_\delta=\varepsilon (1+\delta)$.
In Fig.~\ref{radcor}, we compare the effective radiation correction
factor calculated as $P_\delta/\varepsilon_0$, where
$\varepsilon_0$ is the detection efficiency for signal events with
$E_{\rm ISR}<0.5$ MeV, with the factor $1+\delta$. It is seen that
the condition $\chi^2_{3\pi}<30$ significantly reduces the radiative
correction in the region between the $\omega$ and $\phi$ resonances.
The effective threshold on $E_{\rm ISR}$ (50\% efficiency for
events with an ISR photon) is about 75 MeV for $\chi^2_{3\pi}<30$
and 135 MeV for $\chi^2_{3\pi}<100$.

To estimate the uncertainty in the radiative correction due to the
fitting procedure, we vary the 
ﬁtted parameters of the nominal model within their errors and calculate 
the shifts of the product $P_\delta$. The sum of the obtained relative shifts
in quadrature is then multiplied by $(1+\delta)$. 
The model dependence of the radiative corrections is studied by  
comparing the radiative corrections obtained in the nominal model and
models 1, 2, and 3. It is significant at five energy points with $E >
1033$ MeV. The model uncertainty is added in quadrature to the
uncertainty estimated above. 

The systematic uncertainty in the detection efficiency due to ISR is
discussed in Sec.~\ref{deteff}.

The final-state radiative correction, which is about 1\% at the
$\omega$ and $\phi$ resonances~\cite{HHK}, is included to the
measured $e^+e^-\to \pi^+\pi^-\pi^0$ cross section.
The emission of a hard photon from the final state results in a
decrease in the detection efficiency due to the condition on
$\chi^2_{3\pi}$. Comparing the detection efficiency with and without
FSR, we find that the efficiency correction is about 0.02\% near
the $\omega$ resonance and about 0.3\% near the $\phi$ resonance,
where the more stringent condition $\chi^2_{3\pi}<30$ is used.
The systematic uncertainty of the FSR efficiency correction simulated
by PHOTOS is estimated to be approximately 30\% based on the $D^0\to K^-\pi^+$
decay study~\cite{FSRsys}. Therefore, we conclude that the
systematic uncertainty due to FSR effects is less than 0.1\% and is
negligible.

The values of the measured Born cross section with statistical and 
systematic errors are listed in Tables~\ref{tab4a}, \ref{tab4b}, and
\ref{tab4c}. The statistical uncertainty of the cross section 
includes the statistical uncertainties of $N_{3\pi}$,
$\varepsilon$, and $IL$, and the uncertainty associated with the beam energy 
measurement, calculated using Eq.~(\ref{desys}).
The systematic uncertainty of the cross section
includes the systematic uncertainties in the luminosity measurement (0.7\%),
number of signal events, detection efficiency, and radiative correction.
This total systematic uncertainty includes an uncorrelated component. The 
uncorrelated uncertainty is discussed above and is 0.52\% in the energy range
763.94--810.58 MeV and 0.78\% outside this range. However, we cannot
confidently exclude it from the total uncertainty to obtain the correlated
component. Therefore, we consider the total systematic uncertainty to be a
conservative estimate of the correlated uncertainty.

\section{Comparison with previous measurements}
\begin{figure}
\centering
\includegraphics[width=0.40\linewidth]{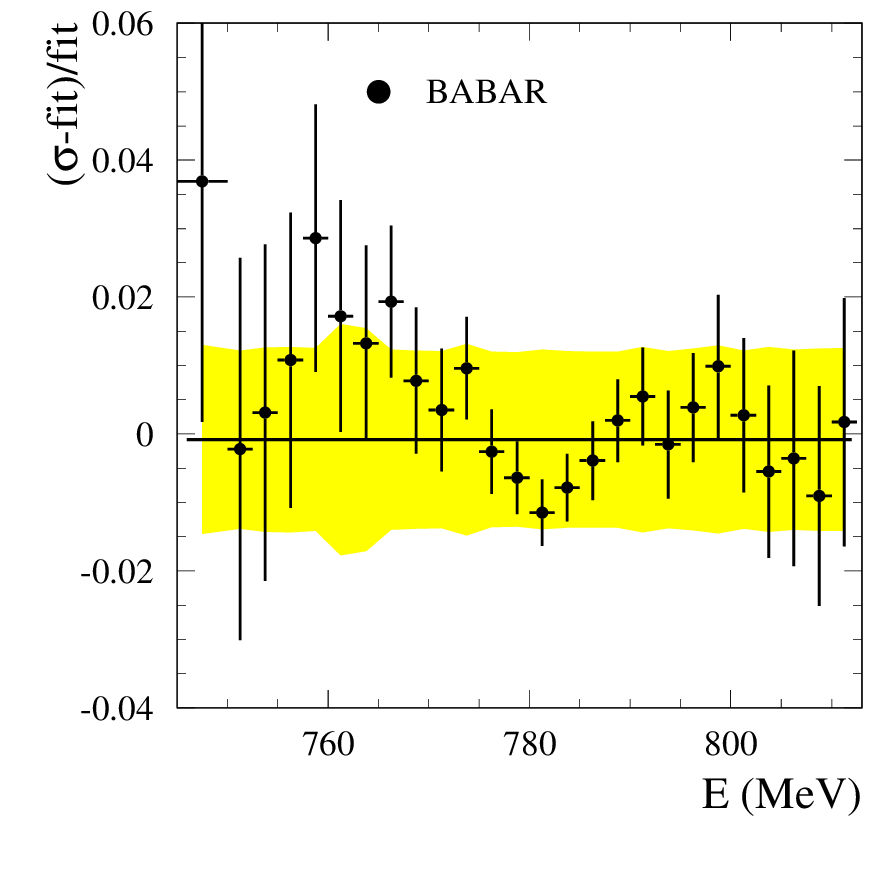}
\includegraphics[width=0.40\linewidth]{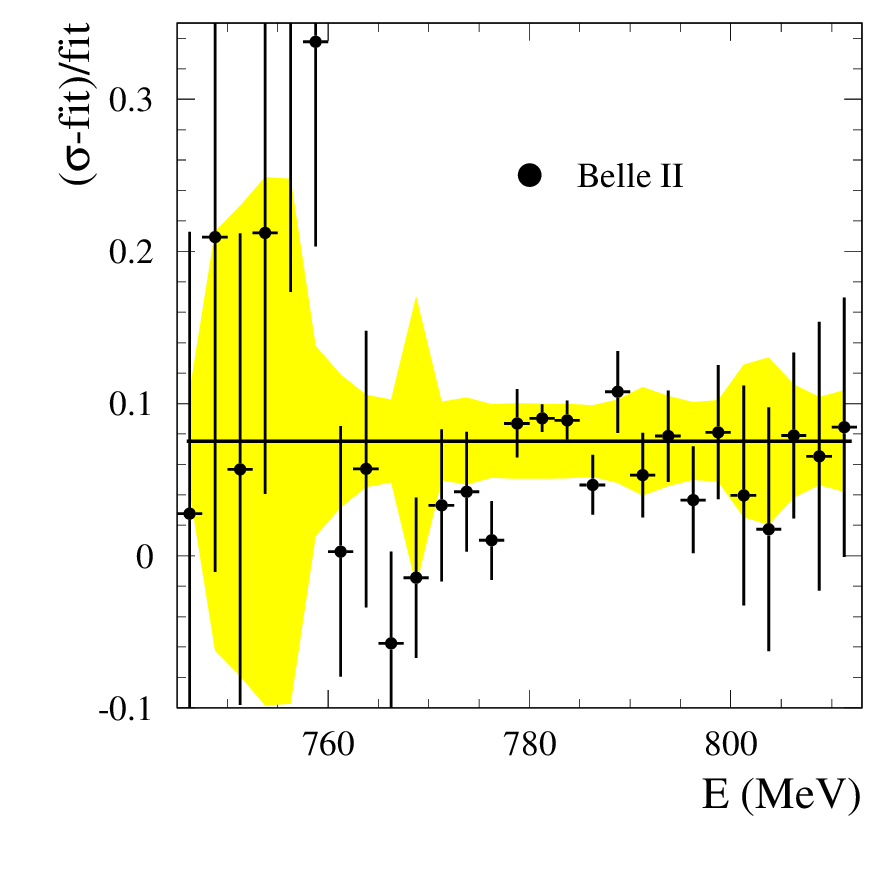}
\includegraphics[width=0.40\linewidth]{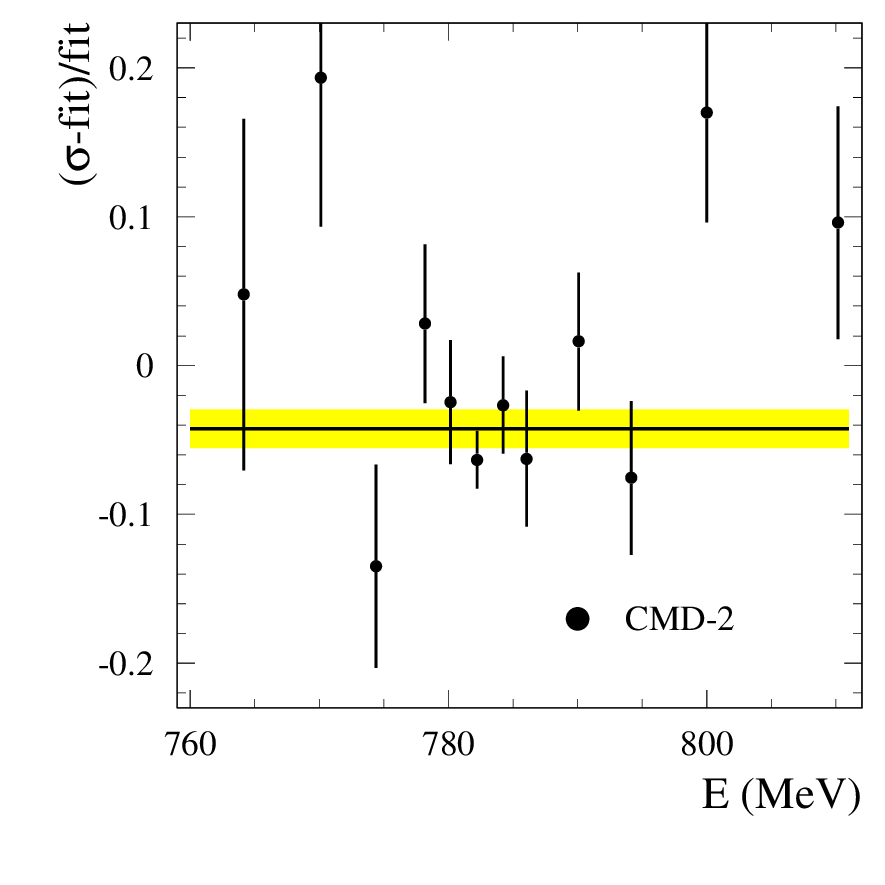}
\includegraphics[width=0.40\linewidth]{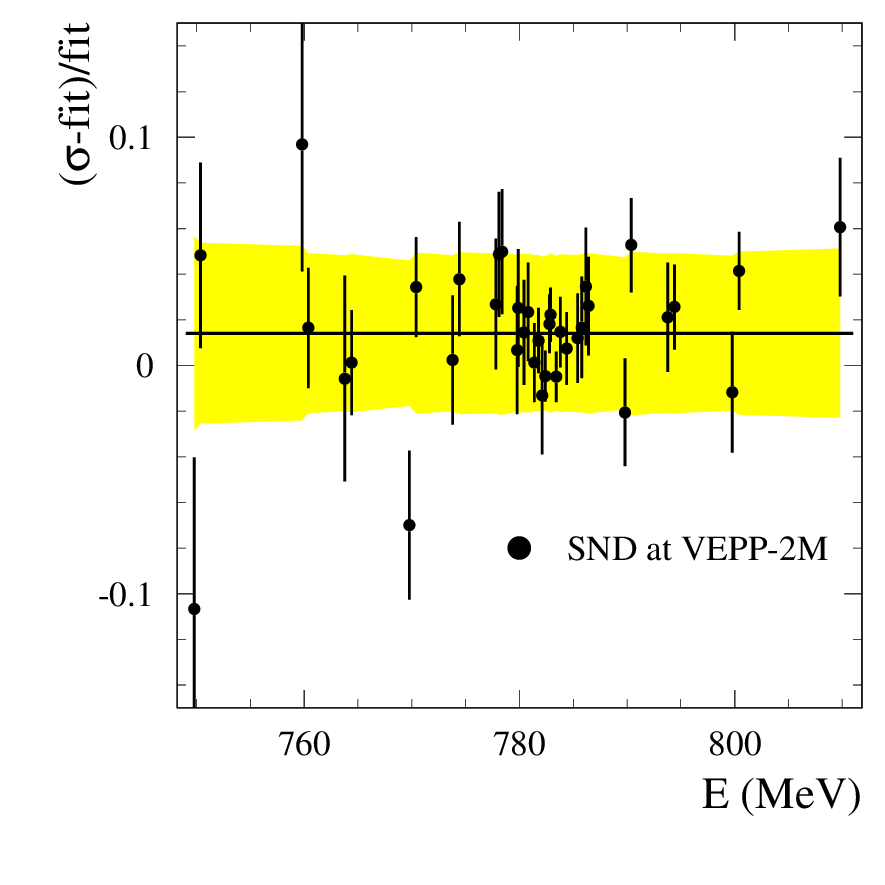}
\caption{The ratios of the cross sections measured in the
BABAR~\cite{BABAR-3pi} (top, left), Belle~II~\cite{belle-3pi} (top,right),
CMD-2~\cite{cmd-3pi-1} (bottom, left), and SND at
VEPP-2M~\cite{snd-3pi-2}(bottom, right) experiments to the cross section
obtained by fitting the data from this work with the VMD model in the
$\omega$ resonance energy region. 
The solid line represents the result of the fit to the data with
a constant. The shaded band shows the systematic uncertainty of the data.
\label{fig16}}
\end{figure}
\begin{figure}
\centering
\includegraphics[width=0.40\linewidth]{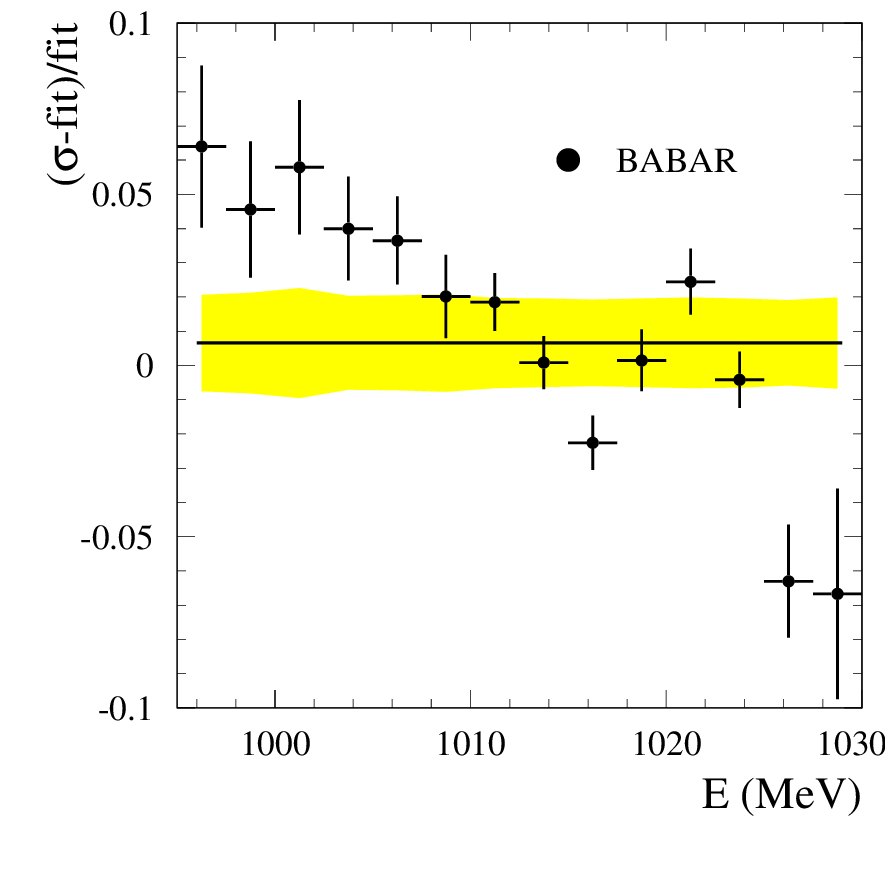}
\includegraphics[width=0.40\linewidth]{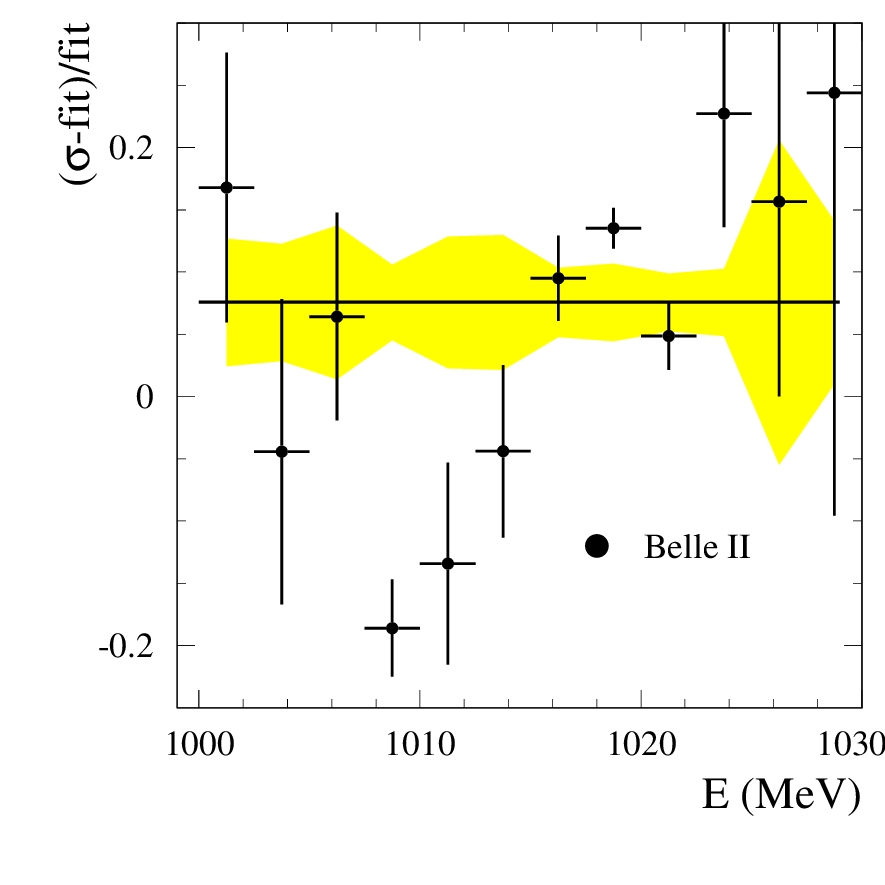}
\includegraphics[width=0.40\linewidth]{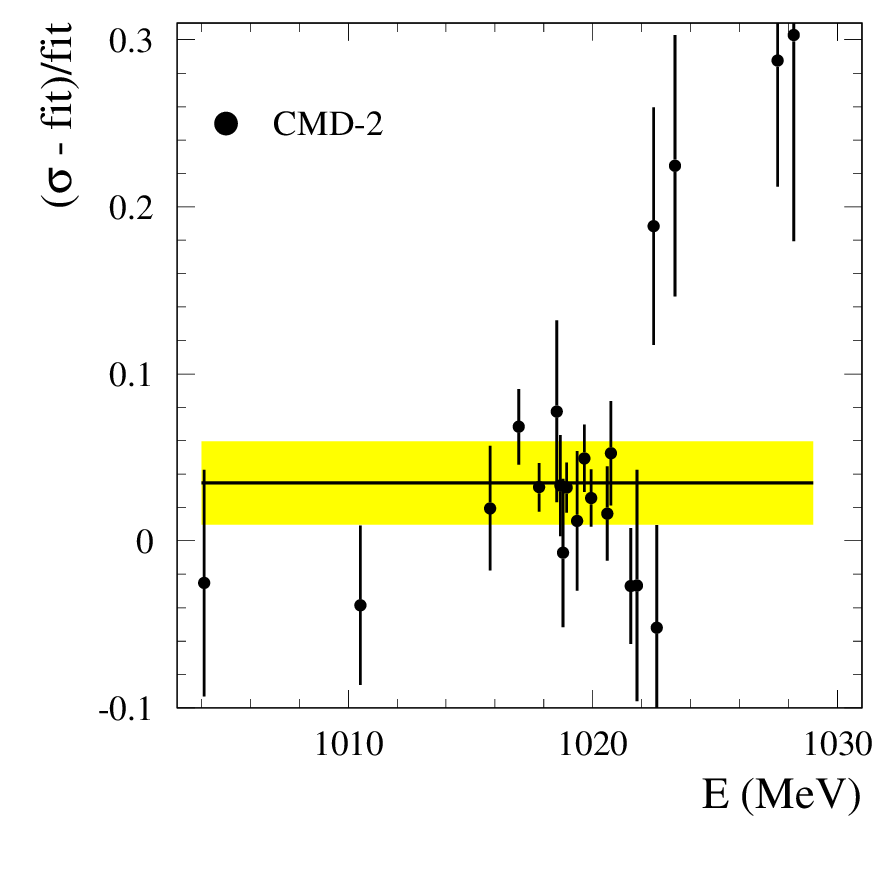}
\includegraphics[width=0.40\linewidth]{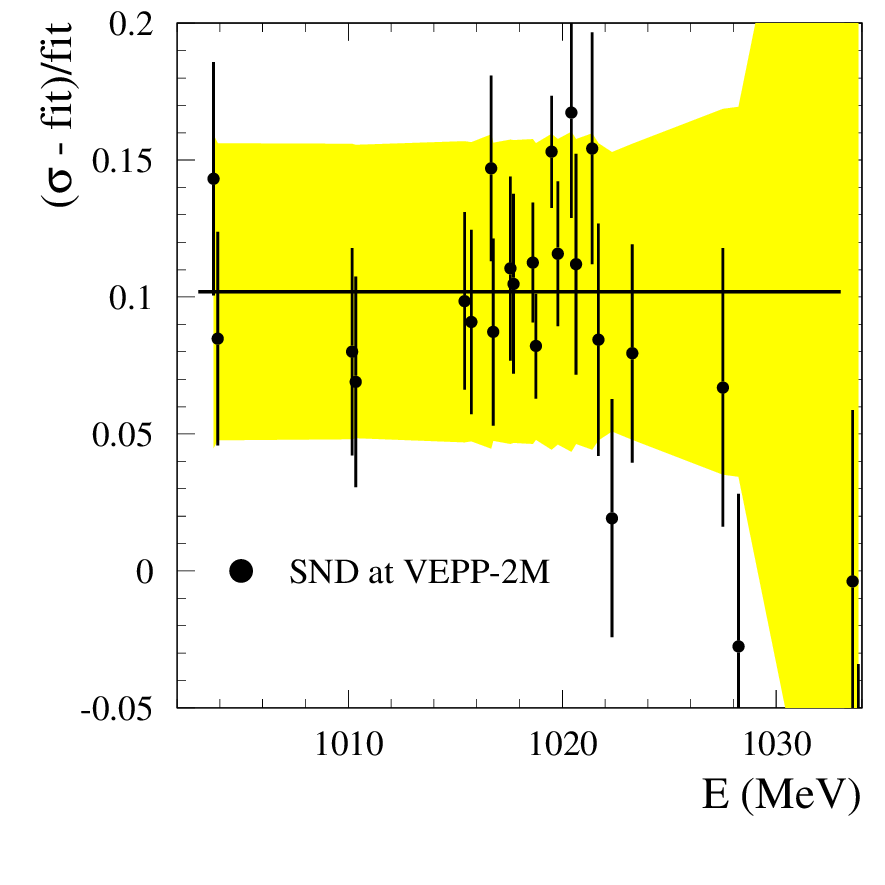}
\caption{The ratios of the cross sections measured in the
BABAR~\cite{BABAR-3pi} (top, left), Belle~II~\cite{belle-3pi} (top,right),
CMD-2~\cite{cmd-3pi-2} (bottom, left), and SND at
VEPP-2M~\cite{snd-3pi-2}(bottom, right) experiments to the cross section
obtained by fitting the data from this work with the VMD model in the
$\phi$ resonance energy region. 
The solid line represents the result of the fit to the data with
a constant. The shaded band shows the systematic uncertainty of the data.
\label{fig17}}
\end{figure}
\begin{figure}
\centering
\includegraphics[width=0.40\linewidth]{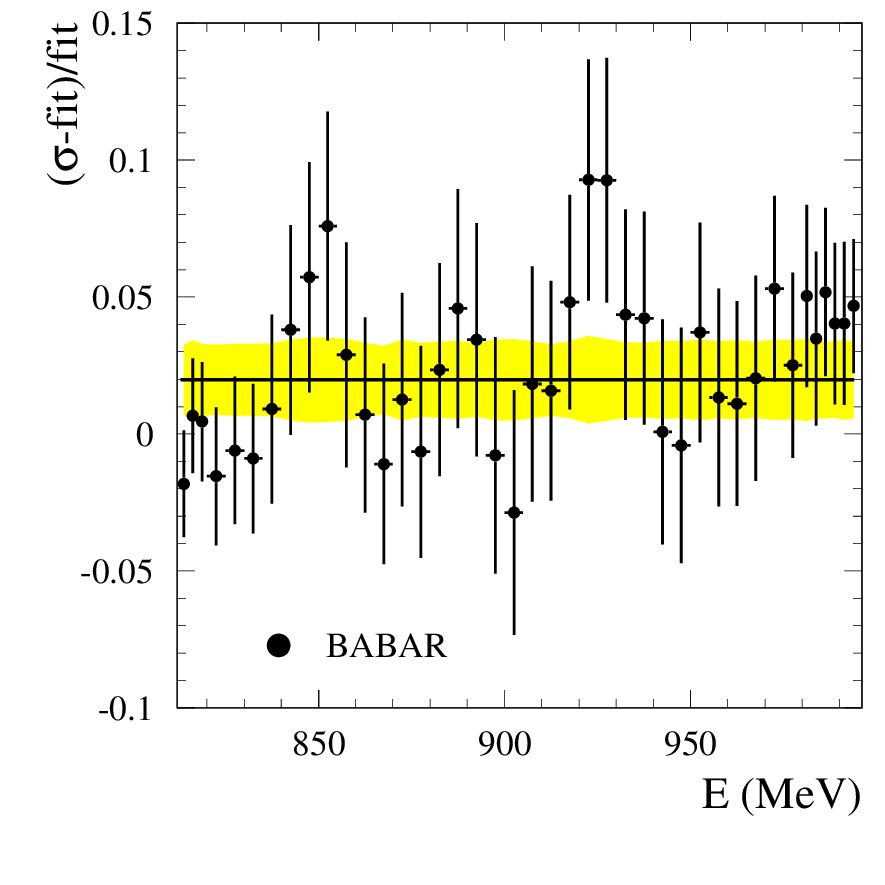}
\includegraphics[width=0.40\linewidth]{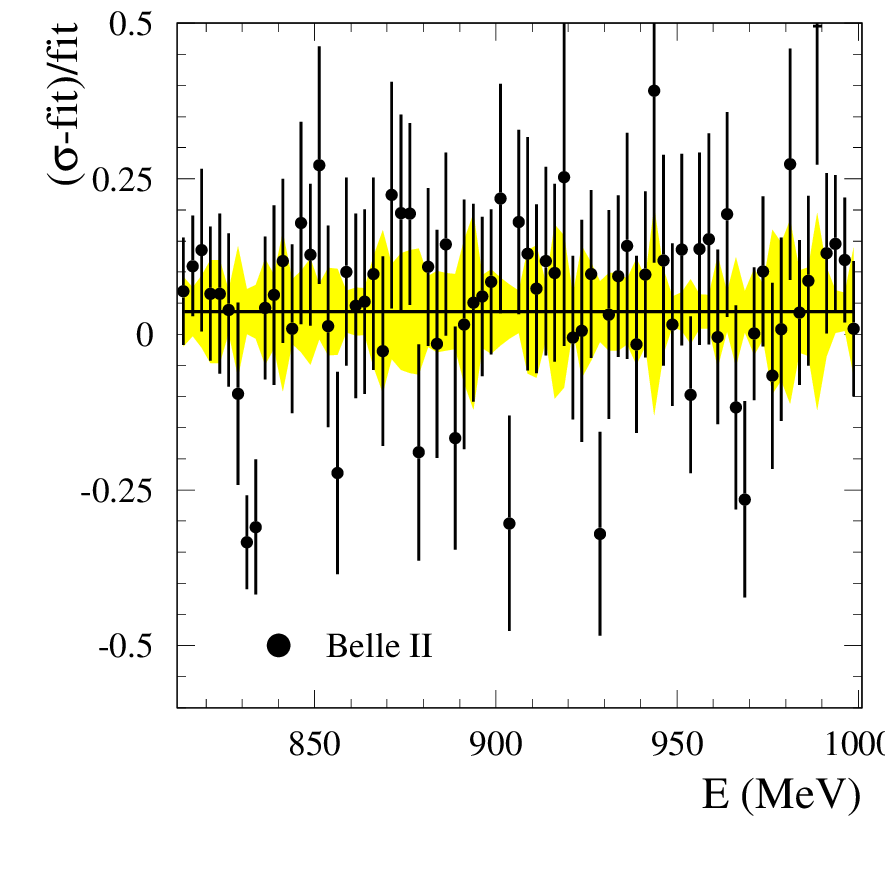}
\includegraphics[width=0.40\linewidth]{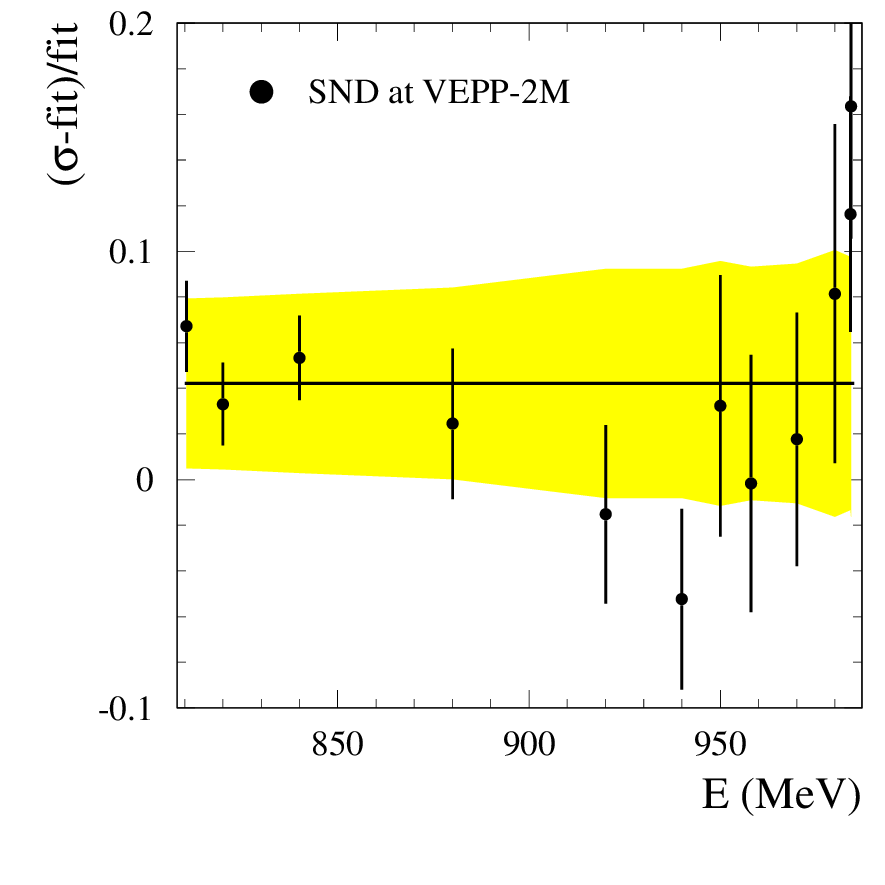}
\includegraphics[width=0.40\linewidth]{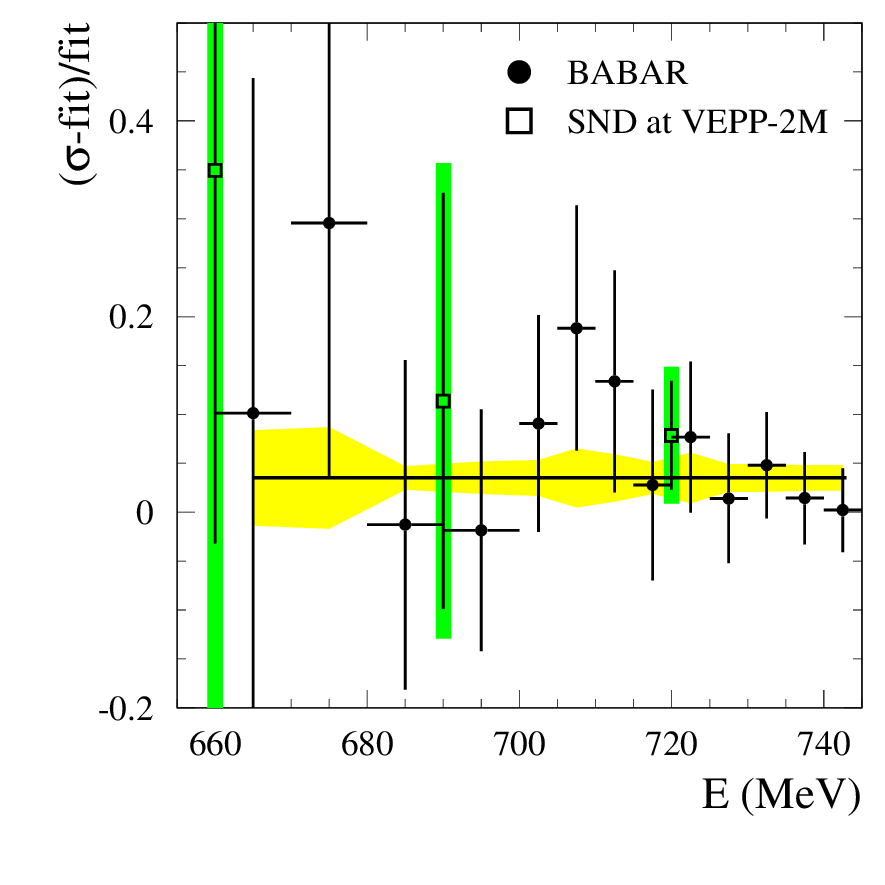}
\caption{The ratios of the cross sections measured in the
BABAR~\cite{BABAR-3pi} (top, left), Belle~II~\cite{belle-3pi} (top,right),
and SND at VEPP-2M~\cite{snd-3pi-2}(bottom, left) experiments to the cross section
obtained by fitting the data from this work with the VMD model in the
energy region between $\omega$ and $\phi$ resonances. The bottom right
panel shows the same plot for the BABAR~\cite{BABAR-3pi} and SND at
VEPP-2M~\cite{snd-3pi-2} data. The solid line represents the result of the fit 
to the data with a constant. The shaded band shows the systematic uncertainty of 
the data.
\label{fig18}}
\end{figure}
A comparison of our results with previous measurements is presented 
in Figs.~\ref{fig16}, \ref{fig17}, and \ref{fig18}, which plot the ratios of
the cross sections measured in
Refs.~\cite{BABAR-3pi,belle-3pi,cmd-3pi-1,cmd-3pi-2,snd-3pi-1,snd-3pi-2} to the
cross section obtained by fitting the data from this work in the VMD model. 
The solid line shows the fit of the ratio energy dependence by a constant. 
Our measurement is in reasonable agreement with the BABAR
data~\cite{BABAR-3pi} over the entire energy range, with the SND at VEPP-2M
data~\cite{snd-3pi-2} below the $\phi$ resonance, with the
Belle-II~\cite{belle-3pi} between $\omega$ and $\phi$, and with the
CMD-2 data~\cite{cmd-3pi-2} near the $\phi$ resonance. The ratio of the
Belle-II measurement to our measurement is about 1.075 near the
$\omega$ and $\phi$ resonances. The significance of this deviation is
about $2\sigma$. In the $\omega$ energy region our measurement about
4\% ($2\sigma$) larger than the CMD-2~\cite{cmd-3pi-1} results, while
in the $\phi$ energy region it is about 10\% ($2\sigma$) lower than the
SND at VEPP-2M measurement.

Using the measured $e^+e^-\to \pi^+\pi^-\pi^0$ Born cross section
($\sigma(s)$) we can calculate the $3\pi$ contribution to the muon anomalous
magnetic moment
\begin{equation}
a_\mu^{3\pi}=\frac{\alpha^2}{3\pi^2}\int_{m^2_\pi}^{\infty}
\frac{K(s)}{s} \frac{\sigma(s)|1-\Pi(s)|^2}{4\pi\alpha^2/s}
\,ds,
\label{amu}
\end{equation}
where the kernel function $K(s)$ can be found in Ref.~\cite{whp} and
the vacuum polarization operator $\Pi(s)$ is tabulated in Ref.~\cite{ignatov}.

Since the $e^+e^-\to \pi^+\pi^-\pi^0$ cross section contains two
narrow resonance peaks, numerical integration of Eq.~(\ref{amu}) over
the data points in Tables~\ref{tab4a}, \ref{tab4b},
and \ref{tab4c} using simple methods, such as the trapezoidal
rule or Simpson's rule, do not provide the required accuracy.
Therefore, the $a_\mu^{3\pi}$ calculation is performed 
using the theoretical cross section [Eq.~(\ref{born})] with fitted
parameters. The result for the energy region 0.61-1.1 GeV is listed in 
Table~\ref{tab5} with statistical and systematic uncertainties. 

The statistical uncertainty in $a_\mu^{3\pi}$ is 
obtained using pseudo-experiments (toy MC), where numbers of 
observed signal events are generated using the fitted function. They are
ﬂuctuated according to the data statistical and uncorrelated
systematic (see Sec.~\ref{xsfit})
uncertainties, and the uncertainties in the collider energy [see
Eq.~(\ref{desys})]. 

The systematic uncertainty in $a_\mu^{3\pi}$ is estimated by simultaneously
increasing or decreasing the number of observed events in all data points 
by the amount of systematic uncertainty calculated from the
cross-section systematic error listed in Tables~\ref{tab4a},
\ref{tab4b}, and \ref{tab4c}. 
We also study the dependence of the $a_\mu^{3\pi}$ result on the 
fitting model. Calculations are performed for all models
considered in Sec.~\ref{xsfit}. Deviations from the
nominal $a_\mu^{3\pi}$ value given in Table~\ref{tab5} do not exceed
$0.005\times 10^{-10}$.

The contribution to $a_\mu^{3\pi}$ from the energy region 1.1--1.975 GeV
is calculated using the SND data obtained in Ref.~\cite{snd-3pi-4}. In
this case the integral (\ref{amu}) is evaluated by using the trapezoidal rule.
The obtained $a_\mu^{3\pi}$ for the energy region 1.1--1.975 GeV and for
region 0.61--1.975 are listed in Table~\ref{tab5}, where they are
compared with the results of previous measurements.
\begin{table}
\caption{Values of $a_\mu^{3\pi}$ for different energy intervals. The
first three rows represent the SND results. The next three show the
BABAR results~\cite{BABAR-3pi}. The seventh row is the Belle II 
result~\cite{belle-3pi}, while the last row is the calculation~\cite{HHK}
based on the fit to $e^+e^-\to \pi^+\pi^-\pi^0$ data obtained before 
2021~\cite{cmd-3pi-1,cmd-3pi-2,snd-3pi-1,snd-3pi-2,BABAR-3pi-0,snd-3pi-3}.
\label{tab5}}
\begin{ruledtabular}
\begin{tabular}{lc}
$E$, GeV & $a_\mu^{3\pi}\times 10^{10}$ \\
\hline
0.62--1.10 (SND)  & $42.96\pm0.06\pm 0.46$ \\
1.10--1.975 (SND) & $2.99\pm 0.02\pm 0.08$ \\
0.62--1.975 (SND) & $45.95\pm0.06\pm 0.47$ \\
\hline
0.62--1.10 (BABAR) & $42.91\pm0.14\pm 0.55 \pm 0.09$ \\
1.10--2.00 (BABAR) & $2.95\pm 0.03\pm 0.16$ \\
$<2.00$    (BABAR)& $45.86\pm0.14\pm 0.58$ \\
\hline
0.62--1.80 (Belle II) & $48.91\pm0.23\pm 1.07$\\
$<1.8$~\cite{HHK} & $45.84\pm 0.58 \pm 0.22$ \\
\end{tabular}
\end{ruledtabular}
\end{table}

The SND values of $a_\mu^{3\pi}$ for $E<1.1$ GeV and $E>1.1$ GeV are in good
agreement with the BABAR measurements~\cite{BABAR-3pi}.
The SND value for $E<1.975$ GeV agrees well with the result of Ref.\cite{HHK}
based on $e^+e^-\to \pi^+\pi^-\pi^0$ data obtained before 
2021~\cite{cmd-3pi-1,cmd-3pi-2,snd-3pi-1,snd-3pi-2,BABAR-3pi-0,snd-3pi-3}
\footnote{This value is obtained by authors of Ref.~\cite{HHK}
basing on the fit results listed Tables 1 and 2 of Ref.~\cite{HHK}.},
but are $2.5\sigma$ smaller than the Belle II
measurement~\cite{belle-3pi}.

\section{Conclusion}
In the SND experiment at the VEPP-2000 collider, the measurement of the 
$e^+e^-\to\pi^+\pi^-\pi^0$ cross section has been performed in the center-of-mass
energy range from 560 to 1100 MeV. Our measurement is the most accurate to date.
Its systematic uncertainty is 0.9\% at maximum of the $\omega$
resonance and 1.2\% at maximum of the $\phi$ resonance. The SND
measurement is in the reasonable agreement with the BABAR
measurement~\cite{BABAR-3pi}, but about 7--8\% smaller than the Belle II data.
The leading-order hadronic contribution to the muon magnetic anomaly, 
calculated using the $e^+e^-\to\pi^+\pi^-\pi^0$ cross section measured
by SND in this work and in Ref.~\cite{snd-3pi-4} is $(45.95\pm0.06\pm
0.47)\times 10^{-10}$ for the energy interval 0.62-1.975 GeV.
Our $a_\mu^{3\pi}$ is in good agreement with the BABAR
result~\cite{BABAR-3pi} and calculation~\cite{HHK}
based on $e^+e^-\to \pi^+\pi^-\pi^0$ data obtained before
2021~\cite{cmd-3pi-1,cmd-3pi-2,snd-3pi-1,snd-3pi-2,BABAR-3pi-0,snd-3pi-3},
but is more precise. The difference with the $a_\mu^{3\pi}$ value
obtained by Belle II~\cite{belle-3pi} is $2.5\sigma$.

From the fit to the cross section data with the vector meson dominance 
model, the parameters of the $\omega$, $\rho$, and $\phi$ resonances
have been obtained. The obtained values of ${\cal B}(\omega\to e^+e^-){\cal
B}(\omega\to3\pi)$, mass and width of the $\omega$ meson, 
and ${\cal B}(\rho\to 3\pi)$
have accuracy better than the PDG world average values~\cite{pdg}.

\end{document}